\documentclass[ twoside]{article}
\usepackage[english]{babel}
\usepackage{color}
\usepackage{hyperref}
\usepackage{amsmath} % for align
\usepackage{graphicx}
\usepackage{enumerate}
\usepackage{amsfonts}
\usepackage{amssymb}
\newcommand{\sss}{\setcounter{equation}{0}}
\newtheorem{theorem}{THEOREM}[section]

\newtheorem{lemma}[theorem]{LEMMA}

\newtheorem{corollary}[theorem]{COROLLARY}

\newtheorem{remark}[theorem]{REMARK}

\newtheorem{prop}[theorem]{PROPOSITION}

\newtheorem{definition}[theorem]{DEFINITION}

\newtheorem{aharonov-bohm-ansatz}[theorem]{Aharonov-Bohm Ansatz}

\newcommand{\ere}{ {\mathbb R}}

\def\p2{\mathcal A_{\Phi,2\pi}(B)}
\def\0p2{\mathcal A_{\Phi,2\pi}(0)}
\def\sp2{\mathcal A_{\Phi,2\pi,\hbox{\rm SR}}(B)}

\def\beq{\begin{equation}}
\def\ene{\end{equation}}

\newcommand{\bull}{\hfill $\Box$}
\def\qed{\ifhmode\unskip\nobreak\fi\ifmmode\ifinner
\else\hskip5pt\fi\fi\hbox{\hskip5pt\vrule width4pt height6pt
depth1.5pt\hskip1pt}}
\def\v{\mathbf v}

\def\hv{\hat{\mathbf v}}

\def\curl{\, \hbox{ \rm curl}\,}
\def\mo{\mathbf p}

\def \tf{\tilde{\phi}}

\def \12{\tau_{1/ \sqrt{2},\sigma_1,\sigma_2}}
\def \32{\tau_{\sqrt{3}/ 2,\sigma_1,\sigma_2}}

\begin{document}
\baselineskip=20 pt
\parskip 6 pt

\title{ {High-Velocity Estimates for Schr\"odinger Operators in Two Dimensions: Long-Range Magnetic Potentials and Time-Dependent 
Inverse-Scattering.}
\thanks{ PACS Classification (2008): 03.65Nk, 03.65.Ca, 03.65.Db, 03.65.Ta.}
\thanks{ AMS Classification (2010): 81U40, 35P25
35Q40, 35R30.}}
 \author{ Miguel Ballesteros and Ricardo Weder  \thanks{Fellows of the Sistema Nacional de Investigadores.}   \\
 Departamento de F\'{\i}sica Matem\'atica. \\
 Instituto de Investigaciones en Matem\'aticas Aplicadas y en Sistemas.\\
  Universidad Nacional Aut\'onoma de M\'exico. Apartado Postal 20-126\\ IIMAS-UNAM, Col. San Angel, C.P. 01000, M\'exico D.F., M\'exico\\
miguel.ballesteros@iimas.unam.mx,  weder@unam.mx}

%\date{}
\maketitle

\begin{center}
\begin{minipage}{5.75in}
\centerline{{\bf Abstract}}
\bigskip

We introduce a general class of long-range magnetic potentials and
 derive high velocity limits for the corresponding scattering operators in quantum mechanics, in the case of two dimensions. 
We analyze the high velocity limits, that we obtain, in the presence of an obstacle and we uniquely reconstruct from them the electric potential and the magnetic field outside the obstacle, that are accessible to the particles. We additionally reconstruct  the inaccessible fluxes (magnetic fluxes produced by fields inside the obstacle) modulo $2 \pi$, what gives a proof of the Aharonov-Bohm effect. For every magnetic potential $A$ in our class we prove that its behavior at infinity ($A_\infty(\hv),  \hv \in \mathbb{S}^1$) can be characterized in a natural way; we call it the long-range part of the magnetic potential. Under very general assumptions  we prove that $A_\infty(\hv) + A_\infty(-\hv)$ can be uniquely reconstructed  for every $\hv \in \mathbb{S}^1$.    We characterize properties of the support of the magnetic field outside the obstacle that permit us to uniquely reconstruct  $A_\infty(\hv)$  either for all $ \hv \in \mathbb{S}^1$ or for $\hv$  in  a subset  of $\mathbb{S}^1$. We also give a wide class of magnetic fields outside the obstacle allowing us to uniquely reconstruct  the total magnetic flux  (and  $A_\infty(\hv)$ for all $ \hv \in \mathbb{S}^1$). This is relevant because, as it is well-known,  in general the scattering operator (even if is known for all velocities or energies) does not define uniquely the total magnetic flux  (and  $A_\infty(\hv)$ ). We analyze additionally injectivity (i.e., uniqueness without giving a method for reconstruction) of the high velocity limits of the scattering operator with respect to $A_\infty(\hv)$. Assuming that the magnetic field outside the obstacle is not identically zero, we provide a class of magnetic potentials for which injectivity is valid. 
  
\end{minipage}
\end{center}

\section{Introduction}\sss
We analyze scattering of {charged particles, for example electrons,} traveling in the exterior of a bounded obstacle $K$. We suppose there is a short-range electric potential $V$ and a magnetic field $B$ in {$\Lambda := \mathbb{R}^2 \setminus K $}. Inside the obstacle $K$ there is a magnetic field producing fluxes on each connected component. These fluxes are enciphered in the magnetic potentials (in $\Lambda$) through circulations around the boundary over each connected component of $K$. \\      
{It is assumed that  $|B(x)| \leq  \frac{C}{(1 +|x|)^{\mu}}$, for some constant $C$ and some $\mu > 2$}. This assumption is physically reasonable because the field produced by a magnetic dipole decays
as {$\frac{c}{|x|^3}$, for some constant $c$}, see \cite{grif}, as $|x| \to \infty$, and there is no magnetic monopole seen in nature.
In the physical world (the three dimensional case), assuming absence of magnetic monopoles, it is always possible to find a short-range magnetic potential satisfying the required circulations over the boundary of the obstacle (see \cite{b-w.1}). A magnetic potential $A$ is said to be short-range if
\beq \label{corto1}
|A(x)| \leq C \frac{1}{|x|^{1 + \epsilon}}
\ene        
as $|x|$ tends to infinity, where $C$ is a constant and $\epsilon > 0$. Otherwise it is long-range. Nevertheless, considering long-range magnetic potentials is also important for the following reason:A big portion of the work in scattering through magnetic potentials is done considering two dimensional models (see for example \cite{ab} and \cite{ruij}; \cite{op}-\cite{pt} for a review up to 1989; more recently \cite{w.1}, \cite{nico1}, \cite{EI2}-\cite{EI3}, \cite{ry1}-\cite{ry2}). These models approximate three dimensional situations in which long straight (finite) solenoids are regarded as infinite, producing translational-invariance in one spacial direction. The translation invariance permits the elimination of one degree of freedom, reducing the number of dimensions. The two dimensional models are frequently easier to analyze because explicit solutions in terms of special functions are available, in the case that another symmetries are assumed. In two dimensions the use of long-range magnetic potentials is unavoidable, unless the total magnetic flux is set to zero.     

In this paper we focus our attention to two dimensions. Our techniques and results are easily applicable to three dimensions using our constructions in \cite{b-w.1}.  \\     
{We use a system of units in which the charge of the electron, the speed of light and the Plank's constant have numerical value $1$: 
\beq \label{units}
e = 1, \hspace{1cm}   c = 1, \hspace{1cm} \hbar = 1.
\ene }
      
We introduce a general, and natural, class of magnetic potentials $A : \mathbb{R}^2 \mapsto \mathbb{R}^2 $ associated to the magnetic field $B$ and the magnetic fluxes over each connected component of the obstacle (see Definitions \ref{mp} and \ref{hvlwso-d.1}). We prove that 
our class of magnetic potentials permits to extract a mathematical object that describes the behavior of every potential ($A$) in this class at infinity. We call it the 
long-range  part ($A_\infty$) of $A$, it is given by: 
\begin{equation}\label{mmm}
A_\infty(\hv) = \lim_{s \to \infty } A(s\hv)s, \hspace{1cm} \forall \hv \in \mathbb{S}^1.  
\end{equation}            
{$A_\infty$ can be viewed in terms of a change of gauge from the Coulomb magnetic potential $A^{(c)}$; see Proposition \ref{coupo} and Corollary \ref{tasanan}: Set $\lambda_{A - A^{(c)}} $ such that $ \nabla \lambda_{A - A^{(c)}} = A - A^{(c)}  $ and define $\lambda_{A - A^{(c)}, \infty}(x) : = \lim_{r\to \infty} \lambda_{A - A^{(c)}} (rx) $ (see Remark \ref{lamda-infty}), then we have 
\beq \label{becio-st}
 A_\infty\big( (\cos(\theta), \sin(\theta)) \big) = 
 \Big (\frac{\Phi_B}{2\pi} + \frac{d}{d\theta}  \lambda_{A - A^{(c)}, \infty}( (\cos(\theta), \sin(\theta)) \Big ) \begin{pmatrix}
 - \sin(\theta) \\ \cos(\theta)
 \end{pmatrix},
\ene
where $\Phi_B$ is the total magnetic flux (see Definition \ref{def.tflux}). Eq. \eqref{becio-st} makes explicit the fact that the long-range part of a magnetic potential can be regarded as a physical quantity (the total flux) plus the gradient of a function, which shows the specific gauge we are working with.     
} 
Denoting by $v$ the average speed of the incoming electrons, we derive expressions for the first and second order contributions (in terms of powers of $\frac{1}{v}$) of the scattering operator applied to the incoming electrons, using time-dependent techniques. 
{These expressions are, respectively, given by the limits \eqref{retira1} 
and \eqref{retira2}, 
which we refer to as high-velocity limits of the scattering operator. The limits into consideration are calculated (with error bounds) in Theorems \ref{reconstruction-formula-I} and \ref{reconstruction-formula-ii-g-m-p}, respectively.} 
Our expressions are deduced first for a specific suitable gauge applying, with slight modifications, the analogous results for three dimensions in \cite{b-w.1}. {The core of our proof is the formula for general magnetic potentials, which uses (as an intermediate step) the estimation for the specific gauge above referred.}
{ Similar expansions are studied in \cite{b-w.1} (for short-range magnetic potentials), in \cite{nico1} for the Coulomb gauge ({using time-dependent methods \cite{Enss-Weder} and the stationary Isozaki-Kitada modifiers}) and 
in \cite{w.1} (where only the first order term is addressed and no electric potential is present). Related results for scattering in all space without magnetic fields are derived in \cite{Enss-Weder}, where the time-dependent inverse-scattering methods are introduced. Many other works using time-dependent inverse-scattering techniques prove analogous expansions (see \cite{wj.1} and  \cite{arians}, for example)}.\\ 
{We {uniquely reconstruct}, from the high-velocity limits  {\eqref{retira1}-\eqref{retira2}} of the scattering operator, information from the magnetic and electric potentials, assuming suitable hypothesis. Under very general conditions for the obstacle and the potentials we {uniquely reconstruct}, for every $\hv \in \mathbb{S}^1$,  
\beq \label{maseta}
A_\infty(\hv) + A_\infty(-\hv).
\ene 
Assuming that the magnetic field decays faster than any rational function (and that the obstacle consists on a bounded convex set $K_1$  plus a finite number of isolated dots) we {uniquely reconstruct} $B$ and the magnetic fluxes modulo $2 \pi$ over the connected components of $K$. This is what so far, to the best of our knowledge, can be recovered without the extra assumptions of knowing $A_\infty$ or $V$. Recovering more information about $A_\infty$ or $V$ requires the knowledge of $V$ or $A_\infty$, respectively. The reason is that in the high-velocity limit  {\eqref{retira2}} the long-range part of the magnetic potential and the electric potential are mixed. This is a subtle problem that appears only in the presence of long-range magnetic potentials. The same difficulty is present in \cite{EI2}-\cite{EI1}, where the high-velocity limits {\eqref{retira1}-\eqref{retira2}} are used to recover the electric potential. However, {it appears that} they were not aware of this problem and assumed that the high-velocity limits {\eqref{retira1}-\eqref{retira2}} determine the electric potential quoting \cite{b-w.1} (where only short-range magnetic potentials are used) and \cite{nico1} (where only the Coulomb gauge is studied and what is reconstructed is $\Phi_B B + V$ and not $V$, being $\Phi_B $ the total magnetic flux). Actually, a complete study of time-dependent high-velocity scattering for general long-range magnetic potentials is done for the first time here. {Under the extra assumption of the knowledge of $V$, the results of \cite{EI2}-\cite{EI1}  are complementary to ours (concerning injectivity).}        }
{We prove that, knowing $A_\infty$, $V$ can be {uniquely reconstructed} from {the high-velocity limits of the scattering operator we consider}, provided that $B$ and $V$ decay faster than any rational function and the obstacle consists on a bounded convex set $K_1$ plus a finite number of isolated dots. These assumptions for the obstacle and $B$ are required in all our reconstruction and injectivity results for $A_\infty$. {Uniquely reconstructing} $A_\infty$, knowing $V$, is more complicated. It is actually an important and subtle problem, because it is well-known that in general $A_\infty$ cannot be recovered from the scattering operator (see the comment below Theorem \ref{insecto1}). In this respect we tackle two different approaches: Injectivity and reconstruction methods. \\
Our results concerning injectivity  are closely related to \cite{EI2}-\cite{EI1}, although actually the problems are different because 
here we prove injectivity using only high-velocity scattering data (i.e. the limits {\eqref{retira1}-\eqref{retira2}}), while in \cite{EI2}-\cite{EI1} all energies (including high-velocity limits) are required.
 The ingredient we need to make the injectivity problem well-posed is the magnetic field not to be identically zero (otherwise injectvity is not true), then the high-velocity limits {\eqref{retira1}-\eqref{retira2}} of the scattering operator are injective
with respect to $A_\infty$ (assuming that it is real-analytic in the angular variable, this is the only result where this restriction is required). In 
\cite{EI2}-\cite{EI1} the additional restriction imposed to well-pose the injectivity problem is the total magnetic flux not to be an integer multiple of $2 \pi$.  They also assume that the obstacle is convex and a different class of magnetic potentials is considered: They do not ask for analyticity of $A_\infty$ but assume, instead, homogeneity (of degree $-1$) of the long-range magnetic potential.  \\               
The reconstruction results we prove require a different approach: We characterize the properties of the support of $B$ that allows us to  {uniquely reconstruct} $A_\infty$ totally or partially. Here we use a class of magnetic potentials that is considerably more general that the ones employed for injectivity in this text and in \cite{EI2}-\cite{EI1}. We prove in particular that if $B \ne 0$, then there is an open set in $\mathbb{S}^1$ where $A_\infty $ can be {uniquely reconstructed}. The class of magnetic fields that allows us to reconstruct $A_\infty$ totally is large. Actually not being able to {uniquely reconstruct} $A_\infty(\hv)$, for some $\hv$, by our method imposes strong restrictions to the magnetic field: It has to be compactly supported in the intersection of cylinders:
\beq
\bigcap_{\hat{\bf w} \in \mathcal{N}_{\hv}} \Big (K_1 + \mathbb{R} \hat{\bf w}\Big ), 
\ene  
for some open neighborhood $ \mathcal{N}_{\hv} $ of $\hv$.  
}

{   
\subsubsection{Description of the Model and Main Results} \label{modell}
\paragraph{\large{Description of the Model}} $\\ $
{\bf The Obstacle:} Here we give the main definitions of the mathematical objects we use related to the obstacle and its complement. In Section \ref{sec.obstacle} we prove some useful topological properties. We define the obstacle $K$ and to each connected component of it we introduce a curve surrounding it. Line integrals over this curves define the fluxes of the magnetic potentials we are interested in.    
\begin{definition}[The Obstacle]\label{o.1}{\rm
We denote by $ K $ the obstacle and by $ \Lambda = \mathbb{R}^2 \setminus K  $. We suppose that $  K $ is a compact subset of $ \mathbb{R}^{2}    $ and that its connected components are either points or closed sets with boundaries 
given by $ C^1 $-curves. We utilize the symbols $  K_{l}  $,  $ l \in \{1 \cdots L \}  $, for the connected components of $K$. We assume that 
$\{1 \cdots L \} = I \cup J$, where for every  $ i \in  I $ there is a point  $x^{(i)} \in \mathbb{R}$ such that  $K_{i}  = \{x^{(i)} \} $  and  for every 
 $ j \in  J $  $  \partial  K_j $  is given by a simple, closed,  $ C^1$-curve, that we denote by 
$  \gamma_{j} : [0, 1] \to \mathbb{R}^{2}  $. We suppose, furthermore, that the curves $ \gamma_j $, $ j \in  J $,  are oriented anti-clockwise. }\\
\end{definition}
Let
\begin{align}\label{dist}
{\bf d} : = \min\Big \{ {\rm d}(K_i, K_j) : i,j \in \{ 1, \cdots, L \}, i \ne j  \Big\},
\end{align}
 where ${\rm d}(\cdot, \cdot)$ denotes the distance between sets.
For all $ l \in I  $ we define the curve  $ \gamma_{l}: [0, 1] \to \mathbb{R}^2  $ by
the following equation:
\beq \label{drc.2}
\forall t \in [0, 1] \ : \:\hspace{3cm} \gamma_{l}(t) := x^{(l)} + \frac{{\bf d}}{4} e^{2 \pi  i t}. 
\ene
{\bf Classes of Magnetic Potentials, the Magnetic Field and the Magnetic Fluxes:}
We introduce general properties of the magnetic field, fluxes for the magnetic potentials on each curve $\gamma_l, \:  l \in \{ 1, \cdots, L \} ,$ and the total magnetic flux. More important, we define the two classes of magnetic potentials we use. The first one is given in Definition \ref{mp}. It is used to derive the first order term (in terms of powers of $\frac{1}{v}$, $v$ being the speed)  for the scattering operator (see Theorem \ref{reconstruction-formula-I}). The second class (Definition \ref{hvlwso-d.1}) is necessary for the second order term (Theorem \ref{reconstruction-formula-ii-g-m-p}). It, furthermore, allows us to prove the existence of the long-range part [see Eq. \eqref{mmm}]. In Section \ref{sectionmagnetic} we state and prove all properties and estimations we need for the magnetic potentials.         
\begin{definition}[Magnetic Field]\label{mf}{ \rm
The magnetic field is a measurable function $  B: \Lambda  \to \mathbb{R} $. 
We suppose that
\begin{align} \label{.}
 | B(x) | \leq C (1 + |x|)^{- \mu},  \hspace{1cm}  |\curl B (x)| \leq  C (  1 + |x| )^{-\mu},
\end{align} 
for some $ \mu > 2 $
and some constant number C. In Eq. \eqref{.} above the derivatives are taken in the distributional sense and	
\begin{align}
\curl B \equiv \nabla \times B   : = \Big (  \frac{ \partial  }{ \partial y } B, - \frac{\partial }{  \partial x } B  \Big ).  
\end{align} 
It is assumed that $ \curl B$ is a bounded measurable function. 
 In the case that 
\begin{align}
\int_{\Lambda} B = 0,
\end{align}
we call $B$  a  {\it {\bf short-range magnetic field}}. 
}  
\end{definition}
\begin{definition}[Flux]\label{def.flux}{\rm
The flux is a function $ \Phi : \{ K_{i}  \}_{i= 1}^{L}  \to \mathbb{R} $. }
Below, the quantity $\Phi(K_l)$ represents the total magnetic flux in the 
interior of the curve $\gamma_l $ (see Definition \ref{o.1} and \eqref{drc.2}). In the case that 
$l \in I$  (see Definition \ref{o.1}), it describes not only the magnetic flux inside $K_l$, but also the flux of the magnetic field $B$ inside the curve $\gamma_l$.
\end{definition}
\begin{definition}[Total Flux]\label{def.tflux}{\rm
For every magnetic field $B$ and every flux function $ \Phi : \{ K_{i}  \}_{i= 1}^{L}  \to \mathbb{R} $, we define  
\beq
\Phi_B : = \sum_{i \in  I} \Big ( \Phi(K_i) - \int_{\rm{int}(\gamma_i) } B \Big ) +
\sum_{j \in  J} \Phi(K_j)   + \int_{\Lambda} B
\ene
}
the total flux associated to the obstacle and the magnetic field $B$ 
(see Definition \ref{o.1}). Notice that $  \Phi(K_i) - \int_{\rm{int}(\gamma_i) } B   $ represents the magnetic flux in $K_i$, for every $i \in I$; see Definition \ref{def.flux}. We recall that $ \rm{int}(\gamma_i) $ is the interior of the curve 
$\gamma_i$.   
\end{definition}
\begin{definition}[First Class of Magnetic Potentials]\label{mp} {\rm
We denote by $  \mathcal{A}_{\Phi}(B) $ the set of functions $ A: \overline{\Lambda}  \to \mathbb{R}^{2}$ that satisfy
the following: 
\begin{itemize}
\item $A$ is continuous in $ \overline{\Lambda} \setminus \{  x^{(i)} \}_{i \in I } $ and it belongs to 
 $ L^1_{{\rm Loc}}( \overline{\Lambda}; \mathbb{R}^2) $.
\item $ |A(x)| \leq C \frac{1}{ 1 + |x|}, |x|\geq r_0$, for some constants $C, r_0 >0 $. 
The function $\zeta  : [0, \infty) \to \mathbb{R}$, defined by
\begin{align}\label{ar}
 \zeta(r) : = \max_{x \in \Lambda, |x|  > r} \Big \{  \Big | A(x) \cdot \frac{x}{|x|} \Big |  \Big \}, 
\end{align}
belongs to
 $ \in L^{1}(r_0, \infty)$, for some $r_0 > 0.$
\item $ \int_{\gamma_{k}}A = \Phi(K_k), $ for all $  k \in \{ 1, \cdots, L   \}$, and $ \nabla \times A = B$,   
\end{itemize} 
where the derivatives are taken in the distributional sense. }
\end{definition}
\begin{definition}\label{Diotaab}
For every $a, b \in [0, \infty)$ with $ a  + b > 2 $, we define the function
$\iota_{a,b}: \mathbb{R}^2 \to \mathbb{R}$: 
\begin{align}\label{iotaab}
\iota_{a,b}(x) : = \begin{cases}  \frac{1}{(1 + |x|)^{\min(a, b)}} + \frac{1}{(1 + |x|)^{a + b - 2}}  , & \text{if}\:\: a,b \neq 2,  \\
\frac{1}{(1 + |x|)^{2}} + \frac{\ln(e + |x|)}{(1 + |x|)^{a + b - 2}}, &  \text{
if  $ \: a= 2$ or $\: b = 2$}.  \end{cases}  
\end{align}
\end{definition}
\begin{definition}[Second Class of Magnetic Potentials]\label{hvlwso-d.1}{\rm
For every vector potential $ A \in \mathcal{A}_{\Phi}(B)$, we designate by $\alpha_A : \Lambda  \to \ere  $
the function
\begin{equation}\label{alphaA}
\alpha_A(x) := A(x) \cdot x , \hspace{2cm} \forall x \in \Lambda. 
\end{equation}
Let $\delta > 1$. We denote by $ \mathcal{A}_{\Phi, \delta}(B) $ the set of vector potentials $A \in \mathcal{A}_\Phi(B) \cap C^2(\Lambda,\ere^2 )$ such that  for every neighborhood $ \mathcal N$ of $K$
 there is a constant $C$ satisfying  
\begin{align}\label{hvlwso-d.1.e1} 
\Big |\frac{\partial }{\partial x_1} A(x) \Big | + \Big| \frac{\partial }{\partial x_2} A (x) \Big | & \leq C \frac{1}{(1 + |x|)^{2}}, \hspace{1cm} |\nabla \alpha_{A}(x) | \leq C \iota_{2, \delta}(x),  \hspace{1cm} |\alpha_{A}(x) | \leq C \iota_{1, \delta}(x),\\ \notag 
\Big |\frac{\partial}{\partial x_i} \frac{\partial}{\partial x_j} \alpha_{A}(x) 
\Big | & \leq C \min \Big( \iota_{3, \delta}(x), \frac{\ln(e + |x|)}{(1 + |x|)^2}\Big),
\end{align}
for all $x \in \Lambda \setminus \mathcal N $ and every $i,j \in \{ 1, 2\}$. }
\end{definition}
{\bf The Hamiltonians:}
We define the free and perturbed operators that we study. The free Hamiltonian $(H_0)$ is just the kinetic free energy for an electron with mass $m$ traveling in $\mathbb{R}^2$:
\begin{equation}\label{H0}
H_0  : = \frac{1}{2m} \mo^2, \hspace{1cm} \text{with} \hspace{.5cm} \mo = -i \nabla,  
\end{equation}
with domain $\bf H^2(\mathbb{R}^2)$, where for every open set $O$ in $\mathbb{R}^2$

{$\bf H^n(O)$}

is the Sobolev space of functions with derivatives up to order $n$ square integrable.    
We assume the presence of an electric potential $V$ satisfying the following:
\begin{definition}[Electric Potential]\label{electric-potential}
The electric potential is a real valued function   $ V \in L^2_{\textrm{loc}}(\Lambda)$  that
satisfies 
$$
\| F(|x | \geq r  ) V \mathcal{J} ( - \triangle + 1 )^{-1}    \| \leq C(1+ r)^{-\alpha}, 
$$
for some constants $ C > 0 $ and $ \alpha > 1 $, and every $ r \geq 0$. The symbol $ F(|x| \geq r) $ denotes the multiplication operator by the
 characteristic function of the set $ \{x : |x| \geq r \} $ and $\mathcal{J}:  L^2(\mathbb{R}^2)  \to  L^{2}(\Lambda)  $ is the multiplication operator
 by the characteristic function of the set $ \Lambda
$. We denote by $\bar{V}$ the extension of $V $ to $\mathbb{R}^2$ defined by 
zero outside $\Lambda$.
\end{definition}
The perturbed operator $H(A)$, for every $A\in \mathcal{A}_{\Phi}(B)$, is densely defined in the Hilbert $L^2(\Lambda)$ by the following:
\beq \label{HA}
H(A): = \frac{1}{2 m} \big(\mo- A\big )^2+  V .
\ene
A precise definition of $H(A)$ as a self-adjoint operator in a certain domain is given in Section \ref{ham}.   \\
{ \bf Wave and Scattering Operators:}
Here we define the wave and scattering operators. The proof of existence of wave operators
is done in Section  \ref{wavescatering}. Additionally we prove, in Section  \ref{wavescatering}, a change of gauge formula for the wave operators that directly leads us to the corresponding formula for the scattering operator in Eq. \eqref{scattering-oparetor-change-gauge}. The wave operators are given by the strong limit   
\beq \label{w-s-o.1}
W_{\pm}(A) \equiv W_{\pm}(A, V) := s-\lim_{t \to \pm \infty} e^{itH(A)} \mathcal{J} e^{-i t H_0}.
\ene 
\begin{definition}[Scattering Operator] \label{scatop}{\rm
The scattering operator is defined by the formula 
\beq\label{scattering-operator}
S(A) \equiv S(A, V) := W_{+}(A)^* W_{-}(A). 
\ene  
For every vector potentials  $A$ and $\tilde{A}$ belonging to $ \mathcal{A}_{\Phi}(B)$, such 
that $\tilde{A} - A = \nabla \lambda \equiv \nabla \lambda_{\tilde A - A} $, the change of gauge 
formula for the scattering operator
\beq \label{scattering-oparetor-change-gauge}
S(\tilde{A}) = e^{i \lambda_{\infty}(\mo)} S(A) e^{-i \lambda_{\infty}(-\mo)} 
\ene   
holds true, where {
$$   \lambda_\infty(x)   \equiv \lambda_{\tilde A - A, \infty}(x)  = \lim_{r \to \infty } \lambda(rx), $$ see Remark \ref{lamda-infty}} }.
\end{definition}

\paragraph{\large{Main Results}} $\\ $
{\bf High-Velocity Limits of the Scattering Operator (Reconstruction Formulae):}
We state our theorems giving asymptotic formulae of first and second order in 
$\frac{1}{v}$ ($v$ is the speed) for the scattering operator. The first order approximation is given in Theorem \ref{reconstruction-formula-I}, whose proof is derived at the end of Section \ref{highmagnetic}. The second order approximation is the content of Theorem \ref{reconstruction-formula-ii-g-m-p}, whose proof is done at the end of Section \ref{highpotential}. These two approximations define the high-velocity limits of the scattering operator we study {[see Eqs. \eqref{retira1}-\eqref{retira2}]}, and from them important information from the potentials can be uniquely reconstructed. We use the time-dependent methods for inverse-scattering initiated in \cite{Enss-Weder}. We introduce first some notations that are necessary to understand the theorems. We define:  
For every vector $ \v  \in \mathbb{R}^2\setminus \{ 0 \} $
\beq
\hv : = \frac{\v}{|\hv|}, \hspace{1cm} v := | \v |, \hspace{1cm}\Lambda_{\hv}:= \{x \in \Lambda: x+\tau \hv \in \Lambda,\, \forall \tau \in \ere\},\label{hvlwso.1}
\ene
and 
\beq \label{a}
a(A, \hv,x)\equiv a(\hv,x):= \int_{-\infty}^\infty \hv \cdot A(x+\tau \hv)  \, d\tau,
\ene
for $x \in \Lambda_{\hv}$.  \\
{
In the theorems below we compute (with error bounds) the following high-velocity limits of the scattering operator: For every $\v \in \mathbb{R}^2 \setminus\{0 \}$ and all compact subset $\Lambda_0$ of $\Lambda_{\hv}$,

\begin{itemize}
\item 
\begin{align}\label{retira1}
\lim_{v \to \infty } e^{-im\v\cdot x}\, S(A,V)\, e^{im\v\cdot x}  \phi_0,
\end{align}
 for all  $\phi_0 \in {\bf H}^2(\ere^2) $ with $ \hbox{\rm supp}\, \phi_0 \subset
\Lambda_0 $. 
\end{itemize}}
{

\begin{itemize}
\item \begin{align} \label{retira2}
\lim_{v \to \infty }v \left(  e^{-i m\v \cdot x } \left[S(A,V)- e^{ia(A, \hv,x)} \right]e^{i m\v \cdot x } \phi_{0}, \psi_{0}\right),%= %&
\end{align} 
for every $\phi_0,\psi_0 \in {\bf H}^6(\ere^2)$ supported 
in $\Lambda_0$. Here $(\cdot, \cdot)$ represents the inner product in $L^2(\mathbb{R}^2)$. 
\end{itemize}
}
\begin{theorem} \label{reconstruction-formula-I}{\bf (Reconstruction Formula I)}
 Let  $\Lambda_0$ be a compact subset 
of $\Lambda_{\hv} $, with
$\v \in \ere \setminus \{0\}$. Then, for all flux $\Phi$ and  all $A \in \mathcal{A}_\Phi(B)$ (see Definition \ref{mp})
there is a constant $C$ such that 
\beq
\left\| \left( e^{-im\v\cdot x}\, S(A,V)\, e^{im\v\cdot x} -
e^{i \int_{-\infty}^\infty \, \hv \cdot A(x+\tau \hv )\,d\tau}
 \right) \phi_0
\right\|_{L^2(\ere^2)}\leq
 C \frac{1}{v} \| \phi_0\|_{{\bf H}^2(\ere^2)},
\label{hvlwso.18}
\ene
\beq
\left\| \left(e^{-im\v\cdot x}\, S(A,V)^\ast\, e^{im\v\cdot x} -  e^{-i \int_{-\infty}^\infty \,
\hv \cdot A(x+\tau \hv )\,d\tau} \right )  \phi_0
\right\|_{L^2(\ere^2)}\leq
 C \frac{1}{v} \| \phi_0\|_{{\bf H}^2(\ere^2)},
\label{hvlwso.19}
\ene
for all  $\phi_0 \in {\bf H}^2(\ere^2) $ with $ \hbox{\rm supp}\, \phi_0 \subset
\Lambda_0 $. \\
\end{theorem}
Equations \eqref{hvlwso.18}-\eqref{hvlwso.19} are previously obtained, under different conditions, in \cite{arians}, \cite{b-w.1} and \cite{w.1}. 

\begin{theorem}\label{reconstruction-formula-ii-g-m-p} {\bf(Reconstruction Formula II. General Magnetic Potentials)} {
Suppose that $ B \in C^{2}(\overline \Lambda)$ is such that 
$ | B(x)| \leq C\frac{1}{(1 +|x|)^\mu} $, $ | \frac{\partial}{\partial x_i} B(x)| \leq C \frac{1}{(1 + |x|)^{\mu + 1}} $, $ | \frac{\partial}{\partial x_j}\frac{\partial}{\partial x_i} B(x)| \leq C \frac{1}{(1 + |x|)^{\mu + 2}} $, for every  $i, j \in \{ 1, 2 \}$ and every $x \in \Lambda$}.  Let $\tilde \delta > 1$ and  $A \in \mathcal{A}_{\Phi, \tilde \delta}(B)$. Set 
$\delta = \min(\mu-1, \tilde \delta) $. 
Let  $\Lambda_0$ be a compact subset of $\Lambda_{\hv}$,
  with $\v \in \ere \setminus \{0\}$, and {  
  $\phi_0,\psi_0 \in {\bf H}^6(\ere^2)$ be supported in $\Lambda_0$ . Then the following estimate holds true: 
\begin{align} \label{e-p-t.1.1}
v \left(  e^{-i m\v \cdot x } \left[S(A,V)- e^{ia(A, \hv,x)} \right]e^{i m\v \cdot x } \phi_{0}, \psi_{0}\right)= &
\left(i e^{i a(A, \hv, x) } A_{\infty}(- \hv)\cdot \frac{\mo }{m}  \phi_0, \psi_0 \right) + 
\left(i e^{i a(A, \hv, x) } \phi_0, A_{\infty}( \hv)\cdot \frac{\mo }{m}   
   \psi_0 \right) \\ \notag &
+ \left(-i e^{ia(A, \hv,x)}\int_{-\infty}^\infty
V(x+\tau\hv)\,d\tau \, \phi_0,\, \psi_0\right)\\ \notag &
+
\left( -i e^{i a(A, \hv, x) }\int_{-\infty}^0\, \Xi_\eta (x+\tau \hv,-\infty)\,d\tau \,\phi_0,\psi_0\right) \\ \notag & +
\left(-i \int_{0}^\infty\, \Xi_\eta (x+\tau \hv,\infty)\,d\tau \, e^{ia(A, \hv, x)} \phi_0,\psi_0\right) \\ \notag &
+ R (\v, \phi_0, \psi_0) +  R_L (\v, \phi_0, \psi_0).
\end{align} }
The expression $\Xi_\eta$ depends only on $B$. It is defined in \eqref{hvlwso.2}-\eqref{hvlwso.21}. 
\begin{align}\label{e-p-t.1.2}
\left|R(\v, \phi_0,\psi_0)\right| \leq C \|\phi_0 \|_{{\bf H}^6(\ere^2)}\, \|\psi_0 \|_{{\bf H}^6(\ere^2)}
 \begin{cases} \frac{1}{v^{ \min(\mu-2, \alpha -1)}}, & \hbox{\rm if}\, \min (\mu-3, \alpha -2)
 <0, \\ \\
\frac{|\ln v|}{v}, & \hbox{\rm if}\, \min (\mu-3, \alpha -2)=0, \\ \\ 
\frac{1}{v}, & \hbox{\rm if}\, \min (\mu-3, \alpha -2) > 0 
\end{cases}
\end{align}
and, for every $q \in (0, 1)$, there is a constant $C$ such that  
\begin{align}  \label{e-p-t.1.3}
\Big |R_L(\v, \phi_0,\psi_0)\Big | \leq  C \|\phi_0 \|_{{\bf H}^6(\ere^2)}\, \|\psi_0 \|_{{\bf H}^6(\ere^2)}    \frac{1}{v^q }.
\end{align}
 In case that $\delta > 2$,
\begin{align}  \label{e-p-t.1.3prima}
\Big |R_L(\v, \phi_0,\psi_0)\Big | \leq  C \|\phi_0 \|_{{\bf H}^6(\ere^2)}\, \|\psi_0 \|_{{\bf H}^6(\ere^2)}    \frac{1}{v},
\end{align}
for some constant $C$.
\end{theorem}
Formula \eqref{e-p-t.1.1} is proved in \cite{b-w.1} (for the three dimensions) in the short-range case and,  
{ using time-dependent methods \cite{Enss-Weder} and the stationary Isozaki-Kitada modifiers}, in \cite{nico1} for the Coulomb magnetic potential, a convex obstacle, and a $C_0^\infty$ magnetic field. After a long computation one verifies that the formula in \cite{nico1}, derived for the Coulomb potential, coincides with ours. Related results for scattering in all space without magnetic fields are derived in \cite{Enss-Weder}, where the time-dependent inverse-scattering methods are introduced. Many other works using time-dependent inverse-scattering techniques prove analogous expansions (see \cite{wj.1},  \cite{arians} and \cite{w.1}, for example).     \\
\noindent {\bf {Unique Reconstruction} of $B$, $V$ and the Fluxes (Modulo $2 \pi$):} 
The next Theorem is proved in Section \ref{recelectric}: 
\begin{theorem}\label{r-e-p}
{We assume that the set $J$ defined in Definition \ref{o.1} equals $ \{ 1 \} $
 and that
$ K_{1} $ is convex. We suppose, furthermore, that $  \mathcal{P}(x)  B(x)$ is
bounded for every polynomial  $   \mathcal{P}(x)   $. Then, for every $A \in \mathcal{A}_{\phi}(B)$,  the high-velocity limit {\eqref{retira1}} of the scattering operator   $S(A,V)$ uniquely determines {(with a reconstruction method)} $B(x)$  for almost every $ x \in \mathbb{R}^{2} \setminus K_1 $ and the fluxes
$\Phi (K_i)  $ modulo $2 \pi$, for every $ i \in \{1, \cdots, L\}  $. \\
Suppose, furthermore, that
  $ 
     \mathcal{P}(x) V(x) (\mo^2 + 1)^{-1}   $ is bounded for every polynomial  $   \mathcal{P}(x)   $
}
and that {$ B \in C^{2}(\overline \Lambda)$ is such that $ |  B(x)| \leq C \frac{1}{(1 + |x|)^\mu} $, $ | \frac{\partial}{\partial x_i}  B(x)| 
\leq C \frac{1}{(1 + |x|)^{\mu + 1}} $, $ | \frac{\partial}{\partial x_j}\frac{\partial}{\partial x_i}  B(x)| \leq C \frac{1}{(1 + |x|)^{\mu + 2}} $, for every 
$i, j \in \{ 1, 2 \}$ and every $x \in \Lambda $}. Let $\tilde \delta > 1$ and  $A \in \mathcal{A}_{\Phi, \tilde \delta}(B)$. Assume additionally that $A_\infty$ is known. Then, the high-velocity limits {\eqref{retira1}-\eqref{retira2}}  of the  scattering operator $ S(A,V) $,  known  for all unit vectors $\hv$ and all $\phi_0\in {\bf H}^6(\ere^2)$ with  ${\rm supp } (\phi_0) \subset   \Lambda_{\hv} $, uniquely determine {(with a reconstruction method)} $V(x)$  for almost every $ x \in \mathbb{R}^{2} \setminus K_1 $. 
\end{theorem}
 {We assume below (until the beginning of Section \ref{notex}) { that $ B \in C^{2}(\overline \Lambda)$ is such that $ | B(x)| \leq C\frac{1}{(1 +|x|)^\mu} $, $ | \frac{\partial}{\partial x_i} B(x)| \leq C \frac{1}{(1+|x|)^{\mu + 1}} $, $ | \frac{\partial}{\partial x_j}\frac{\partial}{\partial x_i} B(x)| 
 \leq C \frac{1}{(1 +|x|)^{\mu + 2}} $, for every  $i, j \in \{ 1, 2 \}$ and every $x \in \Lambda $}.  }
\\
{\bf {Unique Reconstruction} of $  A_{\infty}(\hv) + A_{\infty}(- \hv) $ under General Conditions:}
Theorem \ref{Theorem-Inconsistencies-1} below gives important information from the long-range part of the magnetic potential that we can {uniquely reconstruct} under very general circumstances. It is proved in Section \ref{pmdre}. To our knowledge this is the first time that such a quantity is recovered under the conditions we specify. To obtain more information we need to know the electric potential. The proof of injectivity, with respect to the long-range part of the magnetic potential, is addressed in 
 \cite{EI2}-\cite{EI1}, where the knowledge of the electric potential is also necessary {(see explanation above Section \ref{modell})}.
     
\begin{theorem}\label{Theorem-Inconsistencies-1}

Let $\tilde \delta > 1$ and  {$A \in \mathcal{A}_{\Phi,\tilde  \delta}(B)$}. 

We can {uniquely reconstruct}, from the high-velocity limits {\eqref{retira1}-\eqref{retira2}} of the scattering 
operator , 
\beq \label{masmenos}
 A_{\infty}(\hv) + A_{\infty}(- \hv),
\ene
for every $\hv \in \mathbb{S}^1$.
\end{theorem}
{\bf Injectivity With Respect to the Long-Range Part of the Magnetic Potential $A_\infty$, Knowing $V$:}
{Here we consider the problem of injectivity; namely we prove uniqueness without giving reconstruction methods.}
{We assume below (until the beginning of Section \ref{notex}) that the set $J$ defined in Definition \ref{o.1} equals $ \{ 1 \} $,
$ K_{1} $ is convex and that  $  \mathcal{P}(x)  B(x)$ is bounded for every polynomial  $   \mathcal{P}(x)   $.}  Now we state our Theorem of this part (see Section \ref{label} for the proof, in particular we refer to Theorem \ref{insecto}). This theorem is closely related to the results in \cite{EI2}-\cite{EI1}. In 
  \cite{EI2}-\cite{EI1} the requirement of knowing $V$ is also necessary (see explanation above Section \ref{modell}). The classes of magnetic potentials used here and in \cite{EI2}-\cite{EI1} are different and complementary (see the text at the beginning of Section  \ref{label}  for details). Additionally, in \cite{EI2}-\cite{EI1} the obstacle is assumed to be convex. Here we prove injectivity using only high-velocity scattering data  {[the limits \eqref{retira1},\eqref{retira2}]}, while in \cite{EI2}-\cite{EI1} all energies are required. It is an interesting fact that
  injectivity, with respect to the long-range part of the magnetic potential, is in general not valid. To have injectivity we assume that $B \ne 0$. However, in \cite{EI2}-\cite{EI1} a different assumption is needed: The total flux is not an integer multiple of $2 \pi $.            
\begin{theorem}\label{insecto1}
Let $\delta > 1$ and $A \in \mathcal{A}_{\Phi, \delta}(B), \tilde A \in \mathcal{A}_{\tilde \Phi, \delta}(\tilde B)$ such that $A_\infty - \tilde A_\infty $ is real analytic in the angular variable. Suppose that $B \ne 0$. {If the limits 
\eqref{retira1}-\eqref{retira2} coincide for $S(A, V) $ and $ S(\tilde A, V)$, then $B = \tilde B$, $\Phi_B = \tilde \Phi_{\tilde B}$
 and $A_\infty = \tilde A_\infty$. }
\end{theorem}
{\bf Total and Partial {Unique Reconstruction} of the Long-Range Part of the Magnetic Potential $A_\infty$, Knowing $V$:}
{
It is well-known that, in the absence of magnetic field {$B$ outside the obstacle}, {the scattering operator (even if it is known for all energies) does not uniquely determine the total magnetic flux. Actually, in the case that the obstacle is one point, the explicit calculations in \cite{ruij} and \cite{ab} show that the scattering operator is the identity if the total flux is an even multiple of $2 \pi$ and it is minus the identity if the total flux is an odd multiple of $2 \pi$. Additionally, formula \eqref{becio-st} implies that  the long-range part of the magnetic potential is not uniquely determined by the scattering operator, in general. However, if the magnetic field does not identically vanish, we {uniquely reconstruct} the 
long-range part of the magnetic potential in certain directions (depending on where the magnetic field vanishes).} Moreover, for a big class of magnetic fields, we {uniquely reconstruct} the whole long-range part.  We additionally prove that to every long-range magnetic potential a short-range magnetic potential can be added in order to {uniquely reconstruct} the full long-range part from the corresponding scattering operator.}  The main result in this part is 
Theorem \ref{elprinc}, whose proof is derived in Section \ref{rec}.      
{
\begin{definition} \label{deft}
For every open set $O $ in $\mathbb{S}^1$ we denote by 
\beq \label{eca}
\mathcal{C}(O) : = \bigcap_{\hat{\bf w} \in O}\big ( K_1 + \mathbb{R} \hat{\bf w}\big ).
\ene
We denote by $$\mathcal{D}(B) \subset \mathbb{S}^1$$ the set of vectors $ \hv \in \mathbb{S}^1 $  such that there is an open neighborhood $ \mathcal{N}_{\hv} $ of $\hv$ with
\beq \label{def}
{\rm supp}(B) \subset \mathcal{C}(\mathcal{N}_{\hv}). 
\ene
\end{definition}
\begin{remark}\label{nopu}
{Notice that, for every open set $O$,  $ \mathcal{C}(O) $ is an intersection of closed convex cylinders. It follows that it is 
 convex and compact. In particular $\mathcal{D}(B) \ne \emptyset$ implies that $B$ is compactly supported. Moreover, it is geometrically
clear that if $B$ is not zero (up to a set of zero measure) in a neighborhood of $K_1$ or 
if there is a closed $C^1$-curve, whose interior contains $K_1$, where 
$B$ is not zero (up to a set of zero measure in the curve), then $ \mathcal{D}(B) = \emptyset $. }       
\end{remark}
\begin{theorem}\label{elprinc}
Let $\delta > 1$ and $A \in  \mathcal{A}_{\Phi, \delta}(B)$. Suppose that we know $V$. In the case that $ \mathcal{D}(B) = \emptyset$, we can uniquely reconstruct, from the high-velocity limits {\eqref{retira1}-\eqref{retira2}} of the scattering operator,   $A_\infty $ and $\Phi_B$. If $ \mathcal{D}(B) \ne \emptyset $, 
 the high-velocity limits {\eqref{retira1}-\eqref{retira2}} of the scattering operator uniquely determine {(with a reconstruction method)} partially $A_\infty$: $A_\infty(\hv) $ can be {uniquely reconstructed} for every $\hv \notin \mathcal{D}(B) $. In the case that 
 $B \ne 0$, it is always possible to {uniquely reconstruct} $A(\hv)$ for every $\hv$ in some open set in $\mathbb{S}^1$, from the high-velocity limits {\eqref{retira1}-\eqref{retira2}} of the scattering operator. 
\end{theorem} 
{
As we mention in the lines below Theorem \ref{insecto1}, it is not possible (in general) to recover $A_\infty$  from the scattering operator (even if it is assumed to be known for all energies).\\
Having $ \mathcal{D}(B) \ne \emptyset $ imposes strong restrictions to $B$ (see Definition \ref{deft} and Remark \ref{nopu}): Here we give some particular examples in which $ \mathcal{D}(B) = \emptyset$ (and therefore $A_\infty$ can be {uniquely reconstructed}): 
\begin{itemize}
\item $B$ is not compactly supported.
\item $B$ is not zero (up to a set of zero measure) in a neighborhood of $K_1$. 
\item There is a closed $C^1$-curve, whose interior contains $K_1$, where 
$B$ is not zero (up to a set of zero measure in the curve).
\item There is no cylinder of the form $ K_1 + \mathbb{R} \hv $, for some $\hv \in \mathbb{S}^1$, containing the support of 
 $B$.  
\end{itemize} 
} Additionally, in Proposition \ref{aquien} we prove that it is always possible to add a 
short-range magnetic potential in order to be able to fully reconstruct uniquely $A_\infty$ and $\Phi_B$.    
}
}

\subsection{Some Notation Explanations}\label{notex}
We describe some shorthand notations we use in this paper. We denote by $ C $
a generic non-specified constant. The symbol $C$ might depend on all physical parameters, but it cannot depend on the velocity ${\bf v}$. We denote by $B_r(0)$ the open ball of center zero and radius $r$.
We associate measurable functions $A : \mathbb{R}^2 \mapsto \mathbb{R}^2$ with $1-$differential forms as follows:   
$$
A(x_1, x_2) \equiv A_1(x_1, x_2) dx_1 + A_2(x_1, x_2)dx_2. 
$$
Similarly, we associate measurable functions $B : \mathbb{R}^2 \mapsto R$ with $2-$differential forms:
$$
B(x_1, x_2) \equiv B(x_1, x_2) dx_1 \wedge dx_2.  
$$
Using these identifications, we make sense of integrals of the form
$$
\int_{\mathcal M^1} A,
$$
where $\mathcal M^1$ is a one dimensional sub-manifold of $\mathbb{R}^2$ (or a curve), whenever the integral exists.   

We use frequently in this paper vector operations such as cross products and scalar products between vectors in 
$\mathbb{R}^2$ and scalars in $\mathbb{R}$. The way to understand this is the following: 
We identify vectors in $\mathbb{R}^2$ with vectors in $\mathbb{R}^3$
$$
(A_1, A_2 ) \equiv (A_1, A_2, 0) 
$$
and scalars in $ \mathbb{R} $ with vectors in $\mathbb{R}^3$    
$$
B \equiv  (0, 0, B). 
$$
With the help of these identifications we do vector operations using the equivalent forms in $\mathbb{R}^3$.  After the computations we identify the resulting vector in $\mathbb{R}^3$ with the corresponding one in $\mathbb{R}^2$ or $\mathbb{R}$.      

We use the standard notation $\mathbb{S}^1$ to denote the one dimensional sphere immersed in  $\mathbb{R}^2$. Furthermore, we identify the two dimensional Euclidean space with the complex plane: {

$\mathbb{C} = \mathbb{R} + i  \mathbb{R} \equiv \mathbb{R}^2$.\\  
  
Throughout this paper we denote by 
$\bar B$ a bounded measurable extension of $B$ to $R^2$ with the same fluxes as $B$:
$$
 \int_{K_j}\bar B = 0, \hspace{3cm} \forall j \in J. 
$$
In the case the $B \in C^k(\overline \Lambda)$, for some $k \in \mathbb{N} \cup \{ 0\}$, we assume additionally that $ \bar B \in C^k(\mathbb{R}^2) $. Recall that the existence of such extension is basically the definition of $  C^k(\overline \Lambda) $. \\
For all square integrable function $\phi_0 \in \ L^2(\ere^2)$ with compact support in
$\Lambda_{\hv}$, we define
\beq \label{hvlwso.23}
\phi_{\v} := e^{im\v\cdot x}\phi_0.
\ene }  
\section{The Obstacle}\label{sec.obstacle}\sss

\subsection{De Rham Cohomology of $\Lambda$}

For every $  j \in J $, we choose a fixed point $ x^{(j)}  =   (x^{(j)}_1, x^{(j)}_2)   \in \textrm{int}(K_j)   $, 
where $ \textrm{int}(K_j)  $ denotes the interior of the set $ K_j $ (see Definition \ref{o.1}). Given a point 
$  x  = (x_1, x_2)  \in \mathbb{R}^2\setminus \{ x^{(r)} \} $ and an integer $  r \in \{  1, \cdots,  L   \} $, we define 
\begin{equation}\label{drc.1}
A^{(r)}(x) :=  \frac{1}{2 \pi}  \frac{1}{|x - x^{(r)}|^2}\left( \begin{array}{ccc}  x^{(r)}_2 - x_2 \\ x_1  -  x^{(r)}_1   \end{array} \right).
\end{equation}
It is easy to verify that $  \nabla \times A^{(r)} (x) = \delta(x- x^{(r)}) $, with $\delta(x )$ the Dirac distribution. \\

\begin{lemma}\label{drc.3}
 For every function $ A \in C^{1}( \overline{\Lambda} \setminus \{  x^{(i)}  \}_{i \in I}; \mathbb{R}^2 ) $
 such that 
$  \nabla \times A = 0 $ and 
$$
\int_{\gamma_k} A  = 0, \ \ \ \ \   \forall  k \in \{ 1, \cdots L  \},    
$$
there exists a function $  \lambda \in C^2( \Lambda ) $ satisfying $  A = \nabla \lambda $. 
Moreover, we can take $  \lambda(x)  = \int_{C(x_0, x)} A $, where $ x_0 $ is a fixed point
 in $  \Lambda $ and the integral is taken over any differentiable curve $ C(x_0, x)   $ 
in $ \Lambda $ that connects the point $  x_0 $ with $  x $. 
\end{lemma} 

\noindent{\it Proof:} Let $ \gamma   $ be a simple, closed, differentiable curve in $\Lambda$.  We suppose that it is oriented anti-clockwise. Let 
$\epsilon < \frac{{\bf d}}{4}$
[see \eqref{dist}]  be such that 
$$
{\rm d}(K, \gamma) > \epsilon,
$$  
recall that the symbol ${\rm d(\cdot, \cdot)}$ represents the distance.
For every $ i \in I  $ we define the curve $  \gamma_i^{ \epsilon} : [0, 1]  \to  \mathbb{R}^2 $
by 
$$  \gamma_i^{ \epsilon}  (t)  = x^{(i)} + \epsilon  e^{2 \pi i t}, \hspace{3cm} \forall t \in [0,1].  
$$
Stokes' theorem implies that  
$$  \int_{  \gamma_i^{\epsilon}  } A =  \int_{  \gamma_i  } A = 0,  
$$  for every  $ I \in I $. Using Stokes' theorem again we find that
$$   \int_{\gamma} A = \sum_{  \big \{ j \in J : K_j \subset \textrm{int}(\gamma) \big \}   }    \int_{\gamma_j  } A  + 
 \sum_{  \big \{ i \in I : x^{(i)} \in \textrm{int}(\gamma) \big \}   }    \int_{  \gamma_i^{\epsilon}  } A = 0, 
 $$ 
 where the $  {\rm int} (\gamma) $ is the interior of the curve $\gamma$. Consequently, we can define 
$$  \lambda(x)  = \int_{C(x_0, x)} A, $$ 
where $ x_0 $ is a fixed point
 in $  \Lambda $ and the integral is taken over any differentiable curve $ C(x_0, x)   $ 
in $ \Lambda $ that connects the point $  x_0 $ with $  x $. It is clear that $ \nabla \lambda = A $ and that 
$\lambda$ satisfies the desired properties. 
\bull
\begin{prop}\label{drc.4}
Let $A \in  C^1(\overline{\Lambda} \setminus \{ x^{(i)} \}_{i \in I}; \mathbb{R}^2) $ be such that $  \nabla \times  A(x) = 0 $ for every $x \in \Lambda$.  
There exists a function $ \lambda \in C^2(\Lambda)$ satisfying [see Definition \ref{o.1},  \eqref{dist}-\eqref{drc.2} and \eqref{drc.1}],  
$$
A =   \sum_{r = 1}^{L }  \Big ( \int_{\gamma_r} A\Big ) A^{(r)} + \nabla \lambda. 
$$

\end{prop} 

\noindent{\it Proof:} By Stokes' theorem, for any $ k, r \in \{  1, \cdots L  \} $, 
\begin{equation}\label{drc.5}
\int_{\gamma_k} A^{(r)} = \delta_ {k,r},
\end{equation}
where $  \delta_ {k,r}  = 1 $ if $  k = r $ and it is zero otherwise.  To prove (\ref{drc.5}) we compute the integral explicitly.
 In the case that $ k \in J $, 
we calculate the integral over a small circle around $  x^{(k)} $  and use Stokes' theorem.  
The desired result follows from Lemma \ref{drc.3}, since  $    A  -  \sum_{r = 1}^{L } (\int_{\gamma_r} A) A^{(r)}   $ satisfies the hypotheses
 required by it.

\begin{remark}{\rm
If we identify functions $ A \in C^{\infty}(\Lambda; \mathbb{R}^2) $ with 1-differential forms as

$$
(A_1, A_2) \iff A_1 dx_1 + A_2 dx_2,  
$$ 
then $ \nabla \times A $ is identified with the exterior derivative of the differential form.  Proposition \ref{drc.4} (and its proof) implies that 
$  \{ A^{(r)} \}_{r \in \{ 1, \cdots, L  \}} $ defines a basis of the 1-de Rham Cohomology group of $\Lambda  $ 
(see \cite{dr, wa}). }

\end{remark}

\begin{remark}\label{drc.6}{ \rm
The conclusion of Proposition \ref{drc.4} is valid also if we suppose that $A \in  C^0(\overline{ \Lambda} \setminus \{ x_i \}_{i \in I}; \mathbb{R}^2) $,  
instead of $A \in  C^1(\overline{\Lambda} \setminus \{ x_i \}_{i \in I}; \mathbb{R}^2) $. In this case $ \lambda \in C^1(\Lambda)$. This can be proved using regularization arguments as it is done in 
the proof of Proposition 2.5 in \cite{b-w.1} . Actually we have an explicit formula for $ \lambda $:
$$
\lambda(x) =  \int_{C(x_0, x)} \left( A -  \sum_{r = 1}^{L } (\int_{\gamma_r} A) A^{(r)}  \right), 
$$
where $x_0$ is a fixed point in $    \Lambda  $ and $  C(x_0, x) $ is any curve in $\Lambda$ connecting the point $ x_0 $ with $ x $. }

\end{remark}

\section{The Magnetic Field and the Magnetic Potentials} \label{sectionmagnetic}\sss

\begin{remark} \label{lamda-infty} {\rm
For every $A \in \mathcal{A}_0(0)$ (here the flux $0 $ is the function that associates the number $0$ to every connected component of $K$) we denote by $\lambda_A$ the function constructed in Remark \ref{drc.6} such that 
$A = \nabla \lambda_A$ .
It can be proved (see \cite{w.1})   that for every $ x \in \ere^2 \setminus \{ 0 \}$  the limit 
\begin{align}\label{defilambdainfty}
\lambda_{A, \infty}(x) := \lim_{r \to \infty } \lambda_{A}(r x)
\end{align} 
exists. Clearly, $\lambda_{A, \infty}$ is an homogeneous function, $\lambda_{A, \infty}(\rho \,x)=  \lambda_{A, \infty}(x), \rho >0$.
}
\end{remark}

\begin{prop}\label{mfmp.prop.1} {\bf (The Cone Magnetic Potential)}
For every magnetic field $ B $, every $\hat{w} \in \mathbb{S}^1$ and every $\epsilon \in [0, \frac{\pi}{2}]$, there exists a magnetic potential $ A \in \mathcal{A}_{\Phi}(B)   $ satisfying the following properties: 
\begin{itemize} 
\item $A = A_1 + A_2.$
\item $A_1$ is continuous in $  \overline{\Lambda} $, $ \nabla \cdot A_1 = 0  $ in the distributional sense. 
\item {$| A_1(x) | \leq C \frac{1}{(1 + |x|)^{\min(2- \delta, \mu- 1)}}$,}  for every $ \delta > 0 $. 
\item $A_2 \in  C^{\infty} ( \overline{\Lambda} \setminus \{ x^{(i)}  \}_{i \in I }; \mathbb{R}^2) $.  
\item  The support of $ A_2 $  is contained in the cone 
$$  \mathcal{C}:= \{ x \in \mathbb{R}^2 : (x- Q_0 ) \cdot \hat{w} 
 > | x- Q_0 | \cos(\epsilon) \}, $$ 
for some suitable chosen $ Q_{0} \in \mathbb{R}^2 $.
\end{itemize}    
\end{prop}

\noindent{\it Proof:}
{We take the extension $\bar{B}$ of $B$ defined in Section \ref{notex}. We define $  \beta = \int_{\mathbb{R}^2} \bar{B} $. } Let $ h \in C_0^\infty (B_1(0)) $ be such that 
$ \int_{\mathbb{R}^2} h = \beta $.  We introduce
\begin{equation}\label{tildeB}
 \tilde{B} = \bar{B} - h, 
\end{equation}  
then $$ \int_{\mathbb{R}^2}   \tilde {B} = 0. $$  
Let $A_1$ be the Coulomb potential for $ \tilde{B} $ in $\mathbb{R}^2$ (see \cite{wj.1}):
\beq \label{coulomb-potetial}
A_1 (x) = \int_{\mathbb{R}^2} dy \frac{1}{2 \pi} \frac{1}{|x - y|^2} \Big( \begin{array}{ccc} - (x_2 - y_2) \\ x_1 - y_1  \end{array} \Big) 
 \tilde{B}(y).
\ene
It follows from  \cite{wj.1}, Proposition 2.6 and its proof, that
$ A_1 $ has the required properties and that $ \nabla \times A_1 = \tilde{B}$.  \\ 
Let $ \theta_0  \in [ 0, 2 \pi ) $ be such that 
$$ \hat{w} = (\cos(\theta_0), \sin(\theta_0) ). $$  
We denote
 by $ f \in  C^{\infty}(\mathbb{R}^2 \setminus \{ 0 \}) $ a function with the following properties.
For every $\theta \in [\theta_0, \theta_0 + 2\pi):$
\begin{itemize}
\item $ f(r x) = f(x), \hspace{2.5cm}$ for every $ x \in \mathbb{R}^2 \setminus \{ 0 \}  $ and $ r > 0 $.
\item $  f\big((\cos(\theta), \sin(\theta))\big) = 0, \hspace{1cm} $ if 
$ \theta \in [\theta_0, \theta_0 + \frac{\epsilon}{2}] \cup [\theta_0 + 2\pi - 
\frac{\epsilon}{2}, \theta_0 + 2 \pi)  $.
\item  $  f\big((\cos(\theta), \sin(\theta))\big) = \theta, \hspace{1cm} $ if $ \theta \in [\theta_0 + \epsilon, \theta_0 + 2\pi - \epsilon ]  $.
\end{itemize} 
For  every $ l \in \{ 1, \cdots L \}$  and every $ x \in \mathbb{R}^2 \setminus \{ x^{(l)} \} $, we define [see \eqref{drc.1} and \cite{w.1}],

\begin{equation}\label{mfmp.1}
A^{(l,f)} (x) = A^{(l)}(x) - \frac{1}{2 \pi } \nabla  f (x - x^{(l)}).
\end{equation} 
$ A^{(l,f)}$ is supported in the cone 
$$   \Big \{ x \in \mathbb{R}^2 : (x- x^{(l)}) \cdot \big( \cos(\theta_0), \sin(\theta_0) \big ) 
 \geq | x- x^{(l)} | \cos(\epsilon) \Big \} .$$  
 Furthermore, [see \eqref{drc.5}], 
\begin{equation}\label{mfmp.2}
\int_{\gamma_j} A^{(l,f)} = \delta_{l,j},  
\end{equation}
where $ \delta_{l,j}  $ is the delta of Kronecker.   \\
For every $ Q := (q_1, q_2) \in \mathbb{R}^2  $ and every $ x \in \mathbb{R}^2   $ we define (see \cite{w.1}) 
\begin{equation}\label{mfmp.3}
A^{(Q)} (x) :=  \left( \begin{array}{ccc}  q_2 - x_2 \\ x_1  -  q_1   \end{array} \right) \int_{0}^1 h( \tau x  + (1 - \tau ) Q ) \tau d \tau.
\end{equation} 
If we choose $ Q  $ far enough from the support of $h$, $  A^{(Q)} $ is supported in 
the cone  
$$  \Big  \{ x \in \mathbb{R}^2 : (x- Q) \cdot \big ( \cos(\theta_0), \sin(\theta_0) \big ) 
 > | x- Q | \cos(\epsilon) \Big \}. $$  
We define for $ x  \in \mathbb{R}^2 \setminus \{ x^{(l)}  \}_{l \in \{ 1, \cdots, L \}} $ (see Section
 \ref{sec.obstacle} and Definition \ref{def.flux}, see also \cite{w.1})

\begin{equation}\label{mfmp.4}
A_2(x) := A^{(Q)}(x) + \sum_{i \in \{ 1, \cdots, L \}} \Big ( \Phi (K_l) - \int_{\gamma_l } (A_1 + A^{(Q)} ) \Big )
A^{(l,f)}(x).    
\end{equation}
It is easy to see that we can choose a point $ Q_0  \in \mathbb{R}^2 $ such that $  A_2 $ is supported in the cone  
$$  \mathcal{C} = \Big \{ x \in \mathbb{R}^2 : (x- Q_0 ) \cdot \hat{w} 
 \geq | x- Q_0 | \cos(\epsilon)\Big \} .$$
 A straightforward calculation shows that $A_2$ has the desired properties.   
\begin{remark} \label{remshort}{\rm
In the case that $B$ is short-range (see Definition \ref{mf}) we have that
\begin{equation}\label{extra}
\int_{\mathbb{R}^2} \bar B = 0,
\end{equation}
and we can take $h=0$ in the proof of  Proposition \ref{mfmp.prop.1}. Then, $\tilde{B}= \bar B$ and
  $A_1$ [see \eqref{coulomb-potetial}] is the Coulomb magnetic potential in $\mathbb{R}^2$ associated to $B$: 
$
\nabla \times A_1 = B.
$
The fact that 
$$| A_1(x) | \leq C \frac{1}{ (1 +|x|)^{\min(2- \delta, \mu- 1)}},$$  
for every $ \delta > 0 $, shows that
$A_1$ is a short-range magnetic potential (it decays as $\frac{1}{|x|^{1 + \epsilon}}$, for some $\epsilon > 0$, at infinity). This justifies the name we give to the magnetic field, as 
short-range. It is actually impossible to find 
a short-range magnetic potential associated to $B$ if \eqref{extra} is not satisfied for some  extension of $B$, see \cite{wj.1}. Of course, if at least one of the connected components of $K$ has non-empty interior it is always possible to find an extension of $B$ such that \eqref{extra} is satisfied. }       
\end{remark}

\begin{corollary}\label{transversal-vector-potential}
For every magnetic field $ B $, there exists a magnetic potential
 $ A \in \mathcal{A}_{\Phi}(B)   $ such that 
$  A = A_1 + x \times A_T + \tilde{A}_2 $, where
\begin{itemize} 
\item $A_1$ is continuous in $  \overline{\Lambda} $, $ \nabla \cdot A_1 = 0  $ in the distributional sense. 
\item $| A_1(x) | \leq C \frac{1}{(1 + |x|)^{\min(2- \delta, \mu- 1)}}$,  for every $ \delta > 0 $. 
\item $A_T \in C^{\infty} ( \overline{\Lambda} \setminus \{ x^{(i)}  \}_{i \in I } )$ and 
$\tilde A_2 \in C^{\infty} ( \overline{\Lambda} \setminus \{ x^{(i)}  \}_{i \in I };\mathbb{R}^2)$. 
\item  $ \nabla \cdot (x \times A_T) = \mathcal{O}(\frac{1}{|x|^2}) $,  $ \nabla \cdot  \tilde{A}_2 = \mathcal{O}(\frac{1}{|x|^3}) $.
\end{itemize}  
\end{corollary}
\noindent{\it Proof:}
We use the functions $ A_1 $  and $ h $ defined in the proof of Proposition \ref{mfmp.prop.1}. 
We denote by $ A_{h, T}$ the  transversal gauge of $ h $ (see \cite{wj.1} Section 2.2). 
\begin{equation}\label{mfmp.5}
A_{h,T} (x) : = - x \times \int_0^1 \tau h(\tau x) d \tau. 
\end{equation}
We define
\begin{equation}\label{mfmp.6}
A = A_1 + A_{h,T} + \sum_{l = 1}^{L}\Big ( \Phi(K_l)  - \int_{\gamma_l}(A_1 + A_{h,T})   \Big) A^{(l)}. 
\end{equation}
Then $ A \in \mathcal{A}_{\Phi}(B)  $  and we can take 
\begin{equation}\label{mfmp.7}
A_T (x) =  - \int_0^1 \tau h(\tau x) d \tau - \sum_{l = 1}^{L}\Big ( \Phi(K_l)  - \int_{\gamma_l}(A_1 + A_{h,T})   \Big )\frac{1}{2 \pi} \frac{1}{| x - x^{(l)} |^2}. 
\end{equation}
and 
\begin{equation}\label{mfmp.8}
\tilde{A}_2 (x) =  \sum_{l = 1}^{L}\Big (\Phi(K_l)  - \int_{\gamma_l}(A_1 + A_{h,T})   \Big ) x^{(l)} \times (\frac{1}{2 \pi} \frac{1}{| x - x^{(l)} |^2}). 
\end{equation}
Using that  
$$ A^{(l)}(x) = -x \times (\frac{1}{2 \pi} \frac{1}{| x - x^{(l)} |^2}) + x^{(l)} \times  (\frac{1}{2 \pi} \frac{1}{| x - x^{(l)} |^2}) $$
 and Section 2.2 in \cite{wj.1} we prove that $A_T$ and $\tilde{A}_2$ have the required properties.

{ 

\begin{prop}[The Coulomb Magnetic Potential] \label{coupo}
For every magnetic field $ B $, and every $x^{(0)} \in K$, there exists a magnetic potential $ A^{(c)} \in \mathcal{A}_{\Phi}(B)   $ satisfying the following properties: 
\begin{itemize} 
\item 
\beq \label{cs}
A^{(c)} =  \frac{1}{2 \pi}  \frac{\Phi_B }{|x - x^{(0)}|^2}\left( \begin{array}{ccc}  x^{(0)}_2 - x_2 \\ x_1  -  x^{(0)}_1   \end{array} \right) + A^{(s)}.
\ene
\item $A^{(s)}$ is continuous in $ \Lambda $, $ \nabla \cdot A^{(s)} = 0  $ in the distributional sense. 
\item For every neighborhood $ \mathcal N$ of $K$ there is a constant $C$ such that
\beq \label{estas}
| A^{(s)}(x) | \leq C \frac{1}{(1 + |x|)^{\min(2- \delta, \mu- 1)}}
\ene
 for every $ \delta > 0 $ and every $x \in \mathbb{R}^2 \setminus \mathcal N$. 
\end{itemize} 
\end{prop}
}
{
\noindent \emph{Proof:}
We use the notation of the proof of Proposition \ref{mfmp.prop.1}. We define
\beq \label{AS}
A^{(c)}_h (x) := \int_{\mathbb{R}^2} dy \frac{1}{2 \pi} \frac{1}{|x - y|^2} \Big( \begin{array}{ccc} - (x_2 - y_2) \\ x_1 - y_1  \end{array} \Big) 
 h(y),
 \hspace{1cm} A^{(c)}_{\bar B} (x) := \int_{\mathbb{R}^2} dy \frac{1}{2 \pi} \frac{1}{|x - y|^2} \Big( \begin{array}{ccc} - (x_2 - y_2) \\ x_1 - y_1  \end{array} \Big) 
 \bar B(y).
\ene
Notice that
\beq \label{AS1}
A^{(c)}_{\bar B} = A_1 + A^{(c)}_{h} 
\ene 
is the Coulomb magnetic potential in $\mathbb{R}^2$ associated to the magnetic field $\bar B$. 
We take
\beq \label{AS3}
A^{(c)} :=  A^{(c)}_{\bar B} +  \sum_{l = 1}^{L }\Big ( \Phi(K_l)  - \int_{\gamma_l} A^{(c)}_{\bar B}  \Big )  A^{(l)}
\ene
and 
\beq \label{AS4}
A^{(s)} : =  A^{(c)}  - \frac{1}{2 \pi}  \frac{\Phi_B }{|x - x^{(0)}|^2}\left( \begin{array}{ccc}  x^{(0)}_2 - x_2 \\ x_1  -  x^{(0)}_1   \end{array} \right). 
\ene
{ Eqs. \eqref{drc.1}, \eqref{drc.5} and the well-known properties of the Coulomb gauge (in $\mathbb{R}^2$), see \cite{wj.1}, imply that, restricted to $\Lambda$,
\beq
\nabla \times A^{(c)} = B, \hspace{.5cm} \nabla \cdot A^{(c)} = \nabla \cdot A^{(s)}= 0, \hspace{.5cm} \int_{\gamma_l} A^{(c)} = \Phi(K_l), \hspace{.2cm} \forall l \in \{1, \cdots, L  \}.  
\ene 
We proceed now with the estimates. $A_1$ satisfies the conclusions of Proposition  \ref{mfmp.prop.1}, in particular 
\beq \label{particular}
| A_1(x) | \leq C \frac{1}{(1 +|x|)^{\min(2- \delta, \mu- 1)}}, \hspace{1cm} \text{ for every} \:\:  \delta > 0 . 
\ene
 Thus, we only need to analyze (see Definition \ref{def.tflux} and recall Stokes' theorem)
\begin{align} \label{AS5}
  A^{(c)} - A_1  - \frac{1}{2 \pi}  \frac{\Phi_B }{|x - x^{(0)}|^2}\left( \begin{array}{ccc}  x^{(0)}_2 - x_2 \\ x_1  -  x^{(0)}_1   \end{array} \right) = & 
\int_{\mathbb{R}^2} dy \frac{1}{2 \pi}\Bigg[ \frac{1}{|x - y|^2} \Big( \begin{array}{ccc} - (x_2 - y_2) \\ x_1 - y_1  \end{array} \Big) 
 h(y)  -  \frac{1}{|x - x^{(0)}|^2}\left( \begin{array}{ccc}  x^{(0)}_2 - x_2 \\ x_1  -  x^{(0)}_1   \end{array}  \right)h(y)\Bigg ]
 \notag \\ & + \sum_{l = 1}^{L }\Big ( \Phi(K_l)  - \int_{\gamma_l} A^{(c)}_{\bar B}  \Big )  \Bigg[  A^{(l)} - \frac{1}{2 \pi}  \frac{1}{|x - x^{(0)}|^2}\left( \begin{array}{ccc}  x^{(0)}_2 - x_2 \\ x_1  -  x^{(0)}_1   \end{array}  \right)  \Bigg ],
\end{align}
where we used that  $\int_{\mathbb{R}^2} h = \beta$.  
Eq. \eqref{estas} follows from \eqref{particular} and the fact that for every 
$r > 0$ there is a constant $C$ such that 
\beq \label{that}
\Bigg | \frac{1}{|x - z|^2} \Big( \begin{array}{ccc} - (x_2 - z_2) \\ x_1 - z_1  \end{array} \Big) 
   -    \frac{1}{|x - x^{(0)}|^2}\left( \begin{array}{ccc}  x^{(0)}_2 - x_2 \\ x_1  -  x^{(0)}_1   \end{array}  \right) \Bigg | \leq  
 C\Big (\frac{1 }{ 1 + |x| }\Big )^2, 
\ene
uniformly for every $|x - z| \geq r$ and $|x - x^{(0)}|\geq r$. Notice that the singularity in the integrand in \eqref{AS5} is integrable.}    
\bull
}

\subsection{Estimates for the Magnetic Potentials}

\begin{lemma}\label{Lemp.1}
Let $a, b \in [0, \infty)$ with $ a  + b > 2 $. There is a constant $C$ (depending on $a$ and $b$) such that (recall Definition \ref{Diotaab})
\begin{equation}\label{emp.1}
\int_{\mathbb{R}^2} dy \frac{1}{(1 + |y|)^a}  \frac{1}{(1 + |x-y|)^b} \leq 
C \iota_{a,b}(x),  \hspace{1cm} \forall x \in \mathbb{R}^2.  
\end{equation}
\end{lemma}
 
\noindent{\it Proof:} 
We first suppose that $ a \ne 2, b \ne 2 $. We integrate over the set $|y| \leq | x - y| $
\begin{align}\label{emp.3}
\notag \int_{ |y|\leq |x - y| } \, dy \, \frac{1}{(1 + |y|)^a}  \frac{1}{(1 + |x-y|)^b} \leq &  
\int_{ |y | \leq |x|/2} \, dy  \, \frac{1}{(1 + |y|)^a}  \frac{1}{(1 + |x|/2)^b} +
\int_{ |x|/2 \leq |y|\leq |x - y| }  \, dy  \, \frac{1}{(1 + |y|)^a}  \frac{1}{(1 + |y|)^b}
\\ \notag  \leq & 2 \pi 2^b  \frac{1}{(1 + |x|)^b}  \int_{0}^{|x|/2}  \, dr\,
\frac{1}{(1 + r)^{a -  1}}
+ 2 \pi \int_{ |x|/2 }^\infty  \, dr\,
 \frac{1}{(1 + r)^{a + b - 1}} \notag  \\
\leq  & 2 \pi \Big [ \frac{2^b}{|a-2|} \frac{1}{(1 + |x|)^b} \Big(  
\frac{1}{(1 + |x|/2)^{a-2}}  + 1 \Big )  + \frac{1}{a + b - 2} 
\frac{1}{(1 + |x|/2)^{a + b - 2}}\Big ] .  
\end{align}
Similarly we obtain 
\begin{align}
\label{emp.4}
\int_{ |y|\geq |x - y| } \, dy \, \frac{1}{(1 + |y|)^a}  \frac{1}{(1 + |x-y|)^b} \leq &  
 2 \pi  \Big [ \frac{2^a}{|b-2|} \frac{1}{(1 + |x|)^a} \Big(  
\frac{1}{(1 + |x|/2)^{b-2}}  + 1 \Big )   \\  \notag  & \hspace{3cm} +
 \frac{1}{a + b - 2} 
\frac{1}{(1 + |x|/2)^{a + b - 2}}\Big ] .  
\end{align}
Eqs. \eqref{emp.3}-\eqref{emp.4} imply \eqref{emp.1}, see Definition \ref{Diotaab}. The cases $a = 2$ or $b = 2$ are treated similarly.  

\bull

\begin{lemma}\label{specific-potentials}
{Suppose that $ B \in C^{2}(\overline{\Lambda})$. % and that $ \bar B(x)=B(x), x \in \Lambda$.
 Assume, furthermore, that there is a constant $C$ such that 
\begin{align} \label{Bderivatibes}
|  B(x)| \leq \frac{C}{ (1 + |x|)^\mu}, \hspace{.5cm}  \Big| \frac{\partial}{\partial x_i} B(x)\Big| \leq \frac{C}{ (1+ |x|)^{\mu + 1}},\hspace{.5cm}  \Big| \frac{\partial}{\partial x_j}\frac{\partial}{\partial x_i} B(x)\Big| \leq \frac{C}{(1 + |x|)^{\mu + 2}},  
\end{align}
for every
$i, j \in \{ 1, 2 \}$ and every $x \in \Lambda$.} The magnetic potentials defined in Propositions \ref{mfmp.prop.1} {and \ref{coupo}}, and Corollary \ref{transversal-vector-potential} belong to  
 $\mathcal{A}_{\Phi, \mu -1}(B)$ (see Definitions \ref{Diotaab} and \ref{hvlwso-d.1}).

\end{lemma}
\noindent{\it Proof:} We prove the statement in several steps. We first estimate the magnetic potential $A = A_1 + A_2$ derived in Proposition \ref{mfmp.prop.1}. We analyze separately
$A_1$ and $A_2$. { The results for the Coulomb magnetic potential in Proposition \ref{coupo} are a direct consequence of the analysis for $A_1$; we do not include, therefore, the proof}.  The magnetic potential constructed in  Corollary \ref{transversal-vector-potential} is studied in the last part of our proof. \\      
{\bf Estimations for the magnetic potential $A = A_1 + A_2$ derived in Proposition \ref{mfmp.prop.1}:  }\\
{\bf Analysis of $A_1$:}\\ 
Recall that $A_1$ is the Coulomb potential for $\tilde{B}$ in $\mathbb{R}^2$ [see \eqref{tildeB} and \eqref{coulomb-potetial}]: 
\beq \label{hvlwso.5.1}
A_1 (x) = -\frac{1}{2 \pi} \int_{\mathbb{R}^2} dy  \frac{x-y}{|x - y|^2} \times 
 \tilde{B}(y)  = - \frac{1}{2 \pi}  \int_{\mathbb{R}^2} dy \frac{y}{| y|^2} \times
 \tilde{B}( x - y).
\ene
Let $g \in C_0^\infty (\mathbb{R}^2)$ satisfy $g(y) = 1 $ for $|y| \leq 1 $ and $g(y) = 0$ for $|y| \geq 2$. We define 
\begin{align} \label{hvlwso.5.2} 
A_{1, 0}(x) := &- \frac{1}{2 \pi}  \int_{\mathbb{R}^2} dy \: g(y) \frac{y}{| y|^2} \times
\tilde{B}(x - y),  \\
A_{1, \infty}(x) := & - \frac{1}{2 \pi} \int_{\mathbb{R}^2} dy \: \big(1 - g(x - y)\big)\frac{x-y}{|x - y|^2} \times \notag
\tilde{B}(y), 
\end{align}
for every $x  \in \mathbb{R}^3$.
As  $g$ has compact support, there is a constant $C$ such that 
\begin{align}\label{A10}
\Big |\frac{\partial }{\partial x_1} A_{1, 0}(x) \Big | + \Big| \frac{\partial }{\partial x_2} A_{1,0} (x) \Big | \leq \frac{C}{(1 + |x|)^{\mu + 1}}, \hspace{.5cm} 
\Big | \Big( \frac{\partial}{\partial x_1}\Big )^{a_1} \Big(\frac{\partial}{\partial x_2}
\Big)^{a_2} \alpha_{A_{1,0}}(x)\Big| \leq  \frac{C}{ (1 + |x|)^{\mu - 1+ a_1 + a_2} }, 
\end{align}
for all $x \in \mathbb{R}^2$ and every $a_1, a_2 \in \{0, 1, 2 \}$ with $a_1+ a_2 \leq 2$ [see \eqref{alphaA}]. \\
Estimating $A_{1, \infty}$ is more complicated. 
We introduce a useful identity (see Equation below (43) in \cite{wj.1}):
\beq \label{hvlwso.5.4}
 -  \frac{x - y}{|x - y|^2}  =  - \frac{x}{|x|^2} + \frac{  (x \times y) \times ( x - y )- (x \cdot y) (x - y)}{|x|^2|x-y|^2}. 
\ene
Using Lemma \ref{Lemp.1} and \eqref{hvlwso.5.4} we obtain that there is a constant $C$ such that
\begin{align}
\Big |\frac{\partial }{\partial x_1} A_{1, \infty}(x) \Big | + \Big| \frac{\partial }{\partial x_2} A_{1,\infty} (x) \Big | \leq C \frac{1}{(1 + |x|)^{2}}, 
\end{align}
for all $x \in \mathbb{R}^2$. 
Lemma \ref{Lemp.1}, \eqref{hvlwso.5.4} and the scalar triple product 
\begin{align}\label{triplescalar}
 -  \Big( \frac{x - y}{|x - y|^2} \times \tilde B(y)\Big )\cdot x =  
 \Big( \frac{x - y}{|x - y|^2} \times x \Big ) \cdot \tilde B(y),
\end{align}
imply [see \eqref{iotaab}]:
\begin{align}\label{alphaA1infty}
|\alpha_{A_{1, \infty}}(x) | \leq C \iota_{1, \mu-1}(x),
\end{align}
for all $x \in \mathbb{R}^2$ and some constant $C$. Similarly, taking derivatives, we deduce:
\begin{align}\label{nablaalphaA1infty}
|\nabla \alpha_{A_{1, \infty}}(x) | \leq C \iota_{2, \mu-1}(x),
\end{align}
for all $x \in \mathbb{R}^2$ and some constant $C$. To estimate the second derivatives we additionally use the following: 
\begin{align}\label{dxidxjalphaa1infty}
\frac{\partial}{\partial x_i} \frac{\partial}{\partial x_j} A_{1, \infty}(x) =& 
- \frac{1}{2 \pi} \int_{\mathbb{R}^2} dy \: \Big [\frac{\partial}{\partial x_j} \frac{\partial}{\partial x_i} \big(1 - g(x - y)\big)\frac{x-y}{|x - y|^2} \Big ] \times 
\tilde{B}(y) \\ \notag  =& 
- \frac{1}{2 \pi} \int_{\mathbb{R}^2} dy \: \Big [\frac{\partial}{\partial x_j}  \big(1 - g(x - y)\big)\frac{x-y}{|x - y|^2} \Big ] \times 
\frac{\partial}{\partial y_i}\tilde{B}(y).
\end{align}
We obtain that
\begin{align}\label{ddA0infty}
\Big |\frac{\partial}{\partial x_i} \frac{\partial}{\partial x_j} \alpha_{A_{1, \infty}}(x) 
\Big | \leq C \min \Big( \iota_{3, \mu - 1}(x), \frac{\ln(e + |x|)}{(1 + |x|)^2}\Big),  
\end{align}
for all $x \in \mathbb{R}^2$, some constant $C$ and every $i,j \in \{ 1, 2\}$. 
Eqs. \eqref{A10}-\eqref{ddA0infty} imply that $A_1$ satisfies the estimates  \eqref{hvlwso-d.1.e1}  for $ \forall x \in \mathbb R^2$.

\noindent {\bf Analysis of $A_2$:}\\
We recall that $A_2$ is defined in \eqref{mfmp.4}. It is a linear combination of the vector potentials  $A^Q$ and $A^{(l, f)}$ (for $l \in \{1, \cdots, L \}$).  $A^Q$ and $A^{(l, f)}$ are explicitly given in \eqref{mfmp.3} and \eqref{mfmp.1}, respectively. Using this we prove that $A_2$ satisfies the estimates  \eqref{hvlwso-d.1.e1} .

\noindent {\bf Estimations for the magnetic potential derived in Corollary \ref{transversal-vector-potential}:}\\
The vector potential defined in Corollary \ref{transversal-vector-potential} has the form
$$  A_1 + x \times A_T + \tilde{A}_2, $$ 
where $A_1$ is the vector potential derived in Proposition \ref{mfmp.prop.1}. $A_1$ satisfies  the estimates  \eqref{hvlwso-d.1.e1}  for $ \forall x \in \mathbb R^2$. Proving that $  x \times A_T + \tilde{A}_2 $ satisfies  \eqref{hvlwso-d.1.e1}  is straightforward.
  
 \bull
 \begin{lemma}\label{e-p-l.1}
Let $\delta > 1$ and  $A \in \mathcal{A}_{0, \delta}(0)$. Take $r > 1$ such that $ K \subset B_r(0)$. For every $\rho \in (0, 1)$ there is a constant $C$ such that (recall Definition \ref{hvlwso-d.1}) 
\begin{align}\label{princ1}
\forall x \in \mathbb{R}^2,  |x|> 2r, \:  \forall   z \in \ere^2, |z| < \frac{|x|}{2} \: \: : &
 \\ \notag
\Big | \lambda_{A, \infty}(x + z) -  \lambda_{A, \infty}(x ) - & A(x)\cdot z - \int_1^\infty (\nabla \alpha_A)(\tau x)\cdot z d \tau \Big | \\ \notag   \leq \: & C\Big [ \frac{|z|^2}{(1 + |x|)^2}  +  \frac{\ln(e + |x|)}{(1 +  |x|)^{ 2 \rho + \min(1, \delta - 1) (1 - \rho)}  } |z|^{2\rho} (1 + |z|))^{1 - \rho} \Big ].
\end{align}
In the case that $\delta > 2$, the following holds true: 
\begin{align}\label{princ2}
\forall x \in \mathbb{R}^2,  |x|> 2r, \:  \forall   z \in \ere^2, |z| < \frac{|x|}{2} \: \: : &
 \\ \notag
\Big | \lambda_{A, \infty}(x + z) -  \lambda_{A, \infty}(x ) - & A(x)\cdot z - \int_1^\infty (\nabla \alpha_A)(\tau x)\cdot z d \tau \Big | \leq  C  \frac{|z|^2}{(1 + |x|)^2}.
\end{align}

\end{lemma}

\noindent {\it Proof:}
For every $x, y \in \mathbb{R}^2$, we denote 
\begin{align}
L(x, y) : = \Big \{  x + t (y-x) : t \in [0,1]  \Big \}
\end{align}
the line segment joining $x$ and $y$.\\  
We do first some computations, we take $\tau \geq 1 $. By the mean value theorem 
\beq \label{e-p-l.1.0.1}
\frac{1}{\tau} \Big ( \alpha_A(\tau(x + z)) - \alpha_A (\tau x) - (\nabla \alpha_A  )(\tau x)
 \cdot \tau z \Big ) = \frac{1}{\tau} \Big ( \nabla (\alpha_A)(\xi(\tau))\cdot \tau z - \nabla (\alpha_A  )(\tau x)
 \cdot \tau z \Big ) 
\ene 
for some  $ \xi(\tau)\in L(\tau x, \tau(x + z)) $. Using the mean value theorem again and Definition \ref{hvlwso-d.1}, we have 
\begin{align} \label{e-p-l.1.2}
|\nabla (\alpha_A)(\xi(\tau)) -  \nabla (\alpha_A  )(\tau x)|  
& \leq C \sup \Big \{\Big | \frac{\partial}{\partial x_i}  \frac{\partial}{\partial x_j} 
\alpha_A(y) \Big | : y \in L(\tau x, \tau (x + z)), \{i, j \} \subset \{ 1, 2 \}    \Big \} | \tau z| \\ \notag
& \leq C \frac{\ln(e + \tau |x|)}{(1 + \tau |x|)^{2} } |\tau z|.
\end{align}
We note that, for $|x|, \tau \geq 1$, 
\begin{align}\label{xtau}
1 + \tau |x| \geq \frac{1}{4}(1 + \tau)(1 + |x|), \hspace{.5cm}  \tau  \geq \frac{1}{2}(1 + \tau), \hspace{.5cm} \ln(e + \tau |x|) \leq 2 \ln(e + \tau) \ln(e + |x|)
\end{align}
In \eqref{e-p-l.1.0.1} we use \eqref{e-p-l.1.2} and \eqref{xtau} to obtain: 
\beq \label{e-p-l.1.0.2}
\Big | \frac{1}{\tau} \Big ( \alpha_A(\tau(x + z)) - \alpha_A (\tau x) - (\nabla \alpha_A  )(\tau x)
 \cdot \tau z \Big )\Big | \leq 32   \frac{\tau \ln(e + \tau) \ln(e + |x|)}{(1 + \tau)^2\,(1 + |x|)^{2} } |z|^2.
\ene 
 By Definition, \ref{hvlwso-d.1}  we have that 
\beq \label{e-p-l.1.0.3}
\Big |\frac{1}{\tau} \Big ( \alpha_A(\tau(x + z)) - \alpha_A (\tau x) - (\nabla \alpha_A  )(\tau x)
 \cdot \tau z \Big )\Big | \leq   C \Big (\frac{\ln(e + \tau)}{(1 + \tau )^{\min(2, \delta)} } 
 \frac{\ln(e + |x|)}{(1 + |x| )^{\min(1, \delta-1)} } (1 + |z|) \Big ).
\ene 
Interpolating (\ref{e-p-l.1.0.3}) and (\ref{e-p-l.1.0.2}), we get 
\begin{align} \label{e-p-l.1.0.4}
\Big |\frac{1}{\tau} \Big ( \alpha_A(\tau(x + z)) - \alpha_A (\tau x) - & (\nabla \alpha_A  )(\tau x)
 \cdot \tau z \Big ) \Big |   \\ & \notag \leq   C 
 \frac{\ln(e + \tau)}{(1 + \tau)^{\rho + \min(2, \delta) (1 - \rho)  }}
  \frac{\ln(e + |x|)}{(1 +  |x|)^{ 2 \rho + \min(1, \delta - 1) (1 - \rho)}  } |z|^{2\rho} (1 + |z|))^{1 - \rho}. 
\end{align} 
 Notice that $\rho \in (0, 1)$ implies that $ \rho + \min(2, \delta) (1 - \rho) > 1 $.\\
 By Remark \ref{drc.6} and \eqref{defilambdainfty},
 
 $$
 \lambda_{A, \infty}(x)= \int_{x_0}^x\, A(x)+ \int_{1}^\infty\, \frac{1}{\tau} \, \alpha_A(\tau x)\, d \tau,
 $$
 $$
 \lambda_{A, \infty}(x+z)= \int_{x_0}^x\, A(x)+ \int_0^1\, A(x+\tau z)\cdot  z \,d \tau+ \int_{1}^\infty\, \frac{1}{\tau}\,\alpha_ A(\tau (x+z))\, d \tau.
 $$
  Then, by \eqref{e-p-l.1.0.4}, and using again the mean value theorem, we obtain: 
\begin{align} \label{e-p-l.1.1}
\Big | \lambda_{A, \infty}(x + z) - & \lambda_{A, \infty}(x ) -  A(x)\cdot z - \int_1^\infty (\nabla \alpha_A)(\tau x)\cdot z d \tau \Big | \\ \notag 
 =\:  & \Big |\int_0^1 \Big (A(x+ \tau z) - A(x)\Big )\cdot z d\tau  + \int_1^\infty \frac{1}{\tau} \Big ( \alpha_A(\tau(x + z)) - \alpha_A (\tau x) - (\nabla \alpha_A  )(\tau x)
 \cdot \tau z \Big ) d \tau \Big | \\ \notag 
 \leq \:  &  C\Big [ \sup \Big \{ \Big | \frac{\partial}{ \partial x_j} A_i (x) \Big |:  x \in L(x, x+ z), \{ i, j \} \subset \{ 1, 2 \} \Big \} |z|^2 
  \\ \notag & \hspace{6cm} +   \frac{\ln(e + |x|)}{(1 +  |x|)^{ 2 \rho + \min(1, \delta - 1) (1 - \rho)}  } |z|^{2\rho} (1 + |z|))^{1 - \rho}\Big ] \\ 
  \notag \leq \: & C\Big [ \frac{|z|^2}{(1 + |x|)^2}  +  \frac{\ln(e + |x|)}{(1 +  |x|)^{ 2 \rho + \min(1, \delta - 1) (1 - \rho)}  } |z|^{2\rho} (1 + |z|))^{1 - \rho} \Big ], 
\end{align}
which gives \eqref{princ1}.
The introduction of $\rho$ is used to be able to integrate with respect to 
$\tau$, i.e.,  to have $ \rho + \min(2, \delta) (1 - \rho) > 1 $. In the case that $\delta > 2$ this is not necessary. We estimate using Definition \ref{hvlwso-d.1} as in \eqref{e-p-l.1.0.2}: 
\begin{align} \label{e-p-l.1.0.2tilde}
\Big | \frac{1}{\tau} \Big ( \alpha_A(\tau(x + z)) - \alpha_A (\tau x) - (\nabla \alpha_A  )(\tau x)
 \cdot \tau z \Big ) \Big | & \leq   C \frac{\tau  }{(1 + \tau |x|)^{\min(3, \delta)} } |z|^2
 \\ \notag  & \leq C \Big[ \frac{1}{(1 + \tau )^{\min(2, \delta - 1)} } 
 \frac{1}{(1 + |x| )^{\min(3, \delta)} }\Big ] |z|^2. 
\end{align}
Eq. \eqref{princ2} is a direct consequence of \eqref{e-p-l.1.0.2tilde} 
[we also use \eqref{e-p-l.1.1}].

\bull

\begin{corollary}\label{e-p-c.1}
Let $\delta > 1$ and  $A \in \mathcal{A}_{0, \delta}(0)$. For every $\hv \in \mathbb{S}^1$
the limit 
\beq \label{e-p-c.1.1}
A_\infty(\hv):=\lim_{\tau \to \infty}A(\tau \hv)\tau 
\ene 
exists {and it is continuous in $\mathbb{S}^1$}.
For all $r > 1$ with $ K \subset B_r(0)$ there is a constant $C$  such that for every $s \in [2 r, \infty)$
\begin{align}\label{estimAinfty}
\big | A_\infty(\hv) -  A( s \hv ) s \big | \leq C 
\frac{\ln(e+s)}{(1 + s)^{\min(1, \delta - 1)}}. 
\end{align}
\end{corollary}
\noindent{\it Proof:}
Take $r > 1$ such that $ K \subset B_r(0)$. 
Let  $\hv \in \mathbb{S}^1 $. 
Choose $z \in  \ere^2 $ and $s \in (2r, \infty) $. Suppose that  $v \geq 1 $ is such that $\frac{|z|}{v} < 1/2 $,  by Lemma \ref{e-p-l.1}
\begin{align} \label{e-p-c.1.1prima}
v \Big | \big (\lambda_{A, \infty}( \frac{s}{v} z  + s \hv ) & -\lambda_{A, \infty}(s \hv )\big ) 
- A(s \hv)\cdot \frac{sz}{v} -  \int_1^\infty (\nabla \alpha_A)(\tau s \hv )\cdot \frac{sz}{v}  d \tau \Big | 
\\ & \leq  \notag  C \Big [ 
\frac{1}{v} | z |^{2}   
 +  \frac{1}{v^{2 \rho - 1}}\frac{\ln(e + |s|)}{(1 +  |s|)^{ \min(0, \delta - 2) (1 - \rho)}  } |z|^{2\rho}  \Big].
\end{align}
By Definition \ref{hvlwso-d.1} (integrating by parts),  we have that 
\beq \label{e-p-c.1.2}
\Big | \int_1^\infty (\nabla \alpha_A)(\tau s \hv )\cdot s z d \tau \Big | \leq C |z| 
\frac{\ln(e + s)}{s^{\min(1, \delta - 1)}}.
\ene
Suppose that $ s_1 $ and $s_2$ are real numbers bigger than $2 r$ and $s_1 < s_2$. Since
$\lambda_{A, \infty}( \rho  x  ) = \lambda_{A, \infty}(  x  ) $ for every $\rho > 0$, we have that

$$
\begin{array}{l}
A(s_1 \hat{v})\cdot s_1 z - A(s_2 \hat{v})\cdot s_2 z=  -v(\lambda_{A, \infty}( \frac{s_1}{v} z  + s_1 \hv ) +\lambda_{A, \infty}(s_1 \hv )) 
+ A(s_1 \hv)\cdot s_1 z +  \int_1^\infty (\nabla \alpha_A)(\tau s_1 \hv )\cdot s_1 z d \tau \\\\
+ ( v(\lambda_{A, \infty}( \frac{s_2}{v} z 
 + s_2 \hv ) -\lambda_{A, \infty}(s_2 \hv )) 
- A(s_2 \hv)\cdot s_2z 
- \int_1^\infty (\nabla \alpha_A)(\tau s_2 \hv )\cdot s_2 z d \tau ) \\\\
-\left[\int_1^\infty (\nabla \alpha_A)(\tau s_1 \hv )\cdot s_1 z d \tau - \int_1^\infty (\nabla \alpha_A)(\tau s_2 \hv )\cdot s_2 z d \tau \right].
\end{array}
$$
By (\ref{e-p-c.1.1prima}), (\ref{e-p-c.1.2}) 
\begin{align} \label{e-p-c.1.3}
|A (s_1 \hv)\cdot s_1 z - A(s_2 \hv)\cdot s_2 z |\leq 
 & \: C |z| 
\frac{\ln(e + s_1)}{s_1^{\min(1, \delta - 1)}}   + C \Big [ 
\frac{1}{v} | z |^{2}   
 +  \frac{1}{v^{2 \rho - 1}}\frac{\ln(e + |s_2|)}{(1 +  |s_2|)^{ \min(0, \delta - 2) (1 - \rho)}  } |z|^{2\rho}  \Big].
\end{align}
Taking $v \to \infty$ we get  
\begin{align} \label{e-p-c.1.3prima}
|A (s_1 \hv)\cdot s_1 z - A(s_2 \hv)\cdot s_2 z | \leq 
 & \: C |z| 
\frac{\ln(e + s_1)}{s_1^{\min(1, \delta - 1)}} .
\end{align}
Taking  $z=(1,0)$ and $z=(0,1)$  we prove that  $A_\infty(\hv):=\lim_{\tau \to \infty}A(\tau \hv)\tau$ exists and that \eqref{estimAinfty} holds. {The continuity of $A_\infty$ follows from \eqref{estimAinfty} and the fact that $A(s\hat v )s$ is continuous as a function of $\hat v \in \mathbb{S}^1$.}
\bull
{
\begin{definition}\label{ainftyx}
Let $\delta > 1$ and $ A \in \mathcal{A}_{0, \delta}(0) $. For every $x  \in \mathbb{R}^2 \setminus \{  0\}$, we define 
$$A_\infty(x) : = \frac{1}{|x|} A_\infty \Big(\frac{x}{|x|}\Big ).
$$
\end{definition}}
{
\begin{remark} \label{perp} 
It is a direct consequence of Definition \ref{hvlwso-d.1}  that $A_\infty$ is transverse: 
\begin{align} \label{trans}
A_{\infty}(x) \cdot x = 0, \hspace{3cm} \forall x \ne 0.
\end{align} 
\end{remark} 
}
{
\begin{prop}\label{grad}
Let $\delta > 1$ and  $A \in \mathcal{A}_{0, \delta}(0)$. The function 
$\lambda_{A, \infty} $ is differentiable in $x \in \mathbb{R}^2\setminus \{ 0  \} $. It follows, furthermore, that 
\beq \label{grad1}
\nabla \lambda_{A, \infty}(x) =  A_\infty(x). \ene 
In particular
\beq \label{grad2}
\frac{d}{d \theta} \lambda_{A, \infty}\Big ((\cos(\theta), \sin(\theta))\Big)
 =  A_\infty\Big ((\cos(\theta), \sin(\theta)) \Big) \cdot 
\begin{pmatrix}
- \sin(\theta) \\ \cos(\theta)
\end{pmatrix}. 
\ene
\end{prop} 
\emph{Proof:}
Set $x = r(\cos(\theta), \sin(\theta)) = r \hat x $, $(\hat x)^{\perp} = (- \sin(\theta), \cos(\theta))$ and $ y = x + z = r( \cos(\vartheta), \sin(\vartheta)) = r \hat y$. 
As $\lambda_{A, \infty}$ is homogeneous of degree $0$, it is enough to prove that     { [see \eqref{trans}]
\beq \label{grad3}
\lim_{\vartheta \to \theta}\Big | \frac{1}{\vartheta- \theta} \Big(    \lambda_{A,\infty}(\hat y) - \lambda_{A,\infty}(\hat x)\Big ) - 
A_\infty(\hat x)\cdot (\hat x)^{\perp}      \Big | = 0.
\ene
We calculate
\begin{align} \label{grad4}
\Big | \frac{1}{\vartheta- \theta} \Big(    \lambda_{A,\infty}(\hat y) & - \lambda_{A,\infty}(\hat x)\Big ) - 
A_\infty(\hat x)\cdot (\hat x)^{\perp}      \Big | \\ \notag & 
= \frac{1}{|\vartheta- \theta|}  \Big |   \lambda_{A,\infty}( y) - \lambda_{A,\infty}( x) - 
A( x)\cdot z  + A( x)\cdot z - A_\infty(\hat x)\cdot (\vartheta- \theta)(\hat x)^{\perp} 
\\ \notag & \hspace{4cm} - \int_1^\infty (\nabla \alpha_A)(\tau x)\cdot z d \tau  + \int_1^\infty (\nabla \alpha_A)(\tau x)\cdot z d \tau     \Big | \\ \notag &
\leq \frac{1}{|\vartheta- \theta|}  \Big |   \lambda_{A,\infty}( y) - 
\lambda_{A,\infty}( x) - A( x)\cdot z   - 
\int_1^\infty (\nabla \alpha_A)(\tau x)\cdot z d \tau \Big | \\ \notag & +
\frac{1}{|\vartheta- \theta|}  \Big | A( x)\cdot z - A_\infty(\hat x)\cdot (\vartheta- \theta) (\hat x)^{\perp} \Big | 
 +\frac{1}{|\vartheta- \theta|}  \Big | \int_{1}^{\infty} (\nabla \alpha_{A})(\tau x)\cdot z d \tau     
\Big |. 
\end{align}
Using \eqref{princ1} we deduce (notice that $|z| \leq r |\vartheta - \theta|$). 
\beq \label{grad5}
\frac{1}{|\vartheta- \theta|}  \Big |   \lambda_{A,\infty}( y) - 
\lambda_{A,\infty}( x) - A( x)\cdot z   - 
\int_1^\infty (\nabla \alpha_A)(\tau x)\cdot z d \tau \Big | \leq C |\vartheta - \theta|^{2\rho - 1} r^{1 - \rho}.
\ene
Moreover, \eqref{estimAinfty} implies 
\begin{align}\label{grad6}
\frac{1}{ |\vartheta- \theta| }\big | A_\infty(\hat x )\cdot (\vartheta- \theta) (\hat x)^\perp -  A( x ) \cdot z  \big | 
\leq & \Big | \big( A_\infty(\hat x ) - r A(x)\big )\cdot (\hat x)^{\perp} \Big | 
+ \Big |  A(x)r\cdot  \big ((\hat x)^{\perp} - \frac{1}{r (\vartheta- \theta)} z \big ) \Big | \\ \notag 
\leq & C  \Big [
\frac{\ln(e+r)}{(1 + r)^{\min(1, \delta - 1)}} + |\vartheta - \theta | \Big ]
\end{align}
and \eqref{e-p-c.1.2} implies 
\begin{align} \label{grad7}
\Big | \frac{1}{|\vartheta- \theta|}   \int_{1}^{\infty} (\nabla \alpha_{A})(\tau x)\cdot z d \tau     
\Big | \leq C \frac{\ln(e+r)}{ r^{\min(1, \delta - 1)}}. 
\end{align}
Finally, \eqref{grad3} follows from \eqref{grad4}-\eqref{grad7} choosing 
$r = |\vartheta -\theta|^{-\epsilon}$ for some positive conveniently selected $\epsilon$ and 
$\rho$ close to $1$.} 
\bull
}

\begin{corollary}\label{rem1}
{
    Suppose that $ B \in C^{2}(\overline{\Lambda})$ 
     and that $ |  B(x)| \leq C \frac{1}{(1 +|x|)^\mu} $, $ | \frac{\partial}{\partial x_i} B(x)| \leq C \frac{1}{(1 +|x|)^{\mu + 1}} $, $ | \frac{\partial}{\partial x_j}\frac{\partial}{\partial x_i} B(x)| \leq C \frac{1}{(1 + |x|)^{\mu + 2}} $, for every 
$i, j \in \{ 1, 2 \}$ and every $x \in \Lambda$.}  Let $ \tilde \delta > 1$ and  $A \in \mathcal{A}_{\Phi, \tilde \delta}(B)$. Then, 
$$ A_\infty(\hv):=\lim_{\tau \to \infty}A(\tau \hv)\tau$$  exists {and it is continuous as a function of $\hat v \in \mathbb{S}^1$}. We extend Definition \ref{ainftyx} to this case taking 
$$A_\infty(x) : = \frac{1}{|x|} A_\infty \Big(\frac{x}{|x|}\Big ).
$$
\end{corollary}
\noindent {\it Proof:}
Let $r > 1$ such that $K \subset B_r(0)$. We consider a magnetic potential 
$ \underline{A} = \underline{A}_1 +  \underline{A}_2  \in \mathcal{A}_{\Phi}(B)$,  as in Proposition \ref{mfmp.prop.1} (here $  \underline{A}_i$ corresponds to $A_i$, $i \in \{ 1, 2\}$).\\
By Lemma \ref{specific-potentials}, $  \underline{A} \in \mathcal{A}_{\Phi, \mu - 1} $. 
$\underline{A}$ depends on the parameters $\epsilon $,  $Q_0$  and $\hat{w}$ defined in Proposition \ref{mfmp.prop.1}. The support of $\underline{A}_2$ is contained in the cone $  \mathcal{C} = \{ x \in \mathbb{R}^2 : (x- Q_0 ) \cdot \hat{w} 
 \geq | x- Q_0 | \cos(\epsilon) \} $. We take $ \epsilon$ and $Q_0$
 in such a way that there is a real number $r_0$ such that, for every $ s > r_0   $,
\beq \label{e-p-t.1.4otro}
\underline A_{2}(s\hv) = 0
\ene
(take for example $\epsilon = \frac{\pi}{4}$ and $\hat{w}$ orthogonal to $\hv$). 
We denote by 
\beq \label{e-p-t.1.5otro}
 \tilde A : = A - \underline{A}. 
\ene
Then $\tilde A \in \mathcal{A}_{0, \delta}(0)$, where $\delta = \min(\mu-1, \tilde \delta)$. Proposition \ref{mfmp.prop.1} and
\eqref{e-p-t.1.4otro} imply that
\beq \label{e-p-c.1.1otroprima}
\lim_{\tau \to \infty}\underline{A}(\tau \hv)\tau = 0, 
\ene  
which together with Corollary
\ref{e-p-c.1} give the desired result.

\bull 
{
\begin{corollary}\label{tasanan}
{
    Suppose that $ B \in C^{2}(\overline{\Lambda})$ 
     and that $ |  B(x)| \leq C \frac{1}{(1 +|x|)^\mu} $, $ | \frac{\partial}{\partial x_i} B(x)| \leq C \frac{1}{(1 +|x|)^{\mu + 1}} $, $ | \frac{\partial}{\partial x_j}\frac{\partial}{\partial x_i} B(x)| \leq C \frac{1}{(1 + |x|)^{\mu + 2}} $, for every 
$i, j \in \{ 1, 2 \}$ and every $x \in \Lambda$.}  Let $ \tilde \delta > 1$ and  $A \in \mathcal{A}_{\Phi, \tilde \delta}(B)$. Then, 
\beq \label{becio}
 A_\infty\big( (\cos(\theta), \sin(\theta)) \big) = 
 \Big (\frac{\Phi_B}{2\pi} + \frac{d}{d\theta}  \lambda_{A - A^{(c)}, \infty}( (\cos(\theta), \sin(\theta)) \Big ) \begin{pmatrix}
 - \sin(\theta) \\ \cos(\theta)
 \end{pmatrix},
\ene
see Proposition \ref{coupo}, {where the Coulomb magnetic potential $A^{(c)}$ is introduced}. 
\end{corollary}
\emph{Proof:} The result is a direct consequence of Propositions \ref{coupo} and \ref{grad} using $A = A^{(c)}  + A - A^{(c)}$, with $ A - A^{(c)} \in \mathcal{A}_{0,  \delta}(0) $ for  $\delta = \min(\tilde \delta, \mu-1)$, see Lemma \ref{specific-potentials}.  
\bull
\begin{remark}
Corollary \ref{tasanan} makes explicit the fact that the long-range part of a magnetic potential can be regarded as a physical quantity (the total flux) plus the gradient of a function, which shows the specific gauge we are working with.     
\end{remark}
}

\begin{lemma}\label{e-p-l.2}
Let $\delta > 1$ and  $A \in \mathcal{A}_{0, \delta}(0)$. Suppose that $r > 1$ is such that $ K \subset B_r(0)$. Take $\rho \in(0, 1)$.  \\
For every $x \in \ere^2$ with $ |x|> 2r $ and every 
$ z \in \ere^2$ satisfying $|z| < \frac{|x|}{2}$ there is a constant $C$ such that
\begin{align}\label{est1} 
\Big | e^{  i\lambda_{A, \infty}(x + z)} -e^{ i \lambda_{A, \infty}(x )} & - i e^{ i \lambda_{A, \infty}(x )} A(x)\cdot z - i
e^{ i\lambda_{A, \infty}(x )}  \int_1^\infty (\nabla \alpha_A)(\tau x)\cdot z d \tau \Big | \\ \leq
\notag & 
C\Big [ \frac{|z|^2}{(1 + |x|)^2}  +  \frac{\ln(e + |x|)}{(1 +  |x|)^{ 2 \rho + \min(1, \delta - 1) (1 - \rho)}  } |z|^{2\rho} (1 + |z|))^{1 - \rho} \Big ] .
\end{align}
If $\delta > 2$, then 
\begin{align} \label{est2}
\Big | e^{  i\lambda_{A, \infty}(x + z)} -e^{ i \lambda_{A, \infty}(x )} & - i e^{ i \lambda_{A, \infty}(x )} A(x)\cdot z - i
e^{ i\lambda_{A, \infty}(x )}  \int_1^\infty (\nabla \alpha_A)(\tau x)\cdot z d \tau \Big | \\ \leq
\notag & 
C  \frac{|z|^2}{(1 + |x|)^2}.
\end{align}
\end{lemma}

\noindent{\it Proof:}
The result follows from the next calculations: 
\begin{align} \label{e-p-l.2.1}
\Big | e^{ i \lambda_{A, \infty}(x )} \Big ( e^{  i(\lambda_{A, \infty}(x + z)  -  \lambda_{A, \infty}(x ))  } - 1  - & i A(x)\cdot z - i
 \int_1^\infty (\nabla \alpha_A)(\tau x)\cdot z d \tau \Big ) \Big | \\ \notag
 \leq & \:  \Big | \lambda_{A, \infty}(x + z) -\lambda_{A, \infty}(x ) - A(x)\cdot z - \int_1^\infty (\nabla \alpha_A)(\tau x)\cdot z d \tau \Big | \\ \notag
& + |  \lambda_{A, \infty}(x + z) -  \lambda_{A, \infty}(x ) |^{2}  \sum_{n = 2}^{\infty} \frac{1}{n!}|  \lambda_{A, \infty}(x + z) - 
 \lambda_{A, \infty}(x ) |^{n-2} \\ \notag
\leq & \: 
 C\Big [ \frac{|z|^2}{(1 + |x|)^2}  +  \frac{\ln(e + |x|)}{(1 +  |x|)^{ 2 \rho + \min(1, \delta - 1) (1 - \rho)}  } |z|^{2\rho} (1 + |z|))^{1 - \rho} \Big ] \\ \notag &  
  +  C \Big ( \frac{|z|}{|x|}\Big )^2 e^{  |  \lambda_{A, \infty}(x + z) -  \lambda_{A, \infty}(x ) | }. 
\end{align}
In the last equations we used Lemma \ref{e-p-l.1} and Lemma 3.8 in \cite{b-w.1} using that $  \lambda_{A, \infty}(\tau  w ) =  \lambda_{A, \infty}( w ) $, 
for every $w \in \ere^2 \setminus \{ 0 \}$ and every $\tau > 0$. This proves \eqref{est1}. Eq. \eqref{est2} follows similarly, using \eqref{princ2}.

\bull

\begin{lemma}\label{e-p-l.3}
Let $\delta > 1$ and  $A \in \mathcal{A}_{0, \delta}(0)$. Suppose that $r > 1$ is such that $ K \subset B_r(0)$. Take $\rho \in(0, 1)$. Suppose that  $s \in \ere$ is such that $ s> 2r $ and  
$\phi \in  {\bf H}^2(\ere^2) $. Then, there is a constant $C$ satisfying

\begin{align}\label{e-p-l.3.1} 
\Big \| \Big ( e^{ i\lambda_{A, \infty}(\frac{s\mo }{mv} + s \hv)}  -e^{ i \lambda_{A, \infty}( s\hv )} - & i e^{  i\lambda_{A, \infty}( s \hv )} A(s \hv)\cdot \frac{s\mo }{mv}   
  - i
e^{ i\lambda_{A, \infty}( s \hv )}  \int_1^\infty (\nabla \alpha_A)(\tau s \hv)\cdot \frac{s\mo }{mv} d \tau \Big ) \phi \Big \|_{L^2 (\ere^2) }  \\ \ & \notag \hspace{3cm} \leq 
C \Big( \frac{1}{v^2}  + \frac{1}{v^{2 \rho}} 
\frac{   \ln(e + |s|)(1 + (\frac{s}{v})^{1-\rho})  }{(1 + s)^{ \min(1, \delta - 1)(1 - \rho)} }
\Big ) \| \phi  \|_{{\bf H}^2(\ere^2)}.
\end{align}
In case that $ \delta > 2 $ 
\begin{align}\label{e-p-l.3.1tilde} 
\Big \| \Big ( e^{ i\lambda_{A, \infty}(\frac{s\mo }{mv} + s \hv)}  -e^{ i \lambda_{A, \infty}( s\hv )} - & i e^{  i\lambda_{A, \infty}( s \hv )} A(s \hv)\cdot \frac{s\mo }{mv}   
  - i
e^{ i\lambda_{A, \infty}( s \hv )}  \int_1^\infty (\nabla \alpha_A)(\tau s \hv)\cdot \frac{s\mo }{mv} d \tau \Big ) \phi \Big \|_{L^2 (\ere^2) }  \\ \ & \notag \hspace{3cm} \leq 
C  \frac{1}{v^2}  \| \phi  \|_{{\bf H}^2(\ere^2)}.
\end{align}
\end{lemma}

\noindent {\it Proof:}
Let $g \in C^\infty_0(\ere^2)$ satisfy $g(p)=1, |p| \leq 1, g(p)=0, |p| \geq 2$. We denote by $\bar \phi := g(\frac{4 \mo }{mv})\phi$, then
\beq \label{e-p-l.3.2}
\| \phi - \bar \phi  \|_{L^2(\ere^2)} \leq C \frac{1}{v^2}\| \phi \|_{{\bf H}^2(\ere^2)}.
\ene
Thus, we can use $\bar \phi$ instead of $\phi$ in (\ref{e-p-l.3.1}).\\
For every $\mo$ in the support of the Fourier transform of $\bar \phi$,  $ |\frac{s\mo}{mv} | \leq \frac{|s \hv|}{2}  $. Then, we can use  Lemma \ref{e-p-l.2}, applying the Fourier transform, to obtain (\ref{e-p-l.3.1}) and 
(\ref{e-p-l.3.1tilde}).     
\bull

\section{The Hamiltonian} \label{ham}\sss
 The Schr\"odinger's equation for an electron in $\Lambda$  with electric potential $ V$ (see Definition \ref{electric-potential} ) and magnetic
field $ B$ (see Definition \ref{mf})  is given by
\beq \label{h.1} 
i\frac{\partial}{\partial t} \phi = \Big (\frac{1}{2 m} \big(\mo- A\big )^2+  V\Big )\phi,
\ene
where $A \in \mathcal{A}_{\Phi}(B)$ (see Definition \ref{mp}), $ \mo = - i  \nabla $ is the momentum operator and $ m > 0 $ is the mass of the electron.  Note that Definition \ref{electric-potential} implies that
$\bar{V}$ is $H_0- $bounded with relative bound zero [see (\ref{H0})] 
 and, therefore, there is a constant $c_V$ such that 
(see \cite{r-s.2}, Theorem X.18)
\begin{equation}\label{h.2}
c_V  \langle \phi, \phi \rangle \leq \frac{1}{2} \langle \mo \phi, \mo \phi \rangle + \langle V \phi, \phi \rangle, \ \ \ \ \forall \phi \in C_0^\infty (\Lambda).  
\end{equation} 
It follows from Lemma 1.2 chapter 9 of \cite{m.s} that
$$  c_V  \langle \phi, \phi \rangle \leq \frac{1}{2 m} \langle (\mo - A) \phi, (\mo - A) \phi \rangle  + \langle V \phi,\phi \rangle, \ \ \ \ \forall \phi \in C_0^\infty (\Lambda) .   $$
We define the energy bilinear form by 
\begin{equation}\label{h.3}
q_A(\phi, \psi) = \frac{1}{2m} \langle (\mo - A) \phi, (\mo - A) \phi \rangle  + \langle V \phi, \phi\rangle , \ \ \ \ \forall \phi, \psi \in C_0^\infty (\Lambda).   
\end{equation}

\begin{prop}
For every $ A \in \mathcal{A}_\Phi(B) $ there exists a closed extension  $ \bar{q}_A $ of  $q_A$.  The form $ \bar{q}_A $ is bounded from below by $c_V$.  If $ A $ and $\tilde{A}$ belong to $ \mathcal{A}_\Phi(B) $ and $\tilde{A} - A = \nabla \lambda $  (see Remark \ref{drc.6}), then $ \hbox{\rm Dom} (\bar{q}_{\tilde{A}}) = e^{i\lambda}  \hbox{\rm Dom}(\bar{q}_A ) $ and, for every $\phi, \psi \in \hbox{\rm Dom}(\bar{q}_{\tilde A}) $,  $ \bar{q}_{\tilde{A}}(\phi, \psi)  =  \bar{q}_A (  e^{-i\lambda} \phi, e^{- i\lambda } \psi  ) $, here $\hbox{\rm Dom}(\cdot) $ denotes the domain of the corresponding quadratic form. 
\end{prop}

\noindent{\it Proof:} 
First we take $A$ to be the Coulomb gauge ($ \nabla \cdot A = 0 $).
From  Theorem X.23 \cite{r-s.2} applied to the operator $\frac{1}{2 m} \big(\mo- A\big )^2+  V=  \frac{1}{2 m} \big(\mo\big )^2-\frac{1}{m} A\cdot \mo+\frac{1}{2m}A^2+ V- c_V$ with domain $C^\infty_0(\Lambda)$,  it follows that $ q_A $ is closable
and its closure is bounded from below by $c_V$.  Suppose that $ \tilde{A} \in
\mathcal{A}_\Phi(B)  $ and that  $ \lambda \in C^1(\Lambda)  $ is such that 
$ \tilde{A} - A = \nabla \lambda  $. We define the following bilinear
form with domain $ e^{i \lambda} \textrm{Dom}(\bar{q}_A) $
$$
\tilde{q}(\phi, \psi) := \bar{q}_A(e^{- i \lambda }\phi,e^{- i \lambda }\psi ).
$$
It is not difficult to see that $ C_0^{\infty}(\Lambda) \subset \textrm{Dom}(\tilde{q}) $, that $\tilde{q}
$ restricted to $ C_0^{\infty}(\Lambda)   $ coincides with $ q_{\tilde A} $ and that $
\tilde{q} $ is closed. Furthermore, it can be verified also that $ C_0^\infty (\Lambda) $ is a
form-core of $ \tilde{q}  $. It follows that $  \tilde{q}  = \bar{q}_{\tilde{A}}$.   
\bull

\noindent From Theorem VIII.15 \cite{r-s.1}, $ \bar{q}_A $ is the form associated to a
unique self-adjoint operator that we denote by $ H(A)  $. 
\begin{definition}[Hamiltonians]\label{hamiltonians} {\rm
The Hamiltonian $ H(A) $ is the unique self-adjoint operator associated to the
form $ \bar{q}_A $.  }
\end{definition}
We denote by  $\textrm{Dom}(H(A))  $ the domain of $ H(A) $.
It can be easily verified that if $ \tilde{A} \in \mathcal{A}_\Phi(B) $ is such
that $ \tilde{A} - A = \nabla \lambda $, then $\textrm{Dom}(H(\tilde{A})) = e^{i\lambda}
\textrm{Dom}(H(A))  $ and 
\begin{equation}\label{h.4}
H(\tilde{A}) = e^{i \lambda} H(A) e^{-i \lambda}.
\end{equation}  
The electron evolves freely when there are no fields and when there is no
obstacle. The wave function of the free electron is defined in the
whole space $ \mathbb{R}^2 $; it satisfies the Schr\"odinger equation 
$$
i \frac{d}{dt} \phi = H_{0} \phi,
$$
where we recall that $H_0$ is the free Hamiltonian given by
\beq\label{free-hamiltonian}
H_0 = \frac{1}{2 m} \mo^{2},
\ene
with domain 
the Sobolev space ${\bf H}^2(\mathbb{R}^2)$. 
\section{Wave and Scattering Operators} \label{wavescatering}
\begin{prop}\label{wave-operators}
For every magnetic potential $A \in \mathcal{A}_\Phi(B)$, the limits 
(\ref{w-s-o.1}) exist. If $A, \tilde{A} \in \mathcal{A}_\Phi(B) $ and 
$ \tilde{A} - A = \nabla \lambda $, $ W_{\pm}(\tilde{A})$ and 
 $ W_{\pm}(A)$ are related by the change of gauge formula [recall \eqref{defilambdainfty}]
\beq \label{desired}
W_{\pm}(\tilde{A}) = e^{i \lambda(x) } W_{\pm}(A)e^{-i \lambda_\infty(\pm \mo)}. 
\ene
\end{prop}

\noindent{\it Proof:}
As the proof of the statement is standard, we outline it and give proper references for details.  
Let $\chi \in C^\infty(\mathbb{R}^2, [0, 1])$ be identically zero in a compact neighborhood of 
the obstacle $K$ and $1$ outside another compact neighborhood of $K$. As the operator   
$(1- \chi) (H_0+ I )^{-1} $ is compact 
\beq \label{wzer}
s-\lim_{t \to \pm \infty} e^{itH(A)} (1 - \chi) e^{-i t H_0} = 0.
\ene
Thus, the existence of the limits \eqref{w-s-o.1} follows from the existence of the limits 
\beq \label{w-s-o.2}
 s-\lim_{t \to \pm \infty} e^{itH(A)} \chi e^{-i t H_0}, 
\ene
which is true if the integral  
\beq\label{w-s-o.3}
  i\int_{0}^{\pm \infty} e^{it H(A)}(H(A) \chi - \chi H_0)
e^{-it H_0} \phi  
\ene
converges absolutely (using the fundamental theorem of calculus: Cook's argument). First we choose the vector potential $ A $ to be the one defined in Corollary 
\ref{transversal-vector-potential}.     
We prove that the integrand in (\ref{w-s-o.3}) is bounded by an integrable function using the idea of
Loss and Thaller \cite{l-t}: We notice that the angular momentum operator ${\bf L} = x \times \mo$
 commutes with the free Hamiltonian $H_0$ and that 
\beq \label{anmo}
\big( x \times A_T \big) \cdot p = - A_T \cdot {\bf L}. 
\ene
Eq. \eqref{anmo} together with the stationary phase method are the key ingredients to prove the convergence of the 
integral \eqref{w-s-o.3}. We refer to Section 4.2 of \cite{wj.1} for the details. This proves the existence
$W_{\pm}(A)$ for the specific $A$  defined in Corollary 
\ref{transversal-vector-potential}. Now we prove the existence for a general magnetic potential  
 $ \tilde{A} \in \mathcal{A}_\Phi(B) $. \\
Let $ \lambda  $ be such that 
$\tilde{A} - A = \nabla \lambda $. We follow the proof of Lemma of 5.3 in \cite{b-w.1} 
(see also the proof of Lemma 2.3 of \cite{w.1}). Using Equation (\ref{h.4}) we obtain:
\begin{equation}
\begin{array}{l}\label{w-s-o.4}
W_{\pm}(\tilde{A}) =  e^{i \lambda (x)} s-\lim_{t \to \pm \infty}e^{itH(A)}e^{-i\lambda(x)}
\mathcal{J} e^{-itH_0}
=  e^{i \lambda (x)} s-\lim_{t \to \pm \infty}e^{itH(A)}e^{-i\lambda_\infty(x)}
\mathcal{J} e^{-itH_0}.
\end{array}
\end{equation} 
In \eqref{w-s-o.4} we use that, by the Rellich-Kondrachov theorem, $ e^{- i \lambda(x)} - 
e^{-i \lambda_{\infty}(x)}  $ is a compact operator from $ {\rm Dom}(H_0) \to L^{2}(\mathbb{R}^2) $. \\
Remark \ref{drc.6} and Definition \ref{mp} imply that, for $|x| = 1$,
\begin{align} \label{carai}
| \lambda_{\infty}(x + y) - \lambda_{\infty}(x)| \leq C |y|,  \hspace{3cm } \text{for} \: \: |y| < 1/2, 
\end{align}
see lemma 3.8 of \cite{b-w.1} for a detailed proof. Eq. \eqref{carai} together with the fact that
\begin{align}
e^{it H_0} x e^{-it H_0} = t \Big (\frac{x}{t} +  \frac{\mo}{ m} \Big )  
\end{align}
(notice that $\lambda_\infty$ is homogeneous of degree zero) imply that
\beq \label{nosep}
s- \lim_{t \to \pm \infty}e^{it H_0} e^{e^{-i\lambda_\infty(x)}} e^{-it H_0} = e^{i \lambda_\infty(\pm \mo)} 
\ene
(see Equation (2.29) of \cite{w.1}). Eqs. \eqref{w-s-o.4} and \eqref{nosep} imply \eqref{desired}, which gives the existence of the wave operators and the change of gauge formula.    
\bull

\section{High-Velocity Limits for the Wave and Scattering Operators} \label{ihigh}\sss
\subsection{Notation and Basic Formulae}\label{basic-formulas}
For the readers' convenience we recall the formulae [see \eqref{hvlwso.1}]
\beq
\hv  = \frac{\v}{|\hv|}, \hspace{1cm} v = | \v |, \hspace{1cm}\Lambda_{\hv}= \{x \in \Lambda: x+\tau \hv \in \Lambda,\, \forall \tau \in \ere\},\, \hspace{1cm} \hbox{\rm for}\, \v \neq 0.
\label{hvlwso.1recall}
\ene  
\noindent  For every $ x \in \Lambda_{\hv}$ we set
\beq
L_{A,\hv}(t):= \int_0^t\, \hv\cdot A(x+\tau \hv) d\tau, \hspace{3cm}  -\infty \leq t \leq \infty.
\label{hvlwso.2}
\ene
Given a measurable function ${\mathbf f}: \ere^2\times \ere \rightarrow \ere^2$ with $ \mathbf f_t(x):={\mathbf f}(x,t) \in L^1_{\hbox{\rm loc}}
(\ere^2,\ere^2)$, we define
\beq
\Xi_{\mathbf f}(x,t):= \frac{1}{2m}\chi (x)\left [
-\mo\cdot \mathbf f(x,t)-\mathbf f(x,t)\cdot \mo
+(\mathbf f(x,t))^2\right].
\label{hvlwso.20}
\ene
We designate
\beq
\eta (x,t):= \int_0^t (\hv\times B)(x+\tau\hv)d\tau.
\label{hvlwso.21}
\ene

We denote by 
\beq \label{jota}
  \mathcal{J}_{\hv}: L^{2}(\mathbb{R}^2) \to L^{2}(\Lambda_{\hv} ) 
\ene 
the multiplication operator by the characteristic function of the set $ \Lambda_{\hv} $ and
by 
\beq\label{i}
 {\mathbf I}: L^{2}(\Lambda_{\hv} ) \to  L^{2}(\Lambda)   
\ene
the inclusion operator:
\beq\label{inc}
{\mathbf I}(\phi)(x) = \begin{cases} \phi(x), & \text{if}\:   x \in \Lambda_{\hv}, \\ 0, & \text{otherwise.} \end{cases}
\ene
For every function $ \phi \in
L^{2}(\mathbb{R}^2) $ with support contained in $ \Lambda_{\hv} $, we identify
\beq   {\mathbf I} e^{\pm i L_{A,\hv}(t)}  \mathcal{J}_{\hv}  \phi  \equiv  e^{ \pm iL_{A,\hv}(t)}  \phi. 
\ene    
For all measurable function $f$, the operator $ f (\mo) $ is defined by 
$$ f(\mo) = \mathcal{F}^{-1} f (\cdot) \mathcal{F}, $$ 
where $\mathcal{F}$ is Fourier transform:

$$
\mathcal F \varphi(p):= \frac{1}{2 \pi}\,\int_{-\infty}^\infty\, e^{-i p\cdot x}\, \varphi(x)\, dx.
$$
We Remark that
\beq \label{hvlwso.3}
e^{i \mo \cdot \v t} \, f(x) \, e^{-i \mo \cdot \v t}= f(x+\v t),
\quad e^{-im\v\cdot x}\, f(\mo )\, e^{im\v\cdot x}= f(\mo +m \v ),
\ene
and, in particular,
\beq
e^{-im\v\cdot x}\, e^{-it H_0 } \, e^{im\v\cdot x}= e^{-i mv^2 t/2}\, e^{-i\mo\cdot \v t}\,
 e^{-i tH_0}.
\label{hvlwso.5}
\ene
 
\subsection{High-Velocity Estimates I. The Magnetic Potential}\label{highmagnetic}
\begin{lemma}\label{hvlwso.lem.1}
 Let $\v \in \ere \setminus \{0\}$ and $\Lambda_0$ be a compact subset 
of $\Lambda_{\hv} $. Then, for all flux $\Phi$ and  all $A \in \mathcal{A}_\Phi(B)$ (see Definition \ref{mp}),
there is a constant $C$ such that, for all  $\phi \in {\bf H}^2(\ere^2) $ with $ \hbox{\rm supp}\, \phi \subset
\Lambda_0 $:
\beq
\left\| \left(e^{-im\v\cdot x}\, W_\pm(A)\, e^{im\v\cdot x} -e^{-i L_{A,\hv}(\pm \infty)}\right) \phi
\right\|_{L^2(\Lambda)}\leq
 C \frac{1}{v} \| \phi\|_{ {\bf H}^2(\ere^2)}.
\label{hvlwso.6}
\ene
If, moreover, $ \hbox{\rm div}A \in L^2_{\hbox{\rm loc}}\left(\Lambda\right)$
\beq
\left\| \left(e^{-im\v\cdot x}\, W_\pm^\ast(A)\, e^{im\v\cdot x} -e^{i L_{A,\hv}(\pm \infty)}\right) \phi
\right\|_{L^2(\Lambda)}\leq
 C \frac{1}{v} \| \phi\|_{{\bf H}^2(\ere^2)}.
\label{hvlwso.7}
\ene
 \end{lemma}

 \noindent {\it Proof:}
We prove (\ref{hvlwso.6}) for $ W_+(A)  $; the proof for $ W_-(A)  $ is similar.    
We suppose first that the magnetic potential ($A$) is the one constructed in
Proposition \ref{mfmp.prop.1}. We choose $ \hat{w} $ and $\epsilon$ (see the
statement of Proposition \ref{mfmp.prop.1}) such that, 
\beq\label{hvlwso.8}
A_2(x) F(|x-Q_0 - \tau \hv | \leq |\tau / 4| ) = 0,  \quad
\tau \in \mathbb{R},
\ene  
where  $ F(|x-Q_0 - \tau \hv | \leq |\tau / 4| )$ is the multiplication
operator by the characteristic function of the set $  \{ x \in \mathbb{R}^2 :
|x-Q_0 - \tau \hv | \leq |\tau / 4|  \} $ and $ Q_0 $ is introduced in the statement
of  Proposition \ref{mfmp.prop.1}.\\
The Proof of \eqref{hvlwso.6} is similar to the proofs Lemma of 2.4 of \cite{w.1} and Lemma 5.6 of \cite{b-w.1}. Here we have to do only slight modifications to take into consideration the different aspects that we address in this text:
\begin{itemize}
\item In Lemma 5.6 of \cite{b-w.1} the magnetic potentials are bounded, in contrast with their counterpart in \cite{w.1}  and here that are unbounded.    
\item In  Lemma of 2.4 of \cite{w.1} it is proved that 
$$  \left\| \left(e^{-im\v\cdot
      x}\, W_\pm(A)\,
 e^{im\v\cdot x} -e^{-i L_{A,\hv}(\pm \infty)}\right) \phi
\right\|_{L^2(\ere^2)}  = \mathcal{O}\Big ( \frac{1}{v}\Big )  $$
  as  $ v $ tends to $ \infty $, but 
 the estimate in terms of $   \| \phi\|_{{\bf H}^2(\ere^2)} $ is not
 proved and there is no electric potential. However, from the proof of Lemma of 2.4 of \cite{w.1} we can obtain
 the estimate (\ref{hvlwso.6}) as it is done in the proof of Lemma  5.6 of
 \cite{b-w.1}. 
\end{itemize}
We do not repeat a full (long) proof here, since it follows from \cite{b-w.1} and \cite{w.1}. We, instead, sketch the proof and point out the main ingredients as well as proper references where the missing details can be directly read.\\ 
 By our assumptions, there is a function
 $\chi \in C^\infty(\ere^2)$ such that $\chi \equiv 0$ in a neighborhood of $K$ and 
 $$\chi(x)=1, \hspace{1cm} x \in
 \{ x: x= y+ \tau \hv, y \in \,\hbox{\rm supp}\, \phi , \, \tau \in \ere\} \cup \{ x: |x| \geq M\},$$
  for some
 $M$ large enough. For every $x \in \mathbb{R}^2$, we designate by $ \bar{A}(x)
 := \chi (x) A(x)  $ if $x \in \Lambda$ and $ \bar{A}(x)
 := 0$ otherwise.  \\ 
We use the following notation
\beq \label{hvlwso.9}
 H_1:= \frac{1}{v} e^{-im\v\cdot x}H_0 \,e^{im\v\cdot x},\,\hspace{1cm}  H_2:= \frac{1}{v} e^{-im\v\cdot x}H(A)
 \, e^{im\v\cdot x}.
\ene
We have that 
\beq \label{hvlwso.10}
e^{-im\v \cdot x} W_+(A) e^{im\v \cdot x}  = s-\lim_{t \to \infty} e^{it H_2}
\chi e^{-it H_{1}}. 
\ene
As $ \bar{A} (x + \tau \hv  ) = A(x + \tau \hv) $ for every $ x $ lying in the
support of
$\phi $, 
\beq \label{hvlwso.11}
 \left(e^{-im\v\cdot x}\, W_+(A,V)\, e^{im\v\cdot x} -\chi(x)e^{-i L_{A,\hv}\left( \infty \right)} \right)\phi =
 \hbox{\rm s-}\, \lim_{t \rightarrow \infty}\left[ e^{it H_2}\chi (x)e^{-it H_1}- \chi(x)e^{-iL_{\bar{A},\hv}
 \left(t\right)}
 \right]
 \phi.
\ene
Let $g \in C^\infty_0(\ere^2)$ satisfy $g(p)=1, |p| \leq 1, g(p)=0, |p| \geq 2$. Denote
\beq \label{hvlwso.12}
\tf:= g(\mo/ v^\rho) \,\phi, \, \frac{1}{2} \leq \rho < 1.
\ene
Then
\beq
\left\| \tf- \phi\right\|_{L^2(\ere^2)} \leq \frac{1}{v^{2 \rho}} \|\phi\|_{{\bf H}^2(\ere^2)}.
\label{hvlwso.13}
\ene
By (\ref{hvlwso.13}), it is enough to estimate (\ref{hvlwso.11}) with $
\tilde{\phi}  $, instead of $ \phi  $. 
Denote
\beq
P(t,\tau):= e^{i\tau H_2} i \left[H_{2}e^{-i L_{\bar{A}, \hv}(t-\tau)}\chi(x)-e^{-i L_{\bar{A},\hv}(t-\tau)} \chi(x)
\left(H_1-\hv\cdot\bar{A}(x+(t-\tau)\hv)\right)\right] e^{-i\tau H_1}\tf.
\label{hvlwso.14}
\ene
By Duhamel's formula: 
\beq
\left[ e^{it H_2}\chi (x)e^{-it H_1}- \chi(x)e^{-iL_{\bar{A},\hv}
 \left(t\right)}
 \right]
 \tf = \int_0^t\, d \tau\, P(t,\tau).
 \label{hvlwso.15}
 \ene
We define
\beq \label{b}
b(x,t):= \bar{A}(x + t \hv ) + \int_{0}^{t} (\hv \times \underline{B} ) (x + \tau \hv  ) d \tau,
\ene
with $\underline{B} : = \nabla \times \bar A $. Note that $\nabla \cdot \bar A$ is continuous and that $\nabla \cdot (\hv \times \underline B)= - \v \cdot  \nabla \times \underline B $ is bounded. 
By an explicit calculation, we obtain
\beq \label{p}
P(t, \tau) = T_1 + T_2 + T_3,
\ene
where 
\beq
T_1:= \frac{1}{v} e^{i \tau H_2}i e^{-iL_{\bar A,\hv}(x,t-\tau)}\, \left( \Xi_b(x,t-\tau)+ \chi V(x)\right)e^{-i\tau H_1}\tf ,
\label{5.24}
\ene
\beq
\begin{array}{c}
T_2:= \frac{1}{2mv}e^{i \tau H_2}i e^{-i L_{\bar A , \hv}(x,t-\tau)}\Big\{-(\Delta \chi)+2(\mo \chi)\cdot \mo
-2 b(x, t-\tau) \cdot (\mo \chi) + \\\\ \chi [ - \mo \cdot (A - \bar{A})+ | A |^2 - | \bar{A} |^2 -
2 (A- \bar{A})\cdot (\bar{A}  +  \mo - b(x, t - \tau) ) - 2 (\mo \chi) \cdot  ( A - \bar{A})    ]  
\Big \} e^{-i\tau H_1}\tf,
\label{5.25}
\end{array}
\ene
\beq
T_3:=e^{i \tau H_2}i e^{-i L_{\bar A,\hv}(x,t-\tau)} \left[ (\mo \chi)\cdot \hv  - \chi (A - \bar{A}) \cdot \hv   \right]
e^{-i\tau H_1}\tf.
\label{5.26}
\ene
We prove as in \cite{w.1,b-w.1}, see (5.26), (5.35)-(5.37) in \cite{b-w.1}, that there
exists an integrable function $ {\mathbf P}_\v : [0, \infty) \to \mathbb{R}  $ such
that $\| P(t, \tau)\|_{L^2(\Lambda)} \leq {\mathbf P}_\v(\tau) $ and 
\beq \label{hvlwso.16}
\int_{0}^{\infty} {\mathbf P}_\v(\tau) d \tau  \leq \frac{C}{v} \| \phi  \|_{{\bf H}^2(\mathbb{R}^2)}. 
\ene
Here we use  \eqref{hvlwso.8}.  Note that in \cite{b-w.1} this condition was not necessary because the magnetic potential was of short-range. As in \cite{w.1} we need now \eqref{hvlwso.8} because the magnetic potential is of long-range.  Equation (\ref{hvlwso.6}), for the magnetic potential $A$ constructed in Proposition \ref{mfmp.prop.1},
follows from (\ref{hvlwso.11}) and (\ref{hvlwso.14}-\ref{hvlwso.16}). Let $
\tilde{A}  \in \mathcal{A}_\Phi(B)$ be a general magnetic potential, then we
prove that (\ref{hvlwso.6}) holds for $\tilde{A}$ as in the proof of Lemma 2.4 \cite{w.1}, see (2.66)-(2.67) in \cite{w.1},  using the formulae for change of gauge of Proposition  \ref{wave-operators}. \\ 

Now we outline the proof of (\ref{hvlwso.7}). Note that $ e^{i L_{A, \hv}(\pm \infty)} \phi = e^{i L_{\chi A, \hv}(\pm \infty)} \phi $.
 As $ \chi A $ is bounded and $ \nabla \cdot A \in L^{2}_{{\rm Loc}}(\Lambda) $, we can apply (5.41) and (5.42) of \cite{b-w.1}
 with $ \chi A$, instead of $A$, to conclude that
\beq \label{hvlwso.17}
\| e^{i L_{A,\hv}(\pm \infty)} \phi\|_{{\bf H}^2(\Lambda)} =\| e^{i L_{\chi A,\hv}(\pm \infty)} 
\phi\|_{{\bf H}^2(\Lambda)}  \leq C \|\phi\|_{{\bf H}^2(\ere^2)}.
\ene
Finally, from Equation (\ref{hvlwso.17}) we obtain (\ref{hvlwso.7}) following the procedure of the proof of Equation (5.20) of \cite{b-w.1}; see the lines below (5.43) in \cite{b-w.1}.   
\bull

\noindent {\bf \Large{Proof of Theorem \ref{reconstruction-formula-I}}}\\
\noindent {\it Proof:} The proof is basically an application of Lemma \ref{hvlwso.lem.1}
and the definition of the scattering operator [See Eq. \eqref{scattering-operator}].
See the proof of Theorem 5.7 in \cite{b-w.1} for details.   

\bull
\subsection{High-Velocity Estimates II}\label{highpotential}

\begin{theorem}\label{reconstruction-formula-ii} {\bf (Reconstruction Formula II. The Cone Magnetic Potential)}
  Suppose that the vector potential  $A = A_1 + A_2$ is the one defined in Proposition \ref{mfmp.prop.1}. Suppose, furthermore, that $A_2$ satisfies (\ref{hvlwso.8}). Let  $\Lambda_0$ be a compact subset of $\Lambda_{\hv},
 $ with $\v \in \ere \setminus \{0\}$.  Let $\phi_{\v}, \psi_{\v}$ be defined as in  \eqref{hvlwso.23}   with $\phi_0,\psi_0 \in {\bf H}^6(\ere^2)$ with  support in $\Lambda_0$. Then, the following estimations hold true [recall \eqref{hvlwso.1}-\eqref{hvlwso.23} and Section \ref{basic-formulas}]
\beq
\begin{array}{l}
v \left( \left[S(A,V)- e^{ia(\hv,x)}\right] \phi_{\v}, \psi_{\v}\right)=
\left(-i e^{ia(\hv,x)}\int_{-\infty}^\infty
V(x+\tau\hv)\,d\tau \, \phi_0,\, \psi_0\right)\\\\
+
\left( -i e^{i a(\hv, x) }\int_{-\infty}^0\, \Xi_\eta (x+\tau \hv,-\infty)\,d\tau \,\phi_0,\psi_0\right)+
\left(-i \int_{0}^\infty\, \Xi_\eta (x+\tau \hv,\infty)\,d\tau \, e^{ia(\hv, x)} \phi_0,\psi_0\right)+ R (\v, \phi_0, \psi_0),
\end{array}
\label{hvlwso.24}
\ene
where,
\beq 
\left|R(\v, \phi_0,\psi_0)\right| \leq C \|\phi_0 \|_{{\bf H}^6(\ere^2)}\, \|\psi_0 \|_{{\bf H}^6(\ere^2)} %\frac{1}{v^{ \min(\mu-2, \alpha -1, %1- \iota)}}, \: \: \hbox{for every $\iota >0 $}.
\left\{ \begin{array}{c}\frac{1}{v^{ \min(\mu-2, \alpha -1)}},\, \, \hbox{\rm if}\, \min (\mu-3, \alpha -2)
 <0, \\ \\
\frac{|\ln v|}{v}, \,\,\hbox{\rm if}\, \min (\mu-3, \alpha -2)=0, \\ \\
\frac{1}{v},\,\,\hbox{\rm if}\, \min (\mu-3, \alpha -2) > 0, \\ \\
\end{array}\right.
\label{hvlwso.25}
\ene
for some constant $C$. 
\end{theorem}

\noindent{\it Proof:}
We follow the procedure of the proof of Theorem 5.9 in \cite{b-w.1}. Although the proof here is similar to the one of Theorem 5.9 in \cite{b-w.1}, there are some new features in this text: In \cite{b-w.1} the magnetic potential is bounded and short-range, i.e., it decays as $\frac{1}{|x|^{1 + \epsilon}}$ for some $\epsilon > 0$. In this paper it
is neither bounded nor short-range. Since the proof in \cite{b-w.1} is rather long and there are only a few new ingredients here, we outline the proof pointing out the new aspects and refer to \cite{b-w.1} for full details.   \\
We adopt for this proof the following notation to make formulae shorter:
$$
a \equiv a(\hv,x).
$$
As in the proof of Theorem 5.9 in \cite{b-w.1}, see (5.53)-(5.54) in \cite{b-w.1},
\beq
v \left( \left[S(A,V)- e^{ia}\right] \phi_{\v}, \psi_{\v}\right)=
v \left( e^{-iL_{A,\hv}(-\infty)} \phi_0,  \mathcal R_+ \psi_0\right)+ v\left(  \mathcal R_- \phi_0,
e^{-i L_{A,\hv}(\infty) } \psi_0 \right)+
v\left(  \mathcal R_- \phi_0, \mathcal R_+ \psi_0 \right),
\label{5.40}
\ene
where
\beq \label{rpm}
\mathcal R_\pm:= e^{-im\v\cdot x}W_\pm(A, V) e^{im\v\cdot x}  - e^{-i L_{A, \hv}(\pm \infty)}.
\ene
By Lemma \ref{hvlwso.lem.1}
\beq
v \left| \left(  \mathcal R_- \phi_0, \mathcal R_+ \psi_0 \right)\right|
\leq C \frac{1}{v} \|\phi_0 \|_{{\bf H}^2(\ere^2)}\,
\|\psi_0 \|_{{\bf H}^6(\ere^2)}.
\label{5.41}
\ene
We show below that 
\beq
v \left( e^{-iL_{A,\hv}(-\infty)} \phi_0,  \mathcal R_+ \psi_0\right)=
 \left(-i  \int_{0}^\infty\, (\Xi_\eta (x+\tau \hv,\infty)+ \chi V(x+\tau \hv))\,d \tau \,
e^{ia}\phi_0,\psi_0\right)
+ R_+ (\v, \phi_0, \psi_0),
\label{5.42}
\ene

\beq
v \left(\mathcal R_- \phi_0,  e^{-iL_{A,\hv}(\infty)}   \psi_0\right)=
 \left(-i e^{ia} \int_{-\infty}^0\, (\Xi_\eta (x+\tau \hv,-\infty)+ \chi V(x+\tau \hv))\,d \tau \,\phi_0,
\psi_0\right)
+ R_- (\v, \phi_0, \psi_0),
\label{5.43}
\ene
where $ R_\pm$ satisfies (\ref{hvlwso.25}). Note that (\ref{5.43}) follows from (\ref{5.42}) by time inversion and
charge conjugation in the magnetic potential, i.e., by taking complex conjugates and changing $A$ to
$- A$. It can also be proved as in the proof of (\ref{5.42}) that we sketch below. \\
We use the notation of the proof of Lemma \ref{hvlwso.lem.1}. For simplicity we denote by $ O(r)$ a term that
satisfies
$$
\left|O(r)\right| \leq C   \|\phi_0 \|_{{\bf H}^6(\ere^2)}\, \|\psi_0 \|_{{\bf H}^6(\ere^2)} \,r.
$$
It can be proved  using the proof of the Lemma 5.8 of \cite{b-w.1}, specifically the proof of (5.58) in \cite{b-w.1}, recall (\ref{5.25})-(\ref{5.26}), that 
\beq 
 \left\| T_2 + T_3\right\|_{L^2(\ere^2)} \leq C_l \frac{ \| \phi_0 \|_{{\bf H}^6(\ere^2)}
   }{v^{3 - \epsilon } (1+ |\tau|)^{l}}, \hspace{3cm} \forall \epsilon > 0, l \in \mathbb{N}.
\label{5.45}
\ene   
The terms $T_2 $ and $T_3$ in \cite{b-w.1} are different, but the bound is estimated in the same way here. \\
It follows from (\ref{hvlwso.11}), (\ref{hvlwso.15})-(\ref{5.26}), \eqref{rpm} and \eqref{5.45} 
 [see equation (5.57) in \cite{b-w.1}] that
\begin{align}
v \left( e^{-iL_{ A,\hv}(-\infty)} \phi_0,  \mathcal R_+ \psi_0\right)= &
 \left( e^{-iL_{\bar A,\hv}(-\infty)} \phi_0, \right. \\ \notag
 & \left. \lim_{t \rightarrow \infty}\int_0^t\, d\tau e^{i\tau H_2} i
e^{-iL_{\bar A,\hv}(t-\tau)} \, [\Xi_b (x,t-\tau )+ \chi V(x)] e^{-i\tau H_1} \tilde{\psi}_0 \right)+
O(1/v),\label{5.44}
\end{align}
using $ \bar{A} (x + \tau \hv  ) = A(x + \tau \hv) $ for every $ x $ lying in the
support of
$\phi_0 $ and every real $\tau$ and \eqref{hvlwso.13}. $  \tilde{\psi}_0  $ is defined as in \eqref{hvlwso.12}: 
\beq
\tilde \psi_0:= g(\mo/ v^\rho) \,\psi_0, \hspace{2cm} \frac{1}{2} \leq \rho < 1.
\ene 
We designate (see the lines below \eqref{b})
\beq
\bar{\eta} (x,t):= \int_0^t (\hv\times \underline{B} )(x+\tau\hv)d\tau .
\label{bareta}
\ene
Following the proof of (5.59) of \cite{b-w.1} we obtain: 
\begin{align} \label{5.46}
\lim_{t \rightarrow \infty}\int_0^t\, d\tau e^{i\tau H_2} i
e^{-iL_{\bar A,\hv}(t-\tau)} \, & [\Xi_b (x,t-\tau )+ \chi V(x)] e^{-i\tau H_1} \tilde{\psi}_0  \\ \notag 
& = \lim_{t \rightarrow \infty}\int_0^t\, d\tau e^{i\tau H_2} i
e^{-iL_{\bar A,\hv}(t-\tau)} \, [\Xi_{\bar{\eta}} (x,t-\tau )+ \chi V(x)] e^{-i\tau H_1} \tilde{\psi}_0.
\end{align}
Eq. (\ref{5.46}) is similar to Eq. (5.59) in \cite{b-w.1} that
 is proved  in the step 2 of the proof of the Theorem 5.9 in 
\cite{b-w.1}.  The only differences between Eq. (\ref{5.46}) and Eq. 
(5.59) in \cite{b-w.1} are that in the first appear $\Xi_{\bar{\eta}}$ and $\bar A$ and in the second
 $ \Xi_\eta $ and $ A $ (or $A_C$ using the notation of \cite{b-w.1}).  We follow the steps 3 to 6 of the proof of Theorem 5.9 of \cite{b-w.1},
 changing each occurrence of $\Xi_\eta$ and $A_C$ in \cite{b-w.1} by $\Xi_{\bar{\eta}}$ and $\bar A$. We obtain \eqref{5.42}-\eqref{5.43}
  (and hence \eqref{hvlwso.24}) with $\Xi_{\bar{\eta}}$, instead of $ \Xi_\eta $, and $\bar A$, instead of $ A$.  Finally, we notice that
  $  \Xi_\eta $ and $A $ coincide with  $ \Xi_{\bar{\eta}} $ and $\bar A$, respectively, in the support $\phi_0$  and $ \psi _0 $, to get (\ref{hvlwso.24}). 
  
  \bull

\noindent{\bf \Large{Proof of Theorem \ref{reconstruction-formula-ii-g-m-p}}} \\
We use the notation of Corollary \ref{rem1} and its proof. We designate 
\beq \label{e-p-t.1.5.1}
S_\v( \underline{A} , V):= e^{-i mv \cdot x}  S(\underline{A}, V) e^{i mv \cdot x}.
\ene
By (\ref{scattering-oparetor-change-gauge}), see also Remark \ref{lamda-infty} and \eqref{hvlwso.3}, for any positive $s$, 
\begin{align}  \label{e-p-t.1.6}
\Big( v S(A, V) \phi_\v , \psi_{\v} \Big) =  & \: \: \: \: \: \left( v  S_\v( \underline{A} , V) ( e^{-i \lambda_{\tilde A, \infty}( -(\frac{s \mo}{mv} + s \hv)  )} -
 e^{-i \lambda_{\tilde A, \infty}( - s \hv  )} ) \phi_0 ,( e^{-i \lambda_{\tilde A, \infty}( \frac{s \mo}{mv} + s \hv) } -
 e^{-i \lambda_{\tilde A, \infty}(  s \hv  )} ) \psi_0 \right) \\ \notag & +
 \left( v S_\v( \underline{A}, V) e^{-i \lambda_{\tilde A, \infty}( - s \hv  )} \phi_0 ,  ( e^{-i \lambda_{\tilde A, \infty}( \frac{s \mo}{mv} + s \hv) } -
 e^{-i \lambda_{\tilde A, \infty}(  s \hv  )} ) \psi_0 \right)  \\ \notag & + \left( v   e^{-i \lambda_{\tilde A, \infty}( - s \hv  )}  S_\v(\underline{A}, V) \phi_ 0,
 e^{-i \lambda_{\tilde A, \infty}(  s \hv  )} \psi_0 \right)  \\ \notag & +
 \left( v S_\v( \underline{A}, V) ( e^{-i \lambda_{\tilde A, \infty}( -(\frac{s \mo}{mv} + s \hv)  )} -
 e^{-i \lambda_{\tilde A, \infty}( - s \hv  )} ) \phi_0 , e^{-i \lambda_{\tilde A, \infty}(  s \hv  )} \psi_0 \right).  
\end{align}
We denote by 
\beq \label{e-p-t.1.7} 
 R_{1,\v} : = \left( v  S_\v( \underline{A} , V) ( e^{-i \lambda_{\tilde A, \infty}( -(\frac{s \mo}{mv} + s \hv)  )} -
 e^{-i \lambda_{\tilde A, \infty}( - s \hv  )} ) \phi_0,( e^{-i \lambda_{\tilde A, \infty}( \frac{s \mo}{mv} + s \hv) } -
 e^{-i \lambda_{\tilde A, \infty}(  s \hv  )} ) \psi_0 \right)  .
\ene
We use Lemma 3.8 of \cite{b-w.1} to obtain [see also \eqref{e-p-l.3.2}]
\beq \label{esta}
| R_{1, \v} | \leq \frac{C}{v}\| \phi_0 \|_{{\bf H}^1(\ere^2)} \| \psi_0  \|_{{\bf H}^1(\ere^2)} .
\ene
We use  (\ref{hvlwso.18}) and (\ref{hvlwso.19}) to estimate the second and fourth terms of the right hand side of equation  (\ref{e-p-t.1.6}), respectively. We, furthermore, recall that 
[see Remark \ref{lamda-infty} and \eqref{a}]
\begin{align}\label{aquella} 
 a( \underline{A}, \hv, x) + \lambda_{\tilde A, \infty}(  \hv  ) - \lambda_{\tilde A, \infty}( - \hv  ) 
 = \int_{-\infty}^\infty   \underline{A}  (x + \tau \hv)\cdot \hv d \tau + \lambda_{\tilde A, \infty}(  \hv  ) - \lambda_{\tilde A, \infty}( - \hv  )   
 =   a( A, \hv, x) 
\end{align} 
and use Lemma 3.8 of \cite{b-w.1} to obtain: 
\begin{align}  \label{e-p-t.1.8}
\Big( v S(A, V) \phi_\v , \psi_{\v} \Big) =  & R_{1, \v} +  R_{2,\v}  \\ \notag &  +
 \left( v e^{i a(A, \hv, x)}   e^{-i \lambda_{\tilde A, \infty}(  \hv  )} \phi_0 ,  ( e^{-i \lambda_{\tilde A, \infty}( \frac{s \mo}{mv} + s \hv) } -
 e^{-i \lambda_{\tilde A, \infty}(  s \hv  )} ) \psi_0 \right) \\ \notag & +  \left( v   e^{-i \lambda_{\tilde A, \infty}( - s \hv  )}   S_\v( \underline{A}, V)  \phi_ 0,
 e^{-i \lambda_{\tilde A, \infty}(  s \hv  )} \psi_0 \right)  \\ \notag & +
 \left( v e^{i a(A, \hv, x)}  e^{i \lambda_{\tilde A, \infty}(  -\hv  )}  ( e^{-i \lambda_{\tilde A, \infty}( -(\frac{s \mo}{mv} + s \hv)  )} - e^{-i \lambda_{\tilde A, \infty}( - s \hv  )} ) \phi_0 , \psi_0 \right),  
\end{align}
where 
$$
\begin{array}{l}
R_{2,\v} := v \left( \left[\Big( S_\v(\underline{A}, V) -e^{i \int_{-\infty}^\infty\, \hv \cdot \underline{A}(x+\tau \hv)\cdot \hv d \tau} \Big ) \,
e^{-i \lambda_{\underline{A}}(- s \hv)}\right] \phi_0,\,( e^{-i \lambda_{\tilde A, \infty}(  \frac{s \mo}{mv} + s \hv) } -
 e^{-i \lambda_{\tilde A, \infty}(  s \hv  )} ) \psi_0 \right) \\\\ \hspace{1cm}+
v \left(  ( e^{-i \lambda_{\tilde A, \infty}( -(\frac{s \mo}{mv} + s \hv)  )} -
 e^{-i \lambda_{\tilde A, \infty}( - s \hv  )} )  \phi_0 ,
 \left[ S_\v( \underline{A}, V)-e^{i \int_{-\infty}^\infty\, \hv \cdot \underline{A}(x+\tau \hv)\cdot \hv d \tau} \right]^* e^{-i \lambda_{\tilde A, \infty}(  s \hv  )} \psi_0 \right),  
\end{array}
$$
satisfies,
\beq  \label{e-p-t.1.9}
|  R_{2,\v}  | \leq \frac{C}{v} \| \phi_0 \|_{{\bf H}^2(\ere^2)} \| \psi_0 \|_{{\bf H}^2(\ere^2)}.
\ene
Theorem \ref{reconstruction-formula-ii} and (\ref{aquella}) imply that
\begin{align}\label{noches}
 \left( v   e^{-i \lambda_{\tilde A, \infty}( - s \hv  )}   S_\v( \underline{A}, V)  \phi_ 0,
 e^{-i \lambda_{\tilde A, \infty}(  s \hv  )} \psi_0 \right) 
 = &  \hspace{.3cm} R (\v, \phi_0, \psi_0)  \\ \notag & + 
\left(-i e^{ia(\hv,x)}\int_{-\infty}^\infty
V(x+\tau\hv)\,d\tau \, \phi_0,\, \psi_0\right)\\ \notag 
& +
\left( -i e^{i a(\hv, x) }\int_{-\infty}^0\, \Xi_\eta (x+\tau \hv,-\infty)\,d\tau \,\phi_0,\psi_0\right) \\ \notag & +
\left(-i \int_{0}^\infty\, \Xi_\eta (x+\tau \hv,\infty)\,d\tau \, e^{ia(\hv, x)} \phi_0,\psi_0\right),
\end{align}
where $ R (\v, \phi_0, \psi_0) $ satisfies
\eqref{hvlwso.25}. 

Using Definition \ref{hvlwso-d.1}, \eqref{estimAinfty}, \eqref{e-p-t.1.5otro} and \eqref{e-p-c.1.1otroprima}, we find
\beq \label{e-p-t.1.8.2}
 \big |  A_\infty ( \hv )  -  \tilde A( s \hv ) s \big | \leq C 
\frac{\ln(e + s)}{(1 + s)^{\min(1, \delta - 1)}}, \hspace{1cm}   \Big \|  \int_1^\infty (\nabla \alpha_{\tilde A})(\tau s \hv)\cdot \frac{s\mo }{mv} d \tau ) \psi_0 \Big \|   \leq C  \frac{1}{v} \frac{\ln(e + s)}{(1 + s)^{\min(1, \delta-1)}}  
   \| \psi_0 \|_{{\bf H}^2(\ere^2)},
\ene
for sufficiently large $s$. \\
As a consequence of Lemma \ref{e-p-l.3} and \eqref{e-p-t.1.8.2}, we get (for every $\rho \in (0, 1)$)
\begin{align}  \label{e-p-t.1.8.3} 
  \Big \| \Big ( v \big [e^{  -i\lambda_{\tilde A, \infty}(\frac{s\mo }{mv} + s \hv)}  -e^{ - i \lambda_{\tilde A, \infty}( s\hv )} \big ] + & i e^{ -i\lambda_{ \tilde A, \infty}( s \hv )} A_\infty( \hv)\cdot \frac{\mo}{m}    
 \Big )  \phi_0 \Big \|_{L^2 (\ere^2) }  \\ \notag & \leq     
C \Big( \frac{1}{v} + \frac{\ln(e + s)}{(1 + s)^{\min(1, \delta-1)}}    + \frac{1}{v^{2 \rho - 1}} 
\frac{   \ln(e + |s|)(1 + (\frac{s}{v})^{1-\rho})  }{(1 + s)^{ \min(1, \delta - 1)(1 - \rho)} }
\Big ) \| \phi_0  \|_{{\bf H}^2(\ere^2)}
\end{align}
and (if $\delta > 2$)
\begin{align}  \label{e-p-t.1.8.3masuno} 
  \Big \| \Big ( v \big [e^{ - i\lambda_{\tilde A, \infty}(\frac{s\mo }{mv} + s \hv)}  -e^{ -i \lambda_{\tilde A, \infty}( s\hv )} \big ] + & i e^{ i\lambda_{ \tilde A, \infty}( s \hv )} A_\infty( \hv)\cdot \frac{\mo}{m}    
 \Big )  \phi_0 \Big \|_{L^2 (\ere^2) }  \\ \notag & \leq     
C \Big( \frac{1}{v} + \frac{\ln(e + s)}{(1 + s)^{\min(1, \delta-1)}}   \Big ) \| \phi_0  \|_{{\bf H}^2(\ere^2)}.
\end{align}
By direct inspection we verify that for every  $q \in (0, 1)$ we can take a big enough $c > 0$ and $\rho$ sufficiently close to $1$  such that, for $s = v^c$
\begin{align}\label{estimate2menos}
\Big( \frac{1}{v} + \frac{\ln(e + s)}{(1 + s)^{\min(1, \delta-1)}}    + \frac{1}{v^{2 \rho - 1}} 
\frac{   \ln(e + |s|)(1 + (\frac{s}{v})^{1-\rho})  }{(1 + s)^{ \min(1, \delta - 1)(1 - \rho)} }
\Big )\leq C \frac{1}{v^q},
\end{align}
where the constant $C$ depends on $q$. Similarly, an election $ s = v^c$ for big enough $c$ gives
\begin{align}\label{estimate2}
\Big( \frac{1}{v} + \frac{\ln(e + s)}{(1 + s)^{\min(1, \delta-1)}}   \Big ) \leq C \frac{1}{v},
\end{align}
if $\delta > 2$. The desired result follows from \eqref{e-p-t.1.8}-\eqref{estimate2}, arguing as in \eqref{e-p-t.1.8.3} to estimate the fourth line in \eqref{e-p-t.1.8}.
  
 \bull

\section{{Unique Reconstruction} of the Magnetic Field and the Electric Potential} \label{Reconstruction-of-the-Magnetic-Field-and-the-Electric-Potential}

\subsection{The Magnetic Field}\label{recmagnetic}
\begin{lemma}\label{r-m-f.1}
Let $A \in \mathcal{A}_{\Phi}(B)$. For every unit vector $\hv \in \mathbb{S}^2:$ 
\beq
\hv \times \nabla \int_{-\infty}^\infty \hv\cdot A(x+\tau \hv) \, d\tau= - \int_{-\infty}^{\infty}  B(x+\tau\hv)
\, d\tau,
\label{6.1}
\ene
in distribution sense in $\Lambda_{\hv}$.
\end{lemma}

\noindent {\it Proof:}
Denote by $ \hv := (\hv_1, \hv_2)  $ and by $  \bar{\hv} :=   (\hv_1, \hv_2, 0)  $. Consider the functions $ \bf{B} : \Lambda \times \mathbb{R} \to \mathbb{R}^3 $ given by 
$\bf{B}(x_1, x_2, x_3) := (0, 0, B(x_1, x_2))$ and $ \bf{A} : \Lambda \times \mathbb{R} \to \mathbb{R}^3 $ given by $ \bf{A}(x_1, x_2, x_3) := (A_1(x_1, x_2), A_2(x_1, x_2), 0)  $. 
 Then,  Equation (\ref{6.1}) is fulfilled if we prove
\beq
\nabla \int_{-\infty}^\infty \bar{\hv}\cdot \bf{A}( \bf{x}+\tau \bar{\hv}) \, d\tau= \int_{-\infty}^{\infty} \bar{\hv}\times B( \bf{x} + \tau  \bar{\hv})
\, d\tau,
\label{6.1.1}
\ene  
in distribution sense in $\Lambda_{\hv} \times \mathbb{R} $.
The last equation is proved in lemma 6.2 in \cite{b-w.1}. 
\bull

\begin{theorem}\label{r-m-p}
{We assume that the set $J$ defined in Definition \ref{o.1} equals $ \{ 1 \} $
 and that
$ K_{1} $ is convex. We suppose, furthermore, that $  \mathcal{P}(x)  B(x)$ is bounded for every polynomial  $   \mathcal{P}(x)   $.  
} Let $A \in \mathcal{A}_{\Phi}(B)$. The high-velocity limit  { \eqref{retira1}} of the scattering operator $ S(A,V) $, known  for all unit vectors $\hv$ and all $\phi_0\in {\bf H}^2(\ere^2)$ with support $\phi_0 \subset   \Lambda_{\hv} $, uniquely determines {(with a reconstruction method)} $B(x)$ for  almost every $ x \in \mathbb{R}^{2} \setminus K_1 $.
\end{theorem}
\noindent{\it Proof.}
For every unitary vector $\hv$ we denote by $  \Lambda_{\hv, 1} $ the set 
\begin{equation} \label{ast}
  \Lambda_{\hv, 1} :=  \big \{ x \in \mathbb{R}^2 \setminus K_{1} : x + t \hv \in \mathbb{R}^2 \setminus K_{1}, \ \forall t \in \mathbb{R}  \Big \}.
\end{equation}   
From the limit (\ref{hvlwso.18}) we uniquely reconstruct
$$
e^{i \int_{-\infty}^\infty \hv\cdot A(x+\tau\hv)\, d\tau}
$$
for all $x \in \Lambda_{\hv}$ and then, we {reconstruct} $ \int_{-\infty}^\infty \hv\cdot A(x+\tau\hv)\, d\tau+
2\pi n(x,\hv)$ with $n(x,\hv)$ an integer that is locally constant. By Lemma \ref{r-m-f.1} we uniquely
reconstruct 
\beq
\int_{-\infty}^{\infty}  B(x+\tau\hv)
\, d\tau
\label{6.4}
\ene
for a.e. $x \in \Lambda_{\hv}$. As $ K \setminus K_{1} $  is finite, we {uniquely reconstruct} (\ref{6.4}) for almost every $x \in \Lambda_{\hv, 1}$.\\
We take two fixed functions $ \phi_0 $ and $\psi_0$ belonging to $ C_0^{\infty}(\Lambda_{\hv, 1}) $. We suppose, furthermore, that their support
is contained in a ball $B_\epsilon(q)$, such that $ \overline{B_\epsilon(q)} \subset \mathbb{R}^2 \setminus K_1  $. \\
We define, recall that ${\rm d}(\cdot, \cdot)$ denotes the distance,
\beq \label{r-m-f-e.1.1}
K_{1, \epsilon} := \Big \{ x \in \mathbb{R}^2 : {\rm d}(x, K_1) \leq \epsilon   \Big \}
\ene 
and, for any $ z \in \mathbb{R}^2 $, 
\beq \label{r-m-f-e.1}
\phi(z) : = e^{-iz \cdot \mo}\phi_0, \ \ \ \ \ \ \ \ \ \ \ \ \ \ \ \  \psi(z) : = e^{-iz \cdot \mo}\psi_0. 
\ene
We set
\beq \label{r-m-f-e.2}
F(z) = ( B(x) \phi(z), \psi(z) ). 
\ene
As  $  \mathcal{P}(x)  B(x)$  is bounded for every polynomial  $   \mathcal{P}(x)   $, 
\beq \label{r-m-f-e.3}
| F(z) | \leq C_l(1 + |z|)^{l}, \ \ \ \ \forall \ l \in \mathbb{N}.
\ene
As we reconstruct (\ref{6.4}) from the high-velocity limit {\eqref{retira1}} of the scattering operator for almost every $ x \in \Lambda_{\hv, 1} $, we reconstruct the radon transform (see \cite{h})
\beq \label{r-m-f-e.4}
\tilde{F}(\hv, z) := \int_{- \infty}^{\infty} F(z + \tau \hv) d \tau,
\ene
for every $ z   $ such that $( z + \mathbb{R}\hv ) \cap (K_{1, \epsilon} - q  ) = \emptyset $. As $ K_{1, \epsilon} - q $ is convex, $F$
is continuous and $ \mathcal{P}(z) F(Z) $ is bounded for every polynomial  $ \mathcal{P} $, it follows from Theorem 2.6 and Corollary 2.8 of 
\cite{h} that we can {uniquely reconstruct}, from the high-velocity limits {\eqref{retira1}-\eqref{retira2}} of the scattering operator,  $F(z)$ for every $ z \in \mathbb{R}^2 \setminus \big ( K_{1, \epsilon} - q \big) $. As $ \overline{B_\epsilon(q)} \subset \mathbb{R}^2 \setminus K_1  $, 
$0 \notin K_{1, \epsilon} - q$ and we can take $z = 0$ to recover 
$ F(0)= ( B \phi_0  , \psi_0 ) $. Since we have the freedom to choose $ \phi_0, \ \psi_0, \ \epsilon  $ and $q$, then we can {uniquely reconstruct} $B$ almost 
everywhere in $ \ere^2 \setminus K_{1} $.
{
\begin{remark} 
A fundamental ingredient in the proof of Theorem \ref{r-m-p} is the use of Theorem 2.6 in \cite{h}. We can, actually, use this theorem (that requires convexity of the set $K$) because the radon transform can be extended to a distribution defined in the set of lines with empty intersection with $K_1$.   
\end{remark}
}

\subsection{The Electric Potential}\label{recelectric}

\noindent{\bf \Large{Proof of Theorem \ref{r-e-p}:}}\\
The part concerning the magnetic field $B$ is already proved in Theorem \ref{r-m-p}. 
The {reconstruction} of the fluxes modulo $2 \pi$ is done in Section \ref{recaharonov} below (see Theorem \ref{a-b-e}). We proceed to {uniquely reconstruct} $V$. 
By Theorems \ref{reconstruction-formula-I}, \ref{reconstruction-formula-ii-g-m-p} and \ref{r-m-p} 
[see also \eqref{hvlwso.2}-\eqref{hvlwso.21}], we uniquely reconstruct from the high-velocity 
limits {\eqref{retira1}-\eqref{retira2}} of the scattering operator   
\beq \label{r-e-p.1}
\int_{- \infty}^{\infty}( V(x + \tau \hv) \phi_0, \psi_0),
\ene
for every $ \phi_0 , \ \psi_0 \in  C_0^\infty (\ere ^2)$ with compact support in $ \Lambda_{\hv}$. Since $K \setminus K_1$ is a finite number 
of points and the multiplication operator by $ V(x + \tau \hv)  $ is self-adjoint, we can reconstruct (\ref{r-e-p.1}) for every    
  $ \phi_0 , \  \psi_0   \in C_0^\infty (\ere ^2)$ with compact support in the set $\Lambda_{\hv,1}$ [see \eqref{ast}]: We actually first analyze $\psi_0$ using density arguments and then take $ V(x + \tau \hv) $ to the other side of the inner product to analyze $\phi_0$. 
We take two fixed functions $ \phi_0 $ and $\psi_0$ belonging to $ C_0^{\infty}(\Lambda_{\hv,1}) $. 
We suppose, furthermore, that their support
is contained in a ball $B_\epsilon(q)$, such that $ \overline{B_\epsilon(q)} \subset \mathbb{R}^2 \setminus K_1  $. 
For any $ z \in \mathbb{R}^2 $  we define 
\beq \label{r-m-f-e.1prima}
\phi(z) : = e^{-iz \cdot \mo}\phi_0, \ \ \ \ \ \ \ \ \ \ \ \ \ \ \ \  \psi(z) : = e^{-iz \cdot \mo}\psi_0. 
\ene
We define 
\beq \label{r-m-f-e.2prima}
F(z) = ( V(x) \phi(z), \psi(z) ).
\ene
We proceed as in Equations (\ref{r-m-f-e.1.1} - \ref{r-m-f-e.4}) to {uniquely reconstruct} $ V(x)$ for almost every $x  \in \ere^2 \setminus K_1$.
\bull

\section{{{Reconstruction} of the Fluxes Modulo $2 \pi$ and Injectivity Modulo 
$4 \pi$}} \label{recaharonov}
In this section we suppose that $  \mathcal{P}(x)  B(x)$ is bounded for every polynomial  $   \mathcal{P}(x)   $. We suppose, furthermore, that the set $J$ defined in Definition \ref{o.1} equals $ \{ 1 \} $:
\beq \label{a-b-e.1}
K = K_1 \cup \{ x^{(2)}, x^{(3)}, \cdots, x^{(L)} \},
\ene   
and that $K_1$ is convex. 
For every $ i \in \{ 2, 3, \cdots, L \} $ we choose some fixed points $ z^i  $ and $ y^i $ in $ \ere^2 \setminus K $ and a unitary vector 
 $ \hv^i \in \mathbb{S}^1$. We define the sets,  
\beq \label{a-b-e.2}
D^i := {\rm convex} \big ((z^i + \ere \hv^i)\cup ( y^i +  \ere \hv^i  )\big ), 
\ene
where $ {\rm convex}(\cdot)  $ denotes the convex hull.\\
We assume that $ z^i  $, $y^i$ and $ \hv^i $ are chosen in such a way that $x^{(i)}$  belongs to the interior of $D^i$  and $K \cap D^i = \{x^{(i)}\}$.   \\
By Theorem \ref{reconstruction-formula-I} we  {reconstruct}, from the {high-velocity limit {\eqref{retira1}} of the} scattering operator,  
\beq \label{a-b-e.3}
 \int_{- \infty}^{\infty} A (y^i + \tau \hv^i)\cdot \hv^i d \tau -  \int_{- \infty}^{\infty}  A (z^i + \tau \hv^i)\cdot \hv^i  d \tau 
\ene 
modulo $ 2 \pi$.\\
Let $\varepsilon > 0$ be such that the ball $ B_\varepsilon(x^{(i)})  $ is contained in the interior of $D^i$, for $i \in \{ 2, \cdots, L \}$. By Stokes' theorem 
\beq \label{a-b-e.4}
 \int_{- \infty}^{\infty} A (y^i + \tau \hv^i)\cdot \hv^i d \tau -  \int_{- \infty}^{\infty}  A (z^i + \tau \hv^i)\cdot \hv^i  d \tau  + 
\int_{\partial B_{\varepsilon}(x^{(i)})}A = \int_{D^i \setminus B_{\varepsilon} (x^{(i)})} B. 
\ene 
By Theorem \ref{r-m-p} the {high-velocity limit {\eqref{retira1}} of the} scattering operator gives $  \int_{D^i \setminus B_{\varepsilon} (x^{(i)})} B  $  and, therefore, we {reconstruct}
$ \int_{\partial B_{\varepsilon}(x^{(i)})}A $ modulo $2 \pi$. Using again Stokes' theorem and Theorem \ref{r-m-p} we {reconstruct}
$\int_{\gamma_i} A = \Phi (K_i)$ modulo $2 \pi $ for $i \in \{ 2, \cdots, L \}$.  \\
Finally, we choose some fixed points $ z^1$ and $y^1$ in  the complement of $K$ and a unitary vector $ \hv^1 \in \mathbb{S}^1 $.
We define the set     
\beq \label{a-b-e.5}
D^1 := {\rm  convex} ((z^1 + \ere \v^1)\cup ( y^1 +  \ere \v^1  )). 
\ene
We choose ${\bf d} > 0  $ as in Equation \eqref{dist}.
We suppose that $ z^1$, $y^1$ and $\v^1$ are set in such a way that $ K  $ is contained in the interior of 
$D^{1}$ as well as the balls $B_{{\bf d}/4}(x^{{(i)}})$, for every $i \in I $. \\
By Stokes' theorem  
\beq \label{a-b-e.6}
 \int_{- \infty}^{\infty} A (y^1 + \tau \hv^1)\cdot \hv^1 d \tau -  \int_{- \infty}^{\infty}  A (z^1 + \tau \hv^1)\cdot \hv^1  d \tau  + 
\sum_{i \in \{ 2, \cdots, L \}}  \int_{\partial B_{{\bf d}/4}(x^{(i)})}A + \int_{\partial K_1} A  = \int_{D^1 \setminus (  K_1 \cup \cup_{ i  \in \{ 2, \cdots, L \}} B_{{\bf d}/4} (x^{(i)}))} B. 
\ene 
As we {reconstruct} from the {high-velocity 
limit {\eqref{retira1}} of the} scattering operator  $ \int_{\partial B_{{\bf d}/4}(x^{(i)})}A  $ modulo $2 \pi$ for every $i \in \{ 2, \cdots, L  \}$, it follows from 
Theorem  \ref{r-m-p} that we {reconstruct} $ \int_{\partial K_1} A  = \Phi(K_1)  $ modulo $2 \pi$. Then, we have proved the Theorem:
\begin{theorem} \label{a-b-e}

Let $A \in \mathcal{A}_{\Phi}(B)$.  We can {reconstruct} the fluxes
$\Phi (K_i)  $ modulo $2 \pi$, for every $ i \in \{1, \cdots, L\}  $, from the high-velocity limit  { \eqref{retira1}} of the scattering operator $S(A, V)$.  
\end{theorem}
{
\begin{remark}
The fact that the magnetic field can be {uniquely reconstructed}  from the high-velocity limit  {\eqref{retira1}} of the scattering operator and Theorem \ref{a-b-e} imply (see Definition \ref{def.tflux}) that the total flux $\Phi_B$ modulo $2 \pi$ can be {reconstructed} from the high-velocity limit 
{\eqref{retira1}} of the scattering operator.  
\end{remark} 
It turns out that injectivity with respect to the total flux can be proved not only modulo $2 \pi$ but modulo $4 \pi$; this was already proved in \cite{w.1}, taking $V = 0$, for the case of a connected obstacle and a compactly supported magnetic field. {This is an optimal result because, in the case that $K$ is a point, it is proved in \cite{ab} and \cite{ruij} that the scattering operator is the identity if the total flux is an even multiple of $2 \pi$.}  
The proof in \cite{w.1} can be easily adapted to our case. We obtain: 
\begin{theorem}\label{inj}
Let $A \in \mathcal{A}_{\Phi}(B)$ and $  \tilde A \in \mathcal{A}_{\tilde \Phi}(\tilde B) $.
{Suppose that the high-velocity limit \eqref{retira1} coincide  
for  $ S(A, V)$  and $ S(\tilde A, \tilde V) $}. Then, $B = \tilde B$ and 
\beq \label{estale}
\Phi_{B} =  \tilde \Phi_{\tilde B} \hspace{.5cm}({\rm Modulo} \: 4 \pi).
\ene
\end{theorem}
\emph{Proof:} The fact that $B = \tilde B $ is already proved above.
Set $A^{(c)}$ and $\tilde A^{(c)}$ the Coulomb gauges corresponding to $(B, \Phi )$ and 
$(\tilde B, \tilde \Phi)$, respectively, (see Proposition \ref{coupo}). For $\hat v \in \mathbb{S}^1$ and $ x \in \mathbb{R}^2 $
we set $ x = x_{\hat v} + x_{(\hat v)^\perp} $ the orthogonal decomposition of $x$ in the direction of $\hat v$ and the corresponding orthogonal direction (here we suppose that 
$(\hat v, (\hat v)^\perp)$ is a right oriented frame).  
Theorem \ref{reconstruction-formula-I} and Proposition \ref{coupo} imply that
\begin{align} \label{nops}
1 = \lim_{ x_{(\hat v)^\perp} \to - \infty }e^{i \int_{-\infty}^{^\infty} \hat v \cdot (A(x + \tau \hat v) - \tilde  A(x + \tau \hat v) ) d\tau }  = & e^{ i \Big [ \lambda_{A- A^{(c)},\infty}(\hat v) - \lambda_{A- A^{(c)},\infty}(-\hat v) - \big ( \lambda_{\tilde A - \tilde A^{(c)},\infty}(\hat v) - \lambda_{\tilde A - \tilde A^{(c)}, \infty}(-\hat v)\big ) \Big ] } 
\\ \notag & \cdot 
 \lim_{ x_{(\hat v)^\perp} \to - \infty }e^{i \int_{-\infty}^{^\infty} \hat v \cdot (A^{(c)}(x + \tau \hat v) - \tilde  A^{(c)) }(x + \tau \hat v) ) d\tau } \\ \notag = &
  e^{i\Big[\lambda_{A- A^{(c)},\infty}(\hat v) - \lambda_{A- A^{(c)},\infty}(-\hat v) - \big ( \lambda_{\tilde A - \tilde A^{(c)},\infty}(\hat v) - \lambda_{\tilde A - \tilde A^{(c)}, \infty}(-\hat v)\big )\Big ]} 
 \cdot 
e^{i (\Phi_B - \tilde \Phi_{\tilde B})/2 }.
\end{align} 
Eq. \eqref{nops} and the continuity of $\lambda_\infty$ imply that there is a fixed integer $N$ such that
\beq \label{nops1}
\lambda_{A- A^{(c)},\infty}(\hat v) - \lambda_{A- A^{(c)},\infty}(-\hat v) - \big ( \lambda_{\tilde A - \tilde A^{(c)},\infty}(\hat v) - \lambda_{\tilde A - \tilde A^{(c)}, \infty}(-\hat v)\big ) 
 + (\Phi_B - \tilde \Phi_{\tilde B})/2 = 2 \pi N.
\ene
Substituting $\hat v $ by $- \hat v$ in \eqref{nops1} and subtracting the resulting equation to \eqref{nops1} we get  
 \beq \label{nops2}
\lambda_{A- A^{(c)},\infty}(\hat v) - \lambda_{A- A^{(c)},\infty}(-\hat v) - \big ( \lambda_{\tilde A - \tilde A^{(c)},\infty}(\hat v) - \lambda_{\tilde A - \tilde A^{(c)}, \infty}(-\hat v)\big ) 
 = 0
\ene
and, therefore, 
\beq \label{nops3}
\Phi_B - \tilde \Phi_{\tilde B} = 4 \phi N.
\ene
This is Equation \eqref{estale}.  
\bull
}

\section{{{Unique Reconstruction} of the Long-Range Part of the Magnetic Potential} }\label{recincons}
 {In this section we suppose that $ B \in C^{2}(\overline{\Lambda})$ is such that 
 $ | B(x)| \leq C\frac{1}{(1 + |x|)^\mu} $, $ | \frac{\partial}{\partial x_i} B(x)| \leq C \frac{1}{(1 +|x|)^{\mu + 1}} $, $ | \frac{\partial}{\partial x_j}\frac{\partial}{\partial x_i} B(x)| \leq C \frac{1}{(1 + |x|)^{\mu + 2}} $, for every  $i, j \in \{ 1, 2 \}$ and every $x \in \Lambda$. 

 }
\subsection{ {General Results in the Presence of an Electromagnetic Field}} \label{pmdre}

\noindent{\bf \Large{Proof of Theorem \ref{Theorem-Inconsistencies-1} : }} \\
{In this proof recall Section \ref{basic-formulas}.  For $\hv \in \mathbb{S}^1$
we set $ (\hv)^\perp \in \mathbb{S}^1$ the orthogonal (right oriented) complement of $\hv$. 
For every $\phi_0, \psi_0 \in  \bf{H}^6(\mathbb{R}^2) $, compactly supported in $\Lambda_{\hv}$, we denote:

\begin{align} \label{delo}
\Upsilon(\phi_0, \psi_0)  := &
 \left(-i e^{ia(A, \hv,x)}\int_{-\infty}^\infty
V(x+\tau\hv)\,d\tau \, \phi_0,\, \psi_0\right) 
+
\left( -i e^{i a(A, \hv, x) }\int_{-\infty}^0\, \Xi_\eta (x+\tau \hv,-\infty)\,d\tau \,\phi_0,\psi_0\right) \\ \notag & +
\left(-i \int_{0}^\infty\, \Xi_\eta (x+\tau \hv,\infty)\,d\tau \, e^{ia(A, \hv, x)} \phi_0,\psi_0\right) .  
\end{align}   
Let $R > 0 $ such that $K \subset B_{R}(0) $. Suppose that $\phi $ and $\psi$ are supported in $B_1(0)$ and set, for $\tau > R + 1$, 
\beq \label{de}
\phi_\tau(x) : = \phi(x + \tau (\hv)^\perp), \hspace{1cm}  \psi_\tau(x) := \psi(x + \tau (\hv)^\perp).  
\ene
The decay properties of $V$ and $B$, and \eqref{hvlwso.2}-\eqref{hvlwso.21} imply that
\beq \label{de1}
\lim_{\tau \to \infty} \Upsilon(\phi_\tau, \psi_\tau) = 0. 
\ene
Set $A^{(c)}$ the Coulomb gauge corresponding to $\Phi $ and $B$ (see Proposition \ref{coupo}). Proposition \ref{coupo} implies that
\beq \label{de2} 
\lim_{\tau \to \infty} e^{i a(A, \hv, x) } \phi_\tau -   
e^{i \big ( \lambda_{A - A^{(c)}} (\hv) - \lambda_{A - A^{(c)}} (-\hv) + \Phi_B/2 \big ) } \phi_\tau 
 = 0, 
\ene
and the same formula holds for $\psi_\tau$. With the help of \eqref{de1} and \eqref{de2},  Theorem 
 \ref{reconstruction-formula-ii-g-m-p} 
implies that the high-velocity limits {\eqref{retira1}-\eqref{retira2}} of the scattering operator uniquely determine 
{(with a reconstruction method)}
\begin{align}\label{incon}
\Big(i e^{i \big ( \lambda_{A - A^{(c)}} (\hv) - \lambda_{A - A^{(c)}} (-\hv) + \Phi_B/2 \big ) } & \big ( A_{\infty}( \hv) + A_\infty(- \hv)\big ) \cdot \frac{\mo }{m}  \phi, \psi \Big).
\end{align}
 Theorem \ref{reconstruction-formula-I} and \eqref{de2} imply that the scattering operator gives $ e^{i \big ( \lambda_{A - A^{(c)}} (\hv) - \lambda_{A - A^{(c)}} (-\hv) + \Phi_B/2 \big ) } $. Thus, we can {uniquely reconstruct} 
 from the {high-velocity limits {\eqref{retira1}-\eqref{retira2}} of the} scattering operator
$$
\big ( A_{\infty}(\hv) + A_{\infty}(- \hv)\big) \cdot \mo   \phi,
$$ 
from which we {uniquely reconstruct} $  A_{\infty}(\hv) + A_{\infty}(- \hv) $, selecting an appropriate $\phi$. 
}
\bull

{
The next Theorem was already proved in \cite{nico1}, in the case that $B \in C_0^\infty(\mathbb{R}^2)$ and $K$ is convex, using stationary methods (see Theorem 7 in \cite{nico1}). Here we give a different proof using time-dependent methods for our, more general, class of magnetic fields and obstacles.
\begin{prop}[The Case of the Coulomb Gauge] \label{casec}
We assume that the set $J$ defined in Definition \ref{o.1} equals $ \{ 1 \} $
 and that
$ K_{1} $ is convex. Suppose, furthermore, that  $ 
     \mathcal{P}(x) V(x) (\mo^2 + 1)^{-1}   $ and $   \mathcal{P}(x)  B(x) $ are bounded for every polynomial  $   \mathcal{P}(x)   $.  
Let $A^{(c)} \in \mathcal{A}_{\Phi,  \delta}(B)$, for some $\delta > 1$, be the Coulomb gauge (see Proposition \ref{coupo}). 
We can {uniquely reconstruct}, from the high-velocity limits {\eqref{retira1}-\eqref{retira2}} of the scattering operator $S(A^{(c)}, V)$,
\beq \label{masmenos1}
 \Phi_B B + V
\ene
almost everywhere.
\end{prop}
\emph{Proof:}
From Theorems \ref{reconstruction-formula-I}, \ref{reconstruction-formula-ii-g-m-p}, \ref{r-m-p} and
\ref{Theorem-Inconsistencies-1}, and Eq. \eqref{6.1.1} [recall also \eqref{hvlwso.1}-\eqref{a}] we deduce that the high-velocity limits {\eqref{retira1}-\eqref{retira2}} of the scattering operator give
\begin{align}\label{fanta}
 \Bigg ( i e^{i a(A^{(c)}, \hv, x)} \Big[ A^{(c)}_{\infty}( \hv) \cdot \int_{-\infty}^{\infty} \hv\times B( x + \tau  \hv) &  - V( x + \tau  \hv) \Big ]  \phi_0,  \psi_0 \Bigg )\\ \notag & =
 \Bigg ( -i e^{i a(A^{(c)}, \hv, x)} \Big[ \Phi_B B( x + \tau  \hv)  + V( x + \tau  \hv) \Big ]  \phi_0,  \psi_0 \Bigg )
\end{align}
for all $\phi_0, \psi_0 \in {\bf H}^6(\ere^2)$ with compact support in $\Lambda_{\hv}$ and every $\hv \in \mathbb{S}^1$. In Eq. \eqref{fanta} we use Corollary \ref{tasanan} and 
$$  A^{(c)}_{\infty}( \hv)\cdot \int_{-\infty}^{\infty} \hv\times B( x + \tau  \hv)
= \Big( \int_{-\infty}^{\infty}  B( x + \tau  \hv)\Big )  A^{(c)}_{\infty}( \hv)  \cdot \begin{pmatrix}
\hv_2 \\ - \hv_1
\end{pmatrix} .
$$  
Following the proof of Theorem \ref{r-e-p}, using Eq. \eqref{fanta}, we obtain the desired result. 
\bull
}

\subsection{{Injectivity with Respect to the Long-Range Part, Assuming the Knowledge of $V$}} \label{label}
{In this section we assume that the set $J$ defined in Definition \ref{o.1} equals $ \{ 1 \} $,
$ K_{1} $ is convex and that  $  \mathcal{P}(x)  B(x)$ is bounded for every polynomial  $   \mathcal{P}(x)   $.}

{
{Recall that, in the absence of magnetic field outside the obstacle, the scattering operator is not in general injective with respect to the long-range part of the magnetic potential. % if the total flux is an integer  multiple of $2 \pi$ (see \cite{nico1} , \cite{ruij} and \cite{EI2}-\cite{EI1}). 
Nevertheless, we prove here that if the exterior magnetic field does not identically vanishes, then the injectivity is achieved (uniquely reconstructing the long-range part of the magnetic potential is also possible under different conditions, see Section \ref{rec}).}  \\   
In this section we assume we know $V$ and prove injectivity of the high-velocity limits 
{\eqref{retira1}-\eqref{retira2}} of the scattering operator with respect to the long-range part of the magnetic potential. Similar results are obtained in \cite{EI2}-\cite{EI1}. In \cite{EI2}-\cite{EI1} the knowledge of $V$ is also necessary {(we explain this in the lines above Section \ref{modell})}.   

In this section we restrict our class of magnetic potentials to the functions 
$A \in \mathcal{A}_{\Phi, \delta}(B)$, for $\delta > 1$, such that 
\beq \label{i1}
A_\infty\big ((\cos(\theta), \sin(\theta)) \big ) = f_A(\theta)\begin{pmatrix}
- \sin(\theta) \\ \cos(\theta)
\end{pmatrix} 
\ene      
and  $ f_A$ is real analytic. We assume additionally that $B \ne 0$. In \cite{EI2}-\cite{EI1} different assumptions are required, stronger in some sense and weaker in another sense. There it is assumed that 
$\Phi_B$ is not an integer multiple of $2 \pi$, but the magnetic field can vanish. The class of magnetic potentials considered is also different. In \cite{EI2}-\cite{EI1} the magnetic potentials must be of the form
\beq \label{es}
A_{EI} (r(\cos(\theta), \sin(\theta)) = \frac{1}{r} f_{A_{EI}}(\theta)\begin{pmatrix}
- \sin(\theta) \\ \cos(\theta)  \end{pmatrix} + A_{sr}(r(\cos(\theta), \sin(\theta)),
\ene    
where $A_{sr}$ is an infinitely differentiable short-range magnetic potential, $f_{A_{EI}}$ is $C^\infty$ and {
\beq
\Big| \frac{\partial^{\upsilon}}{\partial x^{\upsilon}} (A_{EI}(x)\cdot x)\Big | \leq C_\upsilon (1 + |x|^2)^{- (1 + |\upsilon|)/2}, 
\ene 
for every multi-index $\upsilon$}. The main difference between the classes is that in \cite{EI2}-\cite{EI1} the long-range term must be homogeneous of degree $- 1$ (while here it is not) and that in this paper 
$f_A$ must be real analytic (while in \cite{EI2}-\cite{EI1} only infinitely differentiability is required). The methods and data are also different in both approaches. In this paper we need only to know the high-velocity limits {\eqref{retira1}-\eqref{retira2}} of the scattering operator, while in \cite{EI2}-\cite{EI1} all energies (including 
the high-velocity limits {\eqref{retira1}-\eqref{retira2}}) are used. Here only 
time-dependent methods are used, while in \cite{EI2}-\cite{EI1} stationary and time-dependent methods are addressed. An additional restriction assumed in \cite{EI2}-\cite{EI1} is the obstacle to be convex.        
}  
{
\begin{theorem}\label{insecto}
Let $\delta > 1$ and $A \in \mathcal{A}_{\Phi, \delta}(B), \tilde A \in \mathcal{A}_{\tilde \Phi, \delta}(\tilde B)$ such that $A_\infty - \tilde A_\infty $ satisfies \eqref{i1}. Suppose that $B \ne 0$. {If the limits 
\eqref{retira1}-\eqref{retira2} coincide for $S(A, V) $ and $ S(\tilde A, V)$, then $B = \tilde B$, $\Phi_B = \tilde \Phi_{\tilde B}$
 and $A_\infty = \tilde A_\infty$. }
\end{theorem}
\emph{Proof:} $B = \tilde B$ is proved in Theorem \ref{r-m-p}.
From Theorems \ref{reconstruction-formula-I}, \ref{reconstruction-formula-ii-g-m-p}, \ref{r-m-p} and
\ref{Theorem-Inconsistencies-1}, and Eq. \eqref{6.1.1} [recall also \eqref{hvlwso.1}-\eqref{a}] we deduce that $S(A, V) = S(\tilde A, V)$ implies 
\begin{align}
 \big ( A_{\infty}( \hv) - \tilde A_{\infty}( \hv)\big )\cdot \int_{-\infty}^{\infty} \hv\times B( {\bf x} + \tau  \hv)   \phi_0 = 0
\end{align}
for all $\phi_0 \in {\bf H}^6(\ere^2)$ with compact support in $\Lambda_{\hv}$ and every $\hv \in \mathbb{S}^1$ and, therefore, 
$$ \big ( A_{\infty}( \hv) - \tilde A_{\infty}( \hv)\big )\cdot \int_{-\infty}^{\infty} \hv\times B( x + \tau  \hv)
= \Big( \int_{-\infty}^{\infty}  B( x + \tau  \hv)\Big )  \big ( A_{\infty}( \hv) - \tilde A_{\infty}( \hv)\big ) \cdot \begin{pmatrix}
\hv_2 \\ - \hv_1
\end{pmatrix}   = 0.
$$  
As $   A_{\infty}( \hv) - \tilde A_{\infty}( \hv) $ is transverse, then it vanishes whenever $ \int_{-\infty}^{\infty}  B( x + \tau  \hv) \ne 0  $. As $B \ne 0$, the support theorem for the radon transform (Theorem 2.6 and Corollary 2.8 in 
\cite{h}) implies that there is a $ x_0  $ and a $\hv_0 $ with $x_0 + \mathbb{R} \hv_0 \subset \Lambda_{\hv_0}$ such that 
\beq \label{caray}
 \int_{-\infty}^{\infty}  B( x_0 + \tau  \hv_0) \ne 0.  
\ene
As this integral is continuous with respect to $\hv_0$, the same holds in a neighborhood of 
$\hv_0$. Therefore, for $\hv $ in this neighborhood, $   A_{\infty}( \hv) - \tilde A_{\infty}( \hv) $
vanishes. It follows from analyticity that
\beq \label{yaa}
 A_{\infty} -\tilde  A_{\infty} = 0. 
\ene   
Once this is proved, $\Phi_B = \tilde \Phi_{\tilde B}$ is a consequence of Corollary \ref{tasanan}, integrating $A_{\infty} -\tilde  A_{\infty}$ over the unit circle. 
}
\subsection{{{Unique Reconstruction} of the Long-Range Part, Asuming the Knowledge of $V$}}\label{rec}
{In this section we assume that the set $J$ defined in Definition \ref{o.1} equals $ \{ 1 \} $,
$ K_{1} $ is convex and that  $  \mathcal{P}(x)  B(x)$ is bounded for every polynomial  $   \mathcal{P}(x)   $. Definition \ref{deft} and Remark \ref{nopu} are frequently used in this part.}

{
\begin{lemma} \label{radon}
Let $O$ be an open set in $\mathbb{S}^1$. Suppose that for every line $\ell \subset \mathbb{R}^2 \setminus \mathcal{C}(O)$ {[recall \eqref{eca}]}
\beq \label{def1}
\int_\ell B = 0.
\ene
Then 
\beq \label{defloc}
{\rm supp}(B) \subset \mathcal{C}(O). 
\ene 
\end{lemma}
\emph{Proof:} The result is a direct consequence of the support theorem for the radon transform (Theorem 2.6 and Corollary 2.8 in \cite{h}), since \eqref{def1} signifies 
that the radon transform of $B$ vanishes in the lines $\ell$ and $ \mathcal{C}(O) $ is compact and convex {(see Remark \ref{nopu})}. 
\bull
}
{
\\
\noindent { \bf \Large{Proof of Theorem \ref{elprinc} :}}  \\ 
From Theorems \ref{reconstruction-formula-I}, \ref{reconstruction-formula-ii-g-m-p}, \ref{r-m-p} and
\ref{Theorem-Inconsistencies-1}, and Eq. \eqref{6.1.1} [recall also \eqref{hvlwso.1}-\eqref{a}] we deduce that the high-velocity limits {\eqref{retira1}-\eqref{retira2}} of the scattering operator give {(with a reconstruction method)}
\begin{align}
  A_{\infty}( \hv) \cdot \int_{-\infty}^{\infty} \hv\times B( \bf{x} + \tau  \hv)   \phi_0 
\end{align}
for all $\phi_0 \in {\bf H}^6(\ere^2)$ with compact support in $\Lambda_{\hv}$ and every $\hv \in \mathbb{S}^1$, and, therefore, they give 
$$  A_{\infty}( \hv)\cdot \int_{-\infty}^{\infty} \hv\times B( x + \tau  \hv)
= \Big( \int_{-\infty}^{\infty}  B( x + \tau  \hv)\Big )  A_{\infty}( \hv)  \cdot \begin{pmatrix}
\hv_2 \\ - \hv_1
\end{pmatrix} .
$$  
As $   A_{\infty}( \hv)  $ is transverse {[see \eqref{trans}]}, we {uniquely reconstruct} it whenever $ \int_{-\infty}^{\infty}  B( x + \tau  \hv) \ne 0  $, for some $x \in \Lambda_{\hv}$. { Suppose that 
$\hv \notin \mathcal{D}(B) $. Lemma \ref{radon} implies that in every neighborhood of $\hv$ there is a 
$\hat{\bf w}_0 \in \mathbb{S}^1$ and a $x_0 \in \Lambda_{\hat{\bf w}_0}$ such that $ \int_{-\infty}^{\infty}  B( x_0 + \tau  \hat{\bf w}_0) \ne 0  $}. We can, therefore, {uniquely reconstruct} $A_\infty(\hat{\bf w}_0)$. As $A_\infty$ is continuous and the referred neighborhood is arbitrary, then we can reconstruct $A_\infty(\hv)$.  We have proved that $ \hv \notin \mathcal{D}(B) $ implies that $A_\infty(\hv)$ can be {uniquely reconstructed}. This certainly assures the reconstruction of $A_\infty$, provided $ \mathcal{D}(B) = \emptyset $. In this case  
$\Phi_B$ is obtained from Corollary \ref{tasanan}, after an integration over the unit circle.  \\
Since for every dense set $\mathcal{S}$ of $\mathbb{S}^1$
\beq \label{dense}
K_1 = \bigcap_{\hv \in \mathcal{S} }\Big ( K_1 + \mathbb{R} \hv \Big ), 
\ene  
then, $\mathcal{D}(B)$ cannot be dense, unless $B = 0$ (recall that $B$ is  {defined in $\Lambda$). Thus, $B \ne 0$ implies that  $A(\hv)$ can be uniquely reconstructed from the high-velocity limits {\eqref{retira1}-\eqref{retira2}} of the scattering operator, for every $\hv$ in some open set in $\mathbb{S}^1$ (this argument does not even use the continuity of $A_\infty$).}

\bull
}

{
In the following proposition we assert that if the magnetic field and the flux function $\Phi$ are such that $A_\infty$ cannot be fully {reconstructed} by our method, we can add a short-range magnetic potential (that does not alter the flux function $\Phi$, nor the long-range part of the magnetic potential and not either $\Phi_B$)
 that allows us to {uniquely reconstruct} $A_\infty$ and $\Phi_B$. Physically, this implies turning on a short-range magnetic field.     
\begin{prop} \label{aquien}
Let $\delta > 1$ and $A \in  \mathcal{A}_{\Phi, \delta}(B)$. Suppose that we know $V$. There exists a short-range magnetic field $B_{sr} $ in the Schwartz space, whose support does not intersect the support of $B$ and $K$, and a short-range magnetic potential $A_{sr} \in \mathcal{A}_{0, \tilde \delta} (B_{sr})$ (for some $\tilde \delta > 1$) such that $A_\infty$
and $\Phi_B$ can be {uniquely reconstructed} from the high-velocity limits 
{\eqref{retira1}-\eqref{retira2}} of $S(A + A_{sr}, V)$. Notice that $ A + A_{sr} $ and $A$ have the same fluxes $\Phi$ and the same long-range part $A_\infty$, as well as the same total flux.    
\end{prop}
\emph{Proof:}
If $ \mathcal{D}(B) = \emptyset $, then we take $B_{sr} = 0$ (and thus $A_{sr} = 0$). If not, set 
$\hv \in \mathcal{D}(B)$. Then $B$ is compactly supported in $ \mathcal{C}(N_{\hv})  $ for some open neighborhood $\mathcal{N}_{\hv}$ of $\hv $ in $\mathcal{S}^1$.  Set $r > 0$ such that $K$ and the support of $B$ are contained in $B_r(0)$.
Take $B_{sr}$ as any non-compactly supported function in the Schwartz space, for example, whose integral is zero and such that its support is contained in 
$\mathbb{R}^2 \setminus B_{r + {\bf d}}(0)$; see \eqref{dist}. We set $  A_{sr}$ the Coulomb gauge for $ B_{sr}$ (in $\mathbb{R}^2$). As $  B + B_{sr} $ is not compactly supported, Theorem \ref{elprinc} and Remark \ref{nopu} imply that $ A_\infty $ is {uniquely reconstructed} from the limits {\eqref{retira1}-\eqref{retira2}} for  $S(A + A_{sr}, V)$. The assertion for 
$\Phi_B$ is a consequence of Corollary \ref{tasanan}, after an integration over the unit circle.    
\bull
}

\section{{Physical Considerations}}

The two dimensional scattering problem that we consider in this paper is important in the context of the Aharonov-Bohm effect  (\cite{ab}, \cite{Franz} and \cite{es}).  This effect  is a fundamental issue in physics that has been
extensively studied in the literature. The issue at stake is what are the fundamental electromagnetic quantities in quantum physics, in particular if the magnetic potentials have a physical significance. The  two dimensional models are an idealization of large solenoids that are considered  as infinitely long, what makes the problem translation invariant along the axis of the solenoids and makes it possible to reduce the problem to a two dimensional one. This is actually the model considered in the original papers  \cite{ab}, \cite{Franz} and \cite{es}. See also \cite{ruij}, and for a complete review up to 1989 see \cite{op} and \cite{pt}. For more recent contributions see, for example, \cite{nico1}, \cite{w.1}, \cite{ry1}, \cite{ry2}, \cite{EI2}, \cite{EI1} and their references. Of course, a physical solenoid will always be finite, and no matter how long it is the space outside it will be simply connected and there will be no Aharonov-Bohm effect. Actually, the Aharonov-Bohm effect only appears in these models in the limit of the infinite solenoid when the problem is reduced to two dimensions and the domain where the electrons propagate is not simply connected. For example, it is the exterior of a disc if the infinite solenoid is a cylinder. Furthermore, the magnetic field will always leak outside of a finite solenoid.  These, and another reasons, motivated the study of the Aharonov-Bohm effect in three dimensions,  in the case when the hidden fluxes are contained in the interior of toroidal magnets, or more generally handle bodies. Note that due to its non trivial topology a torus can contain inside a magnetic field  without any leak. This was done experimentally in  \cite{caprez}, \cite{Tonomura1}-\cite{Tonomura4} and theoretically in   \cite{b-w.1}-\cite{b-w.3}.

According to the complete description of electromagnetism in terms of non-integrable phase factors introduced in \cite{wuyang} (see also \cite{Dirac})  the  physically relevant quantities, that can be measured in experiments,  have to be gauge invariant and the only observable quantities related to inaccessible magnetic fields  are hidden fluxes modulo $ 2 \pi$,  i.e., the mathematical objects describing physically relevant quantities have to remain unchanged if the hidden fluxes are changed by adding an integer multiple of $2 \pi$. In three dimensions short-range magnetic potentials are available as long as there are no magnetic monopoles (see \cite{grif}). Then, it is natural to only consider short-range magnetic potentials, and in this case (see  \cite{b-w.1}-\cite{b-w.3}) the scattering operator is gauge invariant  and it remains unchanged if the hidden fluxes are changed by adding an integer multiple of $2 \pi$. In consequence, scattering theory based in the scattering  operator provides a theoretical framework for the Aharonov-Bohm effect in three dimensions, when the hidden fluxes are contained inside tori, or handle bodies, that is consistent with the complete description of electromagnetism in terms of non-integrable phase factors.           

The situation for the two dimensional models is fundamentally different. In two dimensions there are no short-range magnetic potentials as long as the total magnetic flux does not vanish. The need to use long-range magnetic potentials leads to long-range effects. A consequence of these long-range effects is that the scattering operator is not gauge invariant and it does not remain unchanged if the hidden fluxes are changed by adding an integer  multiple of $2 \pi$  (note however that in the case of the infinitely long straight solenoid studied by Aharonov-Bohm in \cite{ab}, the relative phase shift between electrons that travel to the left and to the right of the solenoid is gauge invariant and  invariant by changing the hidden flux by adding an integer multiple of $2 \pi$, which implies that their prediction contains no contradiction).
 This means that the scattering operator contains more information than what can be measured in experiments. For example, as we have proved, we can uniquely reconstruct from the scattering operator the long-range part, $A_\infty(\hv)$, of  the magnetic potential that depends on the gauge and is not invariant by adding to the flux an integer multiple of $2 \pi$ (see Corollary \ref{tasanan}). 
These problems are due to the fact that the idealization of having infinite long solenoids to reduce the problem to two dimensions  produces conceptual problems. This is the price to pay if we want to reduce by one the number of dimensions.


\newpage



\begin{thebibliography}{99}
\bibitem{ab} Aharonov, Y.; Bohm, D. Significance of electromagnetic potentials in the quantum theory. Phys. Rev. (2) 115 (1959) 485-491.
\bibitem{arians} Arians, S. Geometric approach to inverse scattering for the Schr\"odinger's equation with magnetic and
electric potentials. J. Math. Phys. 38 (1997) 2761-2773. 
\bibitem{b-w.1} Ballesteros, M.; Weder, R.
High-velocity estimates for the scattering operator and Aharonov-Bohm effect in three dimensions.
Comm. Math. Phys. 285 (2009), no. 1, 345-398. 
\bibitem{b-w.2} Ballesteros, M.; Weder R. The Aharonov-Bohm effect and Tonomura et al. experiments: Rigorous results, J. Math. Phys. 50 (2009), no 12, 122108, 54pp.
\bibitem{b-w.3} Ballesteros, M.; Weder, R. Aharonov-Bohm Effect and High-Velocity Estimates of Solutions to the Schr\"odinger Equation. Commun. Math. Phys. 303 (2011) 175-211.
\bibitem{caprez} Caprez A.; Barwick, B.; Batelaan, H. Macroscopic test of the Aharonov-Bohm effect. Phys. Rev. Lett. 99 (2007) 210-401.
\bibitem{cham} Chambers, R.G. Shift of an electron interference pattern by enclosed magnetic flux. Phys. Rev. Lett. 5 (1960) 35.
\bibitem{dr} de Rham, G. Differentiable manifolds. Forms, currents, harmonic forms. 
Translated from the French by F. R. Smith. 
With an introduction by S. S. Chern. Grundlehren der Mathematischen Wissenschaften [Fundamental Principles of Mathematical Sciences]
 266 Springer-Verlag, Berlin, 1984. x+167 pp. 
\bibitem{Dirac}  Dirac, P. Quantized singularities in the electromagnetic field. Proc. R. Soc. A
133 (1931) 60-72.
\bibitem{es} Ehrenberg,W.; Siday, R.E. The refractive index in electron optics and the principles of dynamics. Proc.
Phys. Soc. London B 62 (1949) 8-21.
\bibitem{Enss-Weder} Enss, V.; Weder, R. The geometrical approach to multidimensional inverse scattering. J. Math.
Phys. 36 (2005) 3902-3921.
\bibitem{EI2} Eskin, G.; Isozaki, H.; O'Dell, S. Gauge equivalence and inverse scattering for Aharonov-Bohm effect. Comm. Partial Differential Equations 35 (2010), no. 12, 2164-2194.
\bibitem{EI1} Eskin, G.; Isozaki, H. Gauge equivalence and inverse scattering for long-range magnetic potentials. Russ. J. Math. Phys. 18 (2011), no. 1, 54-63.
\bibitem{EI3} Eskin, G.; Ralston, J. Gauge equivalence and the inverse spectral problem for the magnetic Schr\"odinger operator on the torus. Russ. J. Math. Phys. 20 (2013), no. 4, 413-423. 
\bibitem{Franz} Franz, W. Elektroneninterferenzen im Magnetfeld. Verh. D. Phys. Ges. (3)
20 Nr.2 (1939) 65-66; Physikalische Berichte 21 (1940) 686.
\bibitem{grif} Griffiths, D.J. Introduction to Electrodynamics. Third edition. Prentice-Hall New Jersey, 1999. 576pp.   
\bibitem{h}  Helgason, S. The Radon transform. Second edition. Progress in Mathematics, 5. Birkhauser Boston, Inc., Boston, MA, 1999. xiv+188 pp.
\bibitem{wj.1} Jung, W. Gauge transformations and inverse quantum scattering with medium-range magnetic fields.  Math. Phys. Electron. J.  11  (2005), Paper 5 (electronic). 
\bibitem{l-t} Loss, M.; Thaller B. Scattering of particles by long-range magnetic fields.
 Ann. Phys. 176 (1987) 159-180 .
%\bibitem{nico0} Nicoleau, F. A stationary approach to inverse scattering for Schrödinger operators with first order %perturbation. Comm. Partial Differential Equations 22 (1997), no. 3-4, 527-553.
\bibitem{nico1} Nicoleau, F. An inverse scattering problem with the Aharonov-Bohm effect. J. Math. Phys. 41 (2000), no. 8, 5223-5237.
\bibitem{op} Olariu, S.; Popescu, I.I. The quantum effects of electromagnetic fluxes. Rev. Mod. Phys. 57 (1985)
339-436.
\bibitem{pt} Peshkin, M.; Tonomura, A. The Aharonov-Bohm Effect. Lecture Notes in Phys. 340 Springer-
Verlag, Berlin, 1989.
\bibitem{r-s.1} Reed, M.; Simon, B. Methods of Modern Mathematical Physics. I. Functional Analysis. Academic Press, New York-London, 1972. xvii+325 pp. 
\bibitem{r-s.2} Reed, M.; Simon, B. Methods of Modern Mathematical
  Physics. II. Fourier Analysis, Self-adjointness. Academic Press [Harcourt
  Brace Jovanovich, Publishers], New York-London, 1975. xv+361 pp. 
\bibitem{ruij} Ruijsenaars, S. N. M. The Aharonov-Bohm effect and scattering theory. Ann. Physics 146 (1983), no. 1, 1-34. 
\bibitem{m.s} Schechter, M. Spectra of partial differential operators. North-Holland Series in Applied Mathematics and Mechanics, Vol. 14. North-Holland Publishing Co., Amsterdam-London; American Elsevier Publishing Co., Inc., New York, 1971. xiii+268 pp. 
\bibitem{ry1} Roux, Ph.; Yafaev, D. The scattering matrix for the Schr\"odinger operator with a long-range electromagnetic potential. J. Math. Phys. 44 (2003), no. 7, 2762-2786. 
\bibitem{ry2} Roux, Ph.; Yafaev, D. On the mathematical theory of the Aharonov-Bohm effect. J. Phys. A 35 (2002), no. 34, 7481-7492.
\bibitem{Tonomura1} Tonomura, A. ;   Matsuda, T.; Suzuki, R.; Fukuhara, A.; Osakabe, N. ;  Umezaki, H.; Endo, J.; Shinagawa,
K.; Sugita, Y.; Fujiwara, H.Observation of Aharonov-Bohm effect by electron holography . Phys. Rev. Lett. 48 (1982) 1443-1446.
\bibitem{Tonomura2} Tonomura, A.; Osakabe, N. ; Matsuda, T.; Kawasaki, T.; Endo, J.; Yano, S.; Yamada, H. Evidence for Aharonov-Bohm effect with magnetic field completely shielded from electron wave . Phys. Rev.
Lett. 56 (1986) 792-795.
\bibitem{Tonomura3}  Tonomura, A.; Nori, F. Disturbance without the force. Nature
452-20 (2008) 298-299.
\bibitem{Tonomura4}  Tonomura, A. Direct observation of thitherto unobservable quantum phenomena by using electrons. Proc. Natl. Acad. Sci. U.S.A. 102 (2005) 14952-14959.
\bibitem{wa} Warner, F. W. Foundations of differentiable manifolds and Lie groups. Corrected reprint of the 1971 edition. Graduate Texts in Mathematics 94 Springer-Verlag, New York-Berlin, 1983. ix+272 pp. 
\bibitem{w.1} Weder, R. The Aharonov-Bohm effect and time-dependent inverse scattering theory.  Inverse Problems  18  (2002),  no. 4, 1041-1056.
\bibitem{wuyang} Wu, T. T.; Yang, C. N. Concept of nonintegrable phase factors and global formulation of gauge fields. Phys. Rev. D 3 (1975), no. 12, 3845-3857. 
\end{thebibliography}
\end{document}